\documentclass[useAMS,usenatbib]{mnras}

\pdfoutput=0

\usepackage{amssymb}
\usepackage{amsmath}
\usepackage{times}
\usepackage{graphicx}
\usepackage{wasysym}

\def\aap{A\&A}
\def\apj{ApJ}

\def\apjl{ApJ}
\def\pasa{PASA}
\def\mnras{MNRAS}
\def\araa{ARA\&A}
\def\aj{AJ}

\def\apjs{ApJS}
\def\pasp{PASP}

\newcommand{\mincir}{\raise
-2.truept\hbox{\rlap{\hbox{$\sim$}}\raise5.truept 
\hbox{$<$}\ }}
\newcommand{\magcir}{\raise
-2.truept\hbox{\rlap{\hbox{$\sim$}}\raise5.truept
\hbox{$>$}\ }}
\newcommand{\minmag}{\raise-2.truept\hbox{\rlap{\hbox{$<$}}\raise
6.truept\hbox
{$>$}\ }}

\newcommand{\be}{\begin{equation}}
\newcommand{\ee}{\end{equation}}
\newcommand{\ba}{\begin{eqnarray}}
\newcommand{\ea}{\end{eqnarray}}
\newcommand{\brr}{\begin{array}}
 
\newcommand{\err}{\end{array}}
\newcommand{\bc}{\begin{center}}
\newcommand{\ec}{\end{center}}

\DeclareMathAlphabet{\mathsc}{OT1}{cmr}{m}{sc}
\def\testbx{bx}%
\DeclareRobustCommand{\ion}[2]{%
\relax\ifmmode
\ifx\testbx\f@series
{\mathbf{#1\,\mathsc{#2}}}\else
{\mathrm{#1\,\mathsc{#2}}}\fi
\else\textup{#1\,{\mdseries\textsc{#2}}}%
\fi}

\title[The SFRF and CSFRD of $z \sim 0-8$ galaxies]{The evolution of the star formation rate function in the EAGLE simulations: A comparison with UV, IR and H$\alpha$ observations from  ${\bf{z\sim8}}$ to ${\bf{z\sim0}}$.}

\author[A. Katsianis et al.]{A. Katsianis$^{1}$\thanks{E-mail:
     kata@das.uchile.cl }, G. Blanc$^{1}$ $^{2}$, C. P. Lagos $^3$,  N. Tejos$^4$,  R. G. Bower$^5$, A. Alavi$^6$, \newauthor V. Gonzalez$^7$ $^{8}$, T. Theuns$^5$, M. Schaller$^5$ and S. Lopez$^1$ \\
   \\ $^1$ Department of Astronomy, Universitad de Chile, Camino El Observatorio 1515, Las Condes, Santiago, Chile  \\
   $^2$ Observatories  of  the  Carnegie  Institution  for  Science, Pasadena, CA, USA \\
   $^3$ International Centre for Radio Astronomy Research (ICRAR), M468, University of Western Australia, 35 Stirling Hwy, Crawley, WA 6009, Australia \\
   $^4$ Instituto de F\'isica, Pontificia Universidad Cat\'olica de Valpara\'iso, Casilla 4059, Valpara\'iso, Chile \\
   $^5$ Institute for Computational Cosmology, Department of Physics, University of Durham, South Road, Durham, DH1 3LE, UK \\
   $^6$ Department  of  Physics  and  Astronomy,  University  of California, Riverside, CA 92521, USA \\
   $^7$ Centro de Astrofísica y Tecnologías Afines (CATA), Camino del Observatorio 1515, Las Condes, Santiago, Chile \\
   $^8$ Chinese Academy of Sciences South America Center for Astronomy, China-Chile Joint Center for Astronomy, Camino del Observatorio 1515, Las Condes, Chile} 

\begin{document}

\maketitle

\begin{abstract}

We investigate the evolution of the galaxy Star Formation Rate Function (SFRF) and Cosmic Star Formation Rate Density (CSFRD) of $z\sim 0-8 $ galaxies in the Evolution and Assembly of GaLaxies and their Environments (EAGLE) simulations. In addition, we present a compilation of UV, IR and H$\alpha$ SFRFs and compare these with the predictions from the EAGLE suite of cosmological hydrodynamic simulations. We find that the constraints implied by different indicators are inconsistent with each other for the highest star-forming objects at $z < 2$, a problem that is possibly related to selection biases and the uncertainties of dust attenuation effects. EAGLE's feedback  parameters  were  calibrated  to  reproduce realistic galaxy sizes and stellar  masses  at $z =  0.1$. In this work we test if and why those choices yield realistic Star Formation Rates (SFRs) for $z \sim 0-8$ as well. We demonstrate that SNe feedback plays a major role at setting the abundance of galaxies at all star-forming regimes, especially at high redshifts. On the contrary, Active Galactic Nuclei (AGN) feedback becomes more prominent at lower redshifts and is a major mechanism that affects only the highest star-forming systems. Furthermore, we find that galaxies with SFR $\sim 1-10 \, {\rm M_{\odot} \, yr^{-1}}$ dominate the CSFRD at redshifts $z \le5$, while rare high star-forming galaxies (SFR $\sim 10-100 \,{\rm  M_{\odot} \, yr^{-1}}$) contribute significantly only briefly around the peak era ($z \sim 2$) and then are quenched by AGN feedback. In the absence of this prescription objects with SFR $\sim 10-100 \,{\rm  M_{\odot} \, yr^{-1}}$ would dominate the CSFRD, while the cosmic budget of star formation would be extremely high. Finally, we demonstrate that the majority of the cosmic star formation occurs in relatively rare high mass halos ($ {\rm M_{Halo}} \sim 10^{11-13} \, {\rm M_{\odot}}$) even at the earliest epochs.

\end{abstract}


\section{Introduction}
\label{intro}

Within the last decade, large galaxy surveys have allowed us to obtain good insight for a range of properties of galaxies (e.g. stellar masses, star formation rates, metallicities, circular velocities) in a wide redshift range. These provided us with further understanding on how galaxies evolve, allowing us to constrain theoretical models and simulations. The observed number density of star-forming galaxies as a function of their star formation rate, namely the Star Formation Rate Function (SFRF) can give further comparisons with numerical results. The star formation rate, unlike the stellar mass which is a cumulative property, represents an instantaneous census of star formation. Physical mechanisms (e.g. Super-novae and Active Galactic Nuclei feedback) affect the star formation rates of the simulated objects the moment they become prominent. This makes the SFRF an ideal and sensitive test for studying various physical prescriptions.

Galaxy formation and evolution is a complex process that involves various astrophysical  phenomena,  such as the non-linear evolution of dark matter halos, gas cooling, feedback and star formation. Theoretical studies of the evolution of the SFRF and Cosmic Star Formation Rate Density (CSFRD) require detailed modelling. The above have been investigated both by semi-analytic models \citep{Fontanot2012,Gruppioni2016} and hydrodynamic simulations \citep{Dave2011,TescariKaW2013,Katsianis2016}. Typically, the comparison of the models with observations suggested that simulations overproduce galaxies of all kinds of SFRs. According to the authors, the tension with observables could be due to the fact that there was not an implementation of AGN feedback in their models. \citet{TescariKaW2013} and \citet{Katsianis2016} studied the role of feedback from SNe and AGN in the evolution of the star formation rate function of z $\sim1-7$ galaxies using the set of cosmological hydrodynamic simulations, ANGUS (AustraliaN GADGET-3 early Universe Simulations). The authors found that a key factor for reproducing the observed distribution of SFRs is the presence of SNe feedback which is prominent at high redshifts ($z \ge 4$) and become less efficient with time. \citet{Gruppioni2016}, compared  predictions of state-of-the-art semi-analytic models of galaxy formation and evolution  \citep[e.g.][]{Monaco2007,Henriques2015} with observations of the PACS Evolutionary Probe and Herschel Multi-tiered Extragalactic Survey data sets in the COSMOS and GOODS-South fields. The comparison showed that semi-analytic models underpredict the bright end of the SFRF at $z \ge 2$. According to the authors the cause of this underprediction could be due to improper numerical implementation of AGN or stellar feedback which may be too efficient for the bright star-forming objects in the models. 

The Virgo project Evolution and Assembly of GaLaxies  and  their  Environments  simulations \citep[EAGLE,][]{Schaye2015,Crain2015} is a suite of cosmological hydrodynamical simulations in cubic, periodic volumes ranging from 25 to 100 comoving Mpc per side. They track the evolution of baryonic gas, stars, massive  black  holes  and  non-baryonic dark  matter particles from a starting  redshift  of $z = 127$  down to $z = 0$. The different runs were performed to investigate the effects of resolution, box size and various physical prescriptions (e.g. feedback and metal cooling). The EAGLE reference simulation has $2 \, \times \, 1504^{3}$ particles (an  equal  number of gas and dark matter elements) in an L =  100 comoving Mpc volume box, initial gas particle mass of $m_{g} =  1.81 \times 10^6 \, M_{\odot}$, and mass of dark matter particles of $m_{g} =  9.70 \times 10^6 \, M_{\odot}$. It is one of the highest resolution cosmological hydrodynamic simulations ran in such a large volume \citep{Vogelsberger2014,Schaye2015}. It has been calibrated to reproduce observational datasets, such as the present-day stellar mass function of galaxies, the correlation between the black hole and masses and the dependence of galaxy sizes on mass. Alongside with these key properties, the simulation was able to match many other observed properties of galaxies at various eras, like molecular hydrogen abundances \citep{Lagos2015}, colors and luminosities at $z \sim 0.1$ \citep{Trayford2015}, supermassive black hole mass function \citep{Rosas2016}, angular momentum evolution \citep{Lagos2017}, atomic hydrogen abundances \citep{Crain2017} and sizes \citep{Furlong2017}. Therefore the EAGLE simulations can provide a powerful resource for understanding the star formation rates of galaxies and their evolution across cosmic time. In addition, the SFRF was not used to tune the models, so it can be seen as an independent test for the predictions from the simulations.

In this paper we use the EAGLE cosmological suite of simulations to study the evolution of the star formation rate function and cosmic star formation rate density at $z \sim 0-8$. In section \ref{dustcorrectionlaws} we present the compilation of the observed luminosity functions and dust correction laws used for this work alongside with the methodology employed to obtain the galaxy SFRFs\footnote{In the Appendix \ref{table} we present detailed tables of these constraints.}. In Section \ref{thecode} we present a brief description of the EAGLE simulations along with the subgrid model used to describe star formation.  In section \ref{SFRFEagle} we compare the simulated EAGLE SFRFs and CSFRD with the constrains from the observations\footnote{ In Appendix \ref{ResBox} we focus on the effects of resolution and boxsize}. In section \ref{SFRFhalos}  we investigate the contribution of halos with different masses to the SFRF and CSFRD. In section \ref{Feed} we present the SNe and AGN feedback implementations in EAGLE and explore their effect on the simulated SFRF. Finally, in section \ref{concl6} we summarize our main results and conclusions.

\section{The observed star formation rates from galaxy Luminosities}
\label{dustcorrectionlaws}

There has been a considerable effort to estimate the SFRF and Cosmic Star Formation Rate Density in the literature \citep{Menard2011,smit12,Madau2014,Katsianis2016}. Different groups typically rely on the observed luminosities and Luminosity Functions (LFs) with the commonly tracers being the UV, H$\alpha$ and IR luminosities. Evolutionary synthesis models provide the relations between the SFR per unit mass, luminosity and the integrated color of the population. For the case of the UV luminosity the relation to SFR \citep{Ken98a} is found to be
\begin{eqnarray}
\label{eq:SFRpara3}
{\rm SFR}_{\rm UV}  \, ({\rm M}_{\rm \odot} \, {\rm yr^{-1}}) = 0.77 \times 10^{-28} \, \, {\rm L}_{\rm UV}\, ({\rm ergs} \, {\rm s^{-1}} \, {\rm Hz}^{-1})\, , 
\end{eqnarray}
where ${\rm L}_{\rm UV}$ is the UV luminosity of galaxies. The relation is valid from 1500 to 2800 \AA and assumes a \citet{chabrier03} IMF\footnote{Originally the conversion reported by \citet{kennicutt1998}  assumed a \citet{salpeter55} IMF, which produced 1.8 higher SFRs}. A disadvantage of UV light is that it is subject to dust attenuation effects and thus dust corrections are necessary to obtain the intrinsic luminosities, which can then be used to estimate the intrinsic SFRs \citep{smit12}. UV LFs usually provide information for a large number of galaxies at redshifts $z > 2$. At lower redshifts UV samples give key constrains only for low and intermediate star-forming objects \citep{Katsianis2016}. However, there are other tracers that provide information about dusty, high star-forming systems. For example, the Infra-Red (IR) luminosity originating from dust continuum emission is a star formation indicator which at the same time is a good test of dust physics \citep{Hira03}.  The relation between the SFR and total IR luminosity from the evolutionary synthesis model of \citet{kennicutt1998} is found to be:
\begin{eqnarray}
\label{eq:SFRpara2}
{\rm SFR}_{\rm IR} \, ({\rm M}_{\rm \odot} \, {\rm yr^{-1}} )= 10^{-10} \, \, {\rm L}_{\rm IR} \,/{\rm L}_{\rm \odot},
\end{eqnarray}
where ${\rm L}_{\rm IR}$ is the IR luminosity integrated between 8 and 1100 microns of galaxies assuming a \citet{chabrier03} IMF. IR light is typically used to estimate obscured SF, however, small  galaxies with low metallicities do not have enough dust to reprocess the UV light to IR, so IR LFs do not probe the faint end of the SFRF a problem that is quite notable and proportional to redshift \citep{Katsianis2016}. IR light is usually employed as a probe of the dust corrections that have to be used in the observed dust-attenuated UV luminosities. However, it has been discussed in the literature that IR calibrations may overestimate the SFRs of galaxies and the dust corrections they imply for UV light  \citep[e.g.][]{Papov11,Elbaz10,Bauer11,Fumagalli2014,Utomo2014,Hayward2014,Katsianis2015}. Some of the main reasons for overestimating the SFR measured from IR light could be the following:
\begin{itemize} 
\item  buried AGN that boosts the IR luminosity,
\item dust can be heated by old populations not related to recent star-formation, 
\item  converting the IR luminosity into SFR relies on assumptions that possibly do not hold for all galaxies. For example, the IR luminosity can overestimate the instantaneous SFR during the post-starburst phase by up to two orders of magnitude \citep{Hayward2014}. Even though the instantaneous SFR decreases rapidly after the starburst, the stars that were formed in the starburst can remain dust-obscured and thus produce a significant IR luminosity non related to new born stars,
\item  larger polycyclic aromatic hydrocarbons emission of distant galaxies in 24 $\mu m$ observations.
\end{itemize}
The ongoing Herschel mission is able to sample the IR peak and total IR-SED of numerous galaxy spectra, helping to derive more accurate ${\rm L_{IR}}$ values, thus the above problems could be diminished in the near future.

In addition to UV and IR light, H$\alpha$ photons, which are produced from the gas ionized by the radiation of young and massive stars can be used to trace the intrinsic SFR of an object \citep{Hanish2006,Bell07,Ly11,Sobral2013}. According to the synthesis models of \citet{kennicutt1998}, the relation between SFR and H$\alpha$ luminosity is:
\begin{eqnarray}
\label{eq:SFRpara1}
{\rm SFR}_{\rm H\alpha} \, ({\rm M}_{\rm \odot} \, {\rm yr^{-1}} )= 4.4 \times 10^{-42} \, \, {\rm L}_{\rm H\alpha}\, (ergs \, {\rm s^{-1}})\, ,
\end{eqnarray}
where ${\rm L}_{\rm H\alpha}$ is the H$\alpha$ luminosity of the galaxies and conversion assumes a \citet{chabrier03} IMF. H$\alpha$ light is subject to dust attenuation effects so dust corrections are necessary to calculate the intrinsic SFR of the target \citep{Hopkins01,Sobral2013,Zheng2014}.  We note that all the above methods which employ the UV, IR and H$\alpha$ luminosities to measure star formation do not measure instantaneous SFRs, but instead, measure the time-averaged quantity. The different luminosities have their origins in stars of different masses and differences between the different tracers are expected \citep{Lee2009}. For example, H$\alpha$ traces very massive stars, while UV is tracing lower mass stars ($\sim 3 \, M_{\odot}$). IR light can be even more troublesome as it depends on the dust abundance and composition of the target. Due to the above problems a $50\%$ systematic uncertainty in the calibrations given in equations \ref{eq:SFRpara3}, \ref{eq:SFRpara2} and \ref{eq:SFRpara1} can be expected.

To construct the SFRFs for redshift $z \sim 8.0$ to $z \sim 0$ we use a range of UV, IR, H$\alpha$ and radio LFs. We do so since this combination enable us to study the SFRF in a large range of SFRs and redshifts. In addition, we employ the SFRFs presented by \citet[][$z \sim4-7$]{smit12}, \citet[][$z \sim4-7$]{Duncan2014} and \citet[][$z \sim1-4$]{Katsianis2016}.  \\

UV light is subject to dust attenuation effects. We follow the method described in \citet{smit12} and \citet{Katsianis2016} to correct the UV luminosity bins of the UV luminosity function. Like \citet{smit12}, we assume the infrared excess (IRX)-$\beta$ relation of \citet{meurer1999}:
\begin{eqnarray}
  A_{\rm 1600} = 4.43 + 1.99\,\beta,
  \label{eq_A16}
\end{eqnarray}
where $A_{\rm 1600}$ is the dust absorption at 1600 $\AA$ and $\beta$ is the UV-continuum spectral slope. We assume a linear relation between $\beta$ and the luminosity \citep{bouwens2012}:
\begin{eqnarray}
  \langle\beta\rangle=\frac{{\rm d}\beta}{{\rm d}M_{\rm UV}}\left(M_{\rm
      UV,AB}+19.5\right)+\beta_{M_{\rm UV}=-19.5},
  \label{eq_beta}
\end{eqnarray}
Then, following \citet{HaoKen} we assume
\begin{eqnarray}
  {\rm L_{\rm UV_{OBS}}} = {\rm L_{\rm UV_{corr}}e^{-\tau_{UV}}},
  \label{eq_A17}
\end{eqnarray}
where ${\rm \tau_{UV}}$ is the effective optical depth (${\rm \tau_{UV}}={\rm A_{\rm 1600}/1.086}$). H$\alpha$ emission is also subject to dust attenuation effects. \cite{Sobral2013} used a  1 mag correction at the bins of the luminosity function, while \cite{Ly11} use the SFR dependent dust correction suggested by \cite{Hopkins01}. We present the results of both authors.

Finally to obtain the intrinsic star formation rate functions we convert the luminosity bins of the dust corrected LFs and the Kennicutt relations (eq. \ref{eq:SFRpara3}, \ref{eq:SFRpara2} and \ref{eq:SFRpara1}). We present these distributions alongside with the EAGLE reference SFRFs in Fig. \ref{fig:EvolutionSFRF} and \ref{fig:EvolutionSFRF42}. Tables presenting these determinations can be found in the Appendix \ref{table}.

\section{Star formation in the EAGLE simulations}
\label{thecode}

\begin{table*}
\centering
\resizebox{0.95\textwidth}{!}{%
\begin{tabular}{llccccccc}
  \\ \hline & Run & L & N$_{\rm TOT}$ & m$_{\rm DM}$ & m$_{\rm GAS}$  &
  $\epsilon_{com}$ & Resolution & Feedback \\ & & [Mpc] &
  & [M$_{\rm \odot}$] & [M$_{\rm \odot}$] & [kpc] & (Ref) \\ \hline
  & L100N1504-Ref & 100 & 2 $\times$ $1504^3$ & 9.70$\times10^{6}$ &
  1.81$\times10^{6}$ & 2.66 & 1.0 & AGN $+$ SNe \\ \hline
  & L50N752-Ref & 50 & 2 $\times$ $752^3$ & 9.70$\times10^{6}$ & 1.81$\times10^{6}$ & 2.66 & 1.0 & AGN $+$ SNe  \\ \hline
  & L50N752-NoAGN & 50 & 2 $\times$ $752^3$ & 9.70$\times10^{6}$ &
  1.81$\times10^{6}$ & 2.66 & 1.0 & No AGN $+$ SNe\\ \hline
  & L25N376-Ref & 25 & 2 $\times$ $376^3$ & 9.70$\times10^{6}$ &
  1.81$\times10^{6}$ & 2.66 & 1.0 & AGN $+$ SNe \\ \hline
  & L25N376-WeakSNfb & 25 & 2 $\times$ $376^3$ & 9.70$\times10^{6}$ &
  1.81$\times10^{6}$ & 2.66 & 1.0 & AGN $+$ Weak SNe \\ \hline
  & L25N376-StrongSNfb & 25 & 2 $\times$ $376^3$ & 9.70$\times10^{6}$ &
  1.81$\times10^{6}$ & 2.66 & 1.0 & AGN $+$ Strong SNe\\ \hline
  & L25N752-Ref & 25 & 2 $\times$ $752^3$ & 1.21$\times10^{6}$ &
  2.26$\times10^{5}$ & 1.33 & 8.0 & AGN $+$ SNe\\ \hline
  & L25N188-Ref & 25 & 2 $\times$ $188^3$ & 7.76$\times10^{7}$ &
  1.45$\times10^{7}$ & 5.32 & 0.125 & AGN $+$ SNe\\ \hline
  & L25N752-Recal & 25 & 2 $\times$ $752^3$ & 1.21$\times10^{6}$ &
  2.26$\times10^{5}$ & 1.33 & 8.0 & AGN $+$ SNe Recal\\ \hline
\end{tabular}%
}
\caption{Summary of the different EAGLE simulations used in this work. Column 1, run name; column 2, Box size of the simulation in comoving Mpc; column 3, total number of particles (N$_{\rm TOT} =$ N$_{\rm GAS}$ $+$ N$_{\rm DM}$ with N$_{\rm GAS}$ $=$ N$_{\rm DM}$); column 4, mass of the dark matter particles; column 5, initial mass of the gas particles; column 6, comoving gravitational softening length; column 7, Resolution with respect the reference L100N1504-Ref run (higher ratios correspond to higher resolution); column 8, combination of feedback implemented. The L25N752-Recal run is a configuration in which the feedback prescriptions are re-calibrated to obtain the observed galaxy stellar mass function at $z \sim0$. The run is employed to test the convergence of the EAGLE simulations}
\label{tab:sim_runs}
\end{table*}

In this section we present an overview of the EAGLE set of simulations and how star formation is implemented.  The  cosmological  parameters  assumed in all runs are those reported by the \citet{Planck2014,Planck2014b} with the average densities of matter, dark energy and baryonic matter in units of the critical density at redshift 0 being $\Omega_{m} =  0.307$, $\Omega_{\lambda} =  0.693$ and $\Omega_{b} =  0.04825$ respectively, Hubble parameter ${\rm h=H_{o}/100 \, km \, s^{-1} \, Mpc^{-1}} = 0.6777$,  square root of the linear variance of the matter distribution $\sigma_8 =  0.8288$, scalar power-law index of the power spectrum of primordial adiabatic perturbations $n_s =  0.9611$ and primordial abundance of helium $Y = 0.248$. The initial conditions were generated using the transfer function of the CAMB software \citep{Lewis2000} and the perturbation theory as described by \citet{Jenkins2013}. The simulations  were run using an improved and updated version  of  the N-body  TreePM  smoothed  particle  hydrodynamics code GADGET-3 \citep{Springel2005}. The subgrid routines which describe subgrid physics like star formation and stellar mass loss are updated versions of those used for the  GIMIC \citep{Crain2009},  OWLS  \citep{schaye10}  and cosmo-OWLS \citep{LeBrun2014} projects.  In this study we focus  on `intermediate-resolution  simulations' \cite[as labeled by][]{Schaye2015}  using  volumes of side $L = 25$ , $50$ and $100$ Mpc. We also use low-resolution and high-resolution runs with different box sizes to address the effects of resolution and volume. In Table \ref{tab:sim_runs} we present the EAGLE simulations used for this work.

Star formation occurs in cold ($T \leq 10^4 \, K$), high density gas. Cosmological simulations, at present, lack the resolution and the detailed physics to model the cold, inter-stellar phase. To overcome this limitation, EAGLE employs the star formation recipe of \citet{Schaye2008}. In this scheme, gas with densities exceeding the critical density for the onset of the thermo-gravitational instability ($n_H \sim 10^{-2} - 10^{-1} \, cm^{-3}$) is treated as a multi-phase mixture of cold molecular clouds, warm atomic gas and hot ionized bubbles which are all approximately in pressure equilibrium \citep{Schaye2004}. The above mixture is modelled using a polytropic equation of state  $P = k \rho ^{\gamma _{eos}}$, where P is the gas pressure, $\rho$ is the gas density and $k$ is a constant which is normalized to $P/k = 10^3 \, cm^{-3} \, K$ at the density threshold $n_H^{\star}$ which marks the onset of star formation. The hydrogen number density, $n_H$, is related to the overall gas density, $\rho$, via $n_H = X \, \rho/m_{H}$, where X is the hydrogen mass fraction ($X = 0.752$) and $m_H$ is the mass of a hydrogen atom. The threshold $n_{H}^{\star}$ was set as $0.1 \, cm^{-3}$ in the OWLS simulations in accordance with theoretical considarations \citep{Schaye2008} and is comparable with other work in the literature \citep{springel2003,Vogelsberger2013}. In addition to the above density dependent criterion, the star formation model employed in this work takes into account that the transition from the warm phase to the cold occurs more efficiently in metal-rich environments. Thus, EAGLE simulations adopt the metallicity-dependent star formation threshold proposed in \citet{Schaye2004}, which is:
\begin{eqnarray}
\label{threashold}
 n^{\star}_H (Z) = 0.1 \, cm^{-3} (\frac{Z}{0.002})^{-0.64},
\end{eqnarray}
where Z is the gas metallicity. Besides the density threshold, gas has to fulfill the requirement of being cold. Following \citet{DVecchia2012}, gas is eligible to form stars if its temeperature $log_{10}T \leq log_{10}T_{crit} = log_{10}T_{eos}+0.5$, where $T_{eos}$ is the temperature floor and fulfill the metalicity-dependent density criterion described by eq. \ref{threashold}. Finally, to prevent star formation in low-overdensity gas at high redshift, there is an additional criterion for star-forming gas to have an overdensity $\delta > 57.7$.

After the above criteria determine which gas particles are eligible to be star-forming the \citet{Schaye2008} scheme employ the observed Kenicutt-Schmidt low \citep{Schmidt1959,kennicutt1998} to describe star formation. Under the assumption that the gas is self-gravitating the Kennicutt-schmidt star-forming law can be written as:
\begin{eqnarray}
\label{9}
 \dot{\Sigma_{\star}} = A \, (\frac{\Sigma_{g}}{1 \, M_{\odot} \, pc^{-2}})^n,
\end{eqnarray}
where $\Sigma_{\star}$ and $\Sigma_{g}$ are the surface density of stars and gas, respectively. Assuming a polytropic equation of state \ref{9} can be re-written in a pressure dependent form:
\begin{eqnarray}
\label{SFRpoly}
 \dot{m_{\star}} = m_{g} \, A \, (1 \,  M_{\odot} \, pc^{-2} )^{-n} (\frac{\gamma}{G} \, f_{g} \, P)^{(n-1)/2},
\end{eqnarray}
where $m_{g}$ is the mass of the gas particle for which we are computing $\dot{m_{\star}}$,  $n =1.4 $, $A = 1.515 \times 10^{-4} \dot{m_{\star}} yr^{-1} Kpc^{-2}$ $\gamma = 5/3$ is the ratio of specific heats for a monoatomic gas, G is the gravitational constant, $f_{g}$ is the mass fraction in gas, and P is  the total  pressure.

Equations \ref{threashold} and \ref{SFRpoly} define the algorithm to calculate the rate at which gas is converted into stars. Star particles are to be interpreted as Simple Stellar Populations (i.e. an assembly of coeval, initially chemically homogeneous single stars) with an initial mass, age and chemical composition originating from its progenitor gas particle following a \citet{chabrier03} IMF in the range 0.1 - 100 $M_{\odot}$. The stellar evolution and chemical enrichment are described in \citet{wiersma09b}. We follow the metal recycling by massive stars (Type II SNe, stellar winds), intermediate mass stars (Type Ia SNe) and Asymptotic Giant Branch (AGB) stars, of the 11 elements that contribute significantly to the radiative cooling rates using the nucleosynthetic yields from \citet{Portninari1998} and \citet{Marigo2001}. At each time step and for each stellar particle, the stellar mass reaching the end of the main sequence phase is identified using the metallicity-dependent lifetimes of \citet{Portninari1998}.

In this work, galaxies and their host halos are identified by a friends-of-friends (FoF) algorithm \citep{Davis1985} followed by the SUBFIND algorithm \citep{springel2001,DolagSta2009} which is used to identify substructures or subhalos across the simulation. The star formation rate of each galaxy is defined to be the sum of the star formation rate of all gas particles that belong to the corresponding subhalo and that are within a 3D aperture with radius 30 kpc \citep{Schaye2015,Crain2015}.

\section{The evolution of the simulated and observed star formation rates.}
\label{SFRFEagle}

\subsection{The evolution of the star formation rate function.}
\label{SFRFEAGLE0}

\begin{figure*}
\centering
\includegraphics[scale=0.40]{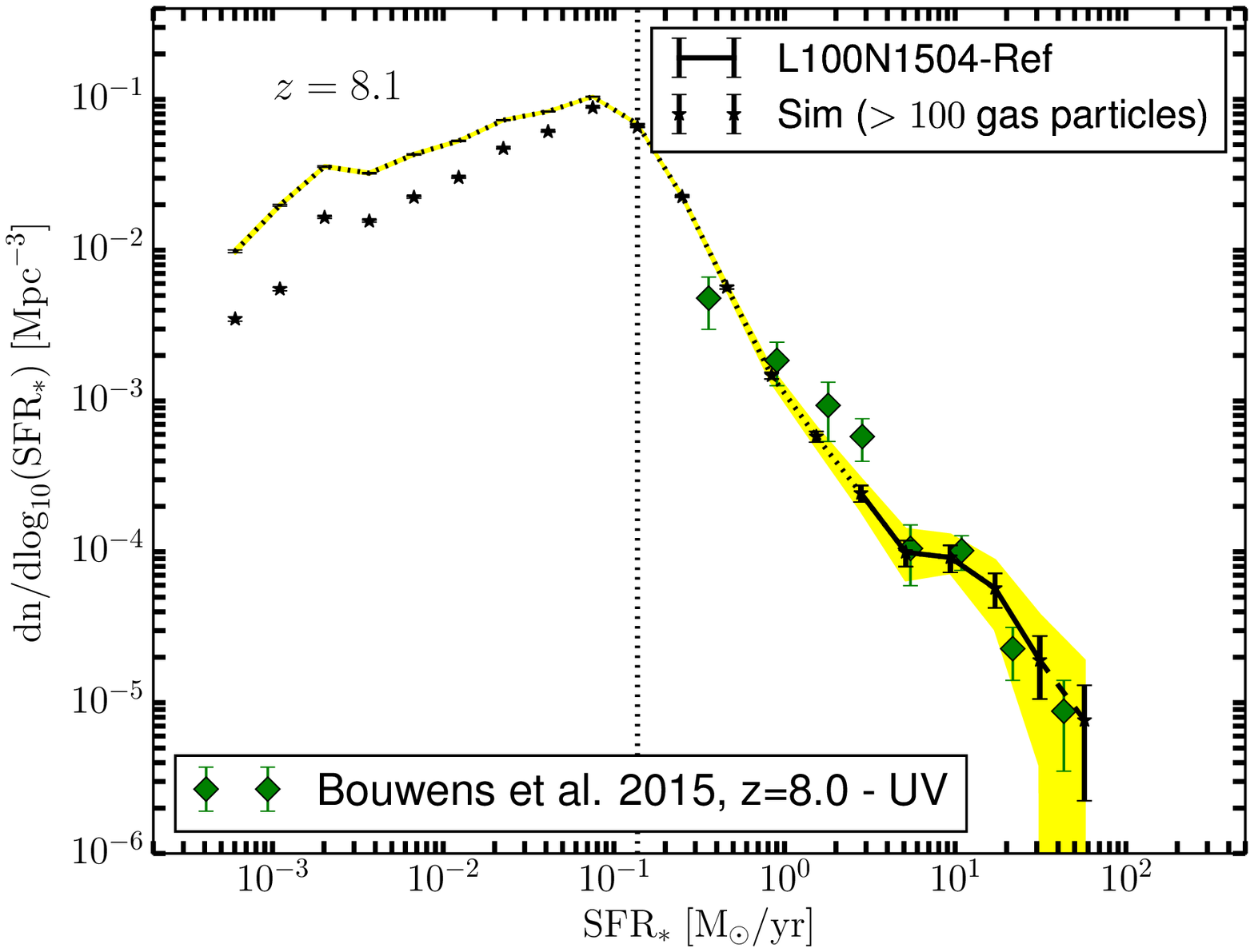}
\includegraphics[scale=0.40]{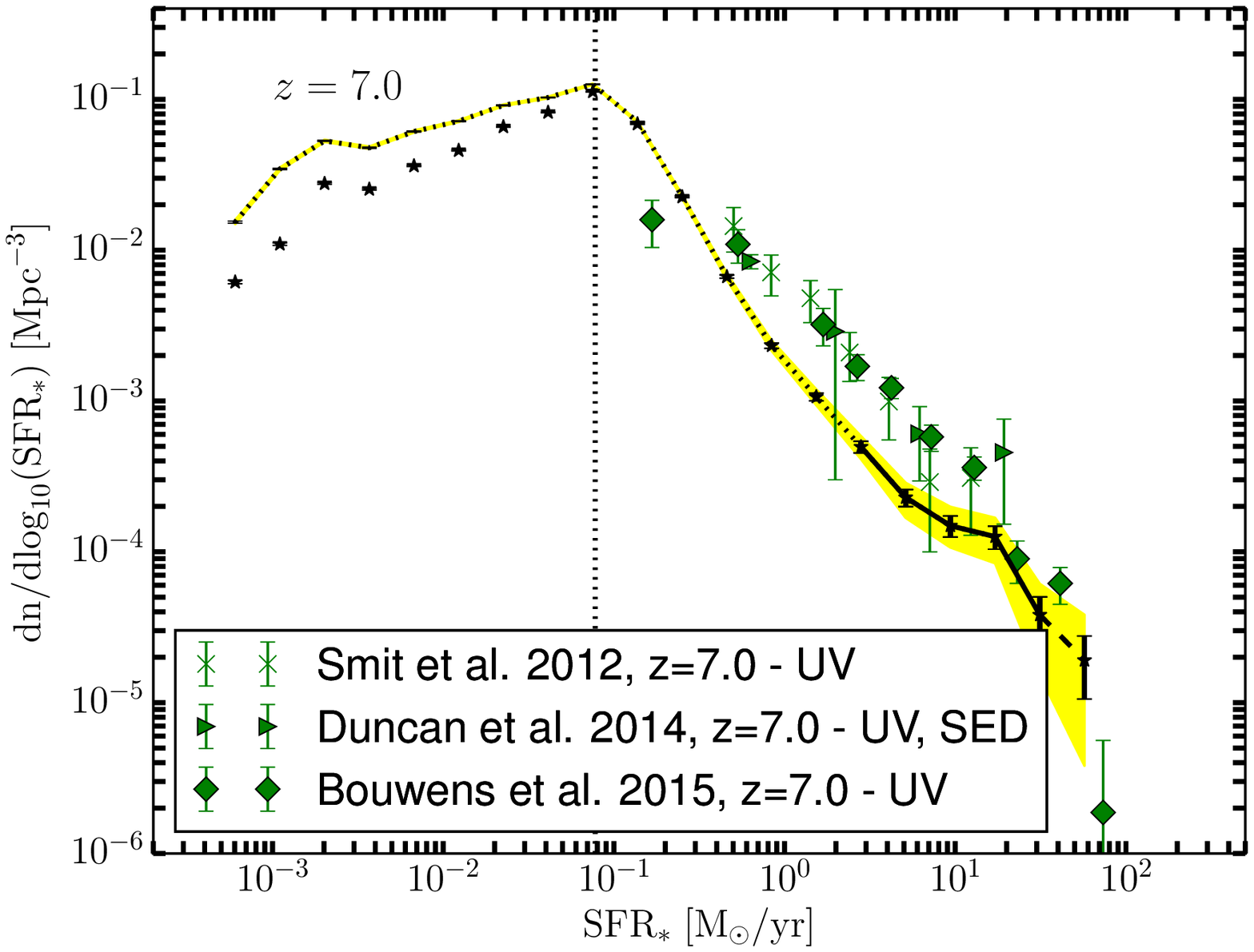}
\includegraphics[scale=0.40]{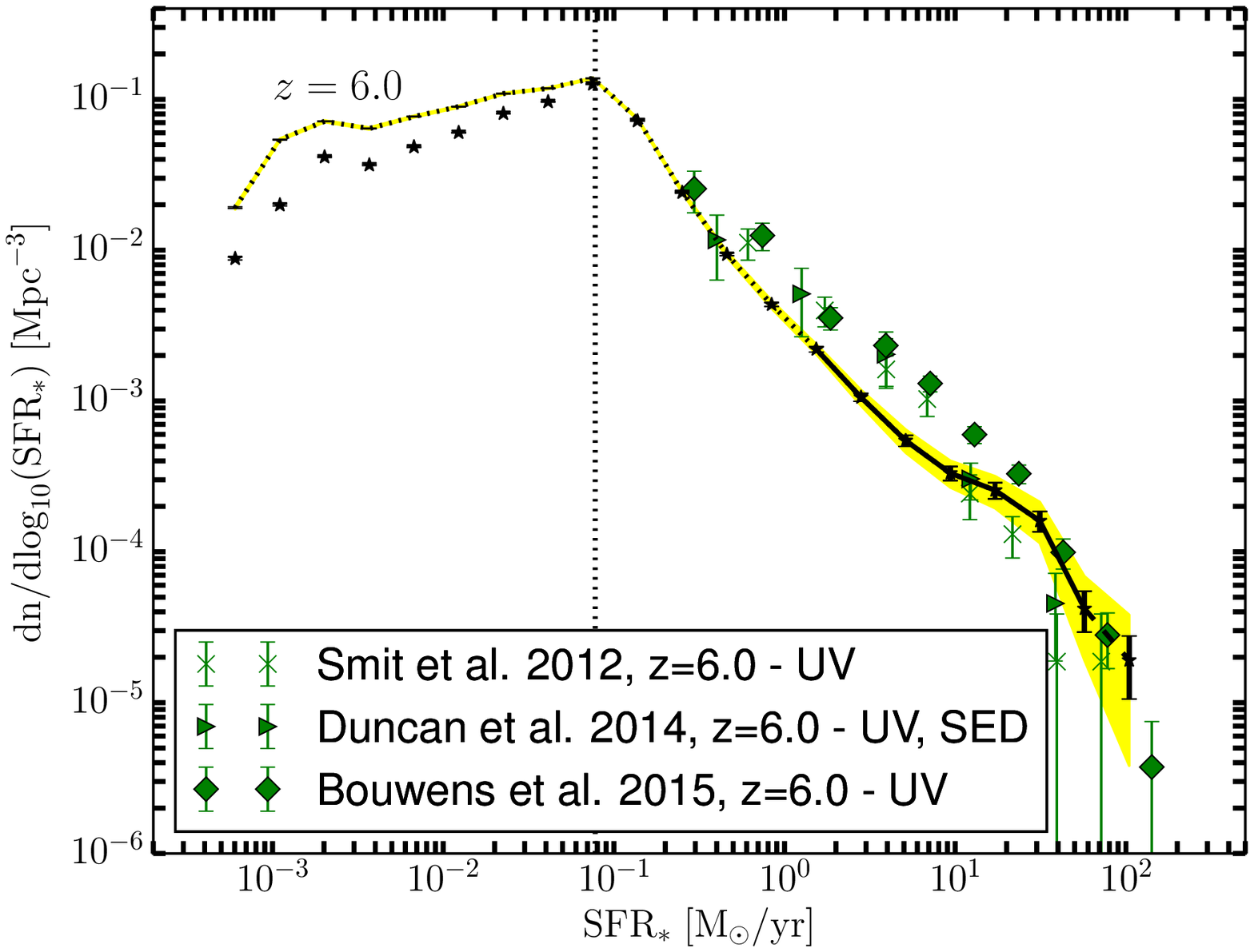}
\includegraphics[scale=0.40]{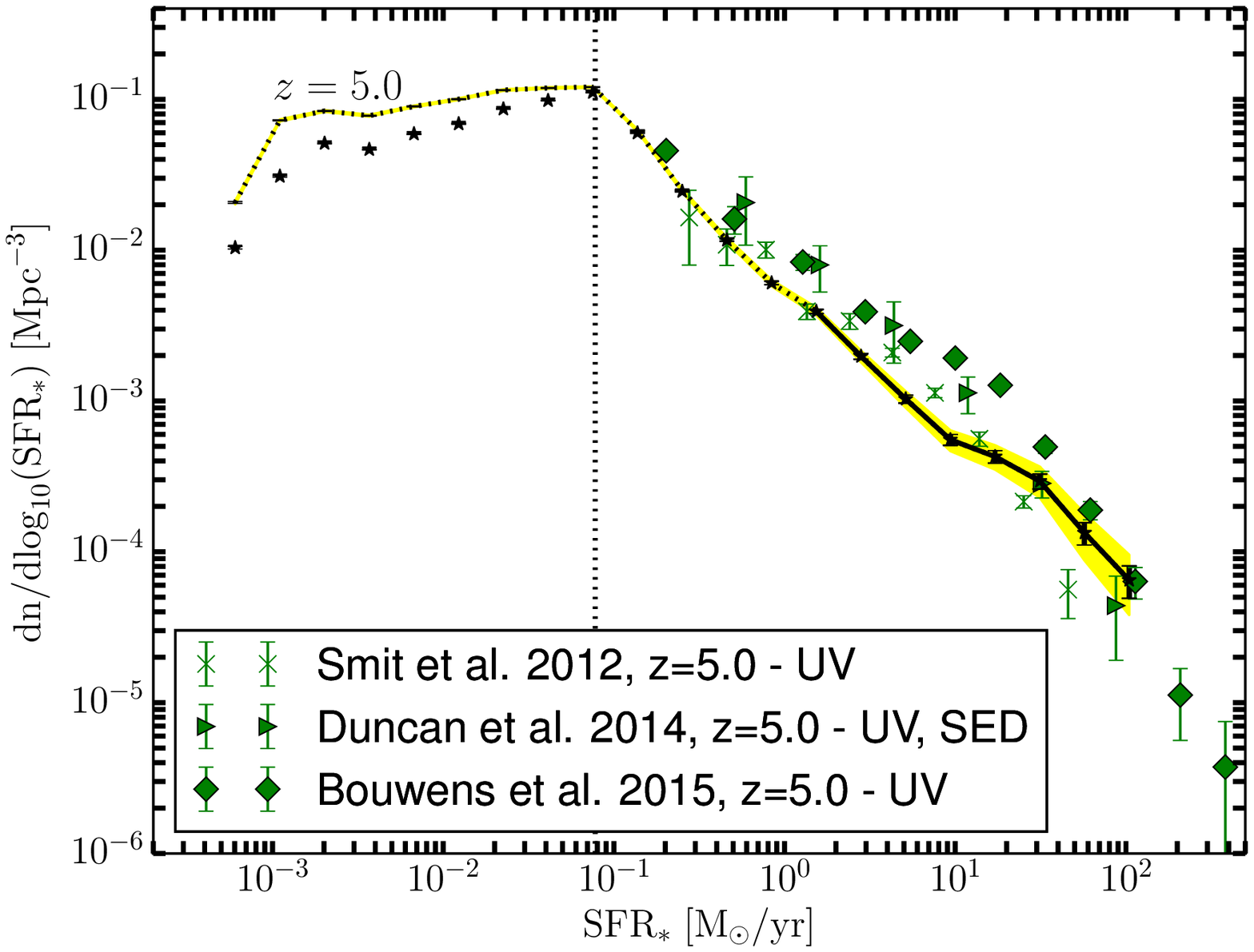}
\includegraphics[scale=0.40]{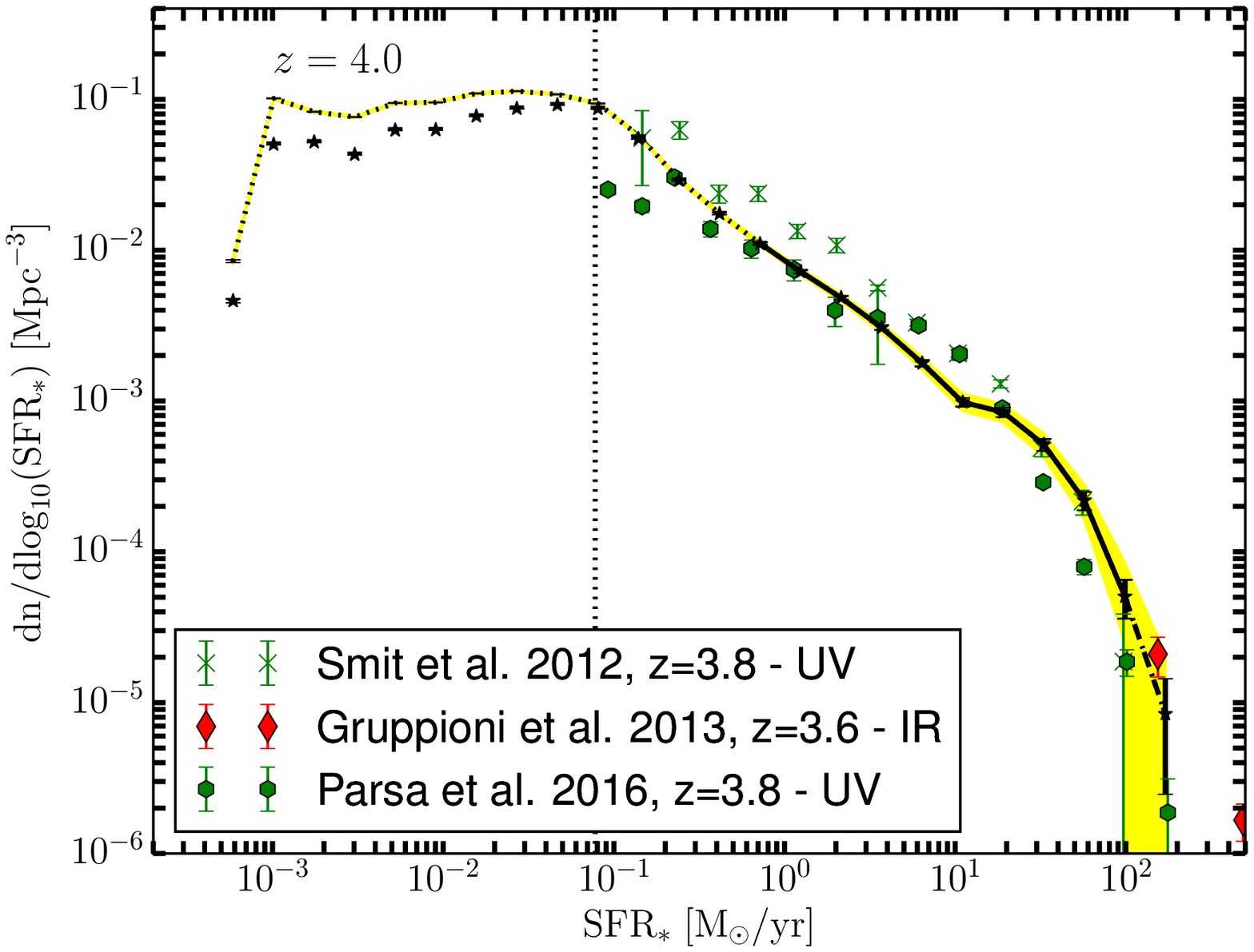}
\includegraphics[scale=0.40]{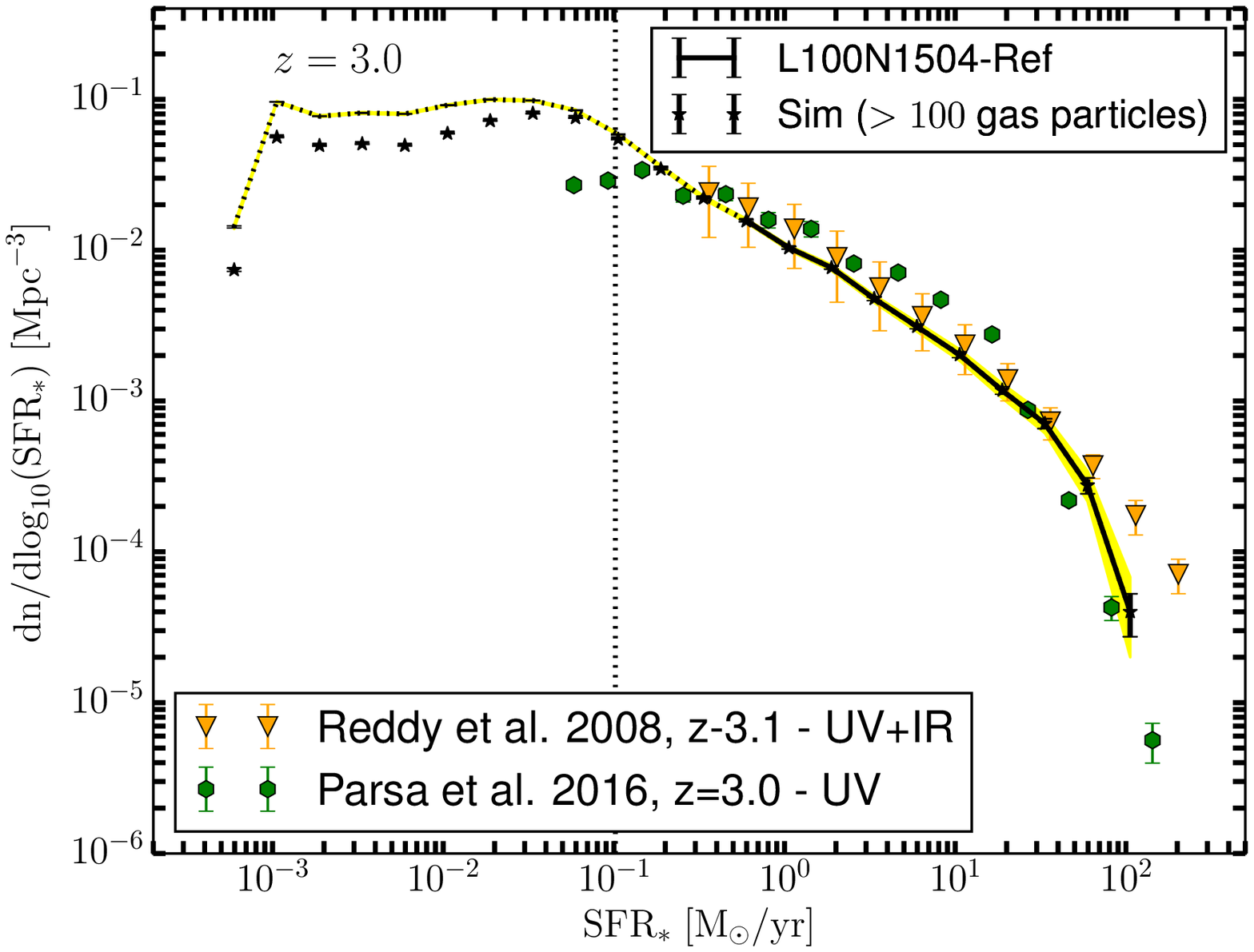}
\caption{Comparison between the L100N1504-Ref SFRF (black dashed line) and observations for redshifts $z \sim 8.0$ (top left panel), $z \sim 7.0$ (top right panel), $z \sim 6.0$ (middle left panel), $z \sim 5.0$ (middle right panel), $z \sim4.0$ (bottom left panel) and $z \sim3.0$ (bottom right panel). For all observations a \citet{chabrier03} IMF and $\Lambda$CDM cosmology same as EAGLE is assumed.  Dark green crosses were taken from the work of \citet{smit12}. The dark green triangles represent the results of \citet{Duncan2014}. We note that an uncertainty of $50 \%$ in the Keniccut calibrations could lead the estimates for the observed SFR to move left or right in the plots by $\sim 0.3$ dex. In the Appendix \ref{table} we present detailed tables of the constraints we obtained using UV, IR and H$\alpha$ LFs. The yellow area represents the 95$\%$ bootstrap confidence interval for 1000 re-samples of the EAGLE SFRs, while the black errorbars represent the 1 sigma poissonian errors. Following the convention by \citet{Schaye2015} when a bin of the EAGLE SFRF contains objects with stellar masses below the mass limit of 100 (initial mass, $m_{GAS}$) baryonic particles curves are dotted, when there are fewer than 10 galaxies curves are dashed. To describe the limits due to poor sampling of gas particles we present the SFRF of objects which contain more than 100 gas particles (black stars+Vertical dotted line). We see that at high redshifts the EAGLE SFRF can give insights mainly for intermediate and high star-forming objects.}
\label{fig:EvolutionSFRF}
\end{figure*}

In this section we present the evolution of the star formation rate function of the EAGLE reference model alongside with constrains from observations.  At high redshifts ($z \gtrsim 4$) usually only the UV-emission from galaxies is observable with the available instrumentation. Thus, the SFRFs at high redshifts rely mostly on UV-selected samples in the literature. For example, \citet{smit12} used the \citet{kennicutt1998} UV-SFR conversion and the luminosity dependent dust corrections of \citet{meurer1999} to transform the UV LFs of \citet[][$z \sim4-6$]{bouwens2007} and \citet[][$z\sim7$]{bouwens2011} into SFRFs. The dark green crosses of Fig. \ref{fig:EvolutionSFRF} represent the results described above. The dark green triangles of Fig. \ref{fig:EvolutionSFRF} represent the SFRFs of \citet{Duncan2014}, which were obtained following the Spectral Energy Distribution (SED) fitting technique \citep{Bruzual2003,Rolling2013}. In addition, we use the UV luminosity functions from \citet[][$z \sim4-10$]{Bouwens2016} and the dust corrections described in subsection \ref{dustcorrectionlaws}. The dark green filled diamonds of Fig. \ref{fig:EvolutionSFRF} show these results. \citet{Bouwens2016} combined the CANDELS, HUDF0$-$9, HUDF1$-$2, ERS, and BoRG/HIPPIES programs to map the evolution of the UV LF. The updated LF determinations reach lower magnitudes $\sim -16$ AB mag and agree well with previous estimates. However, the larger samples and volumes give a more reliable sampling, especially at the characteristic luminosity $L^{\star}$. 

\begin{figure*}
\centering
\includegraphics[scale=0.40]{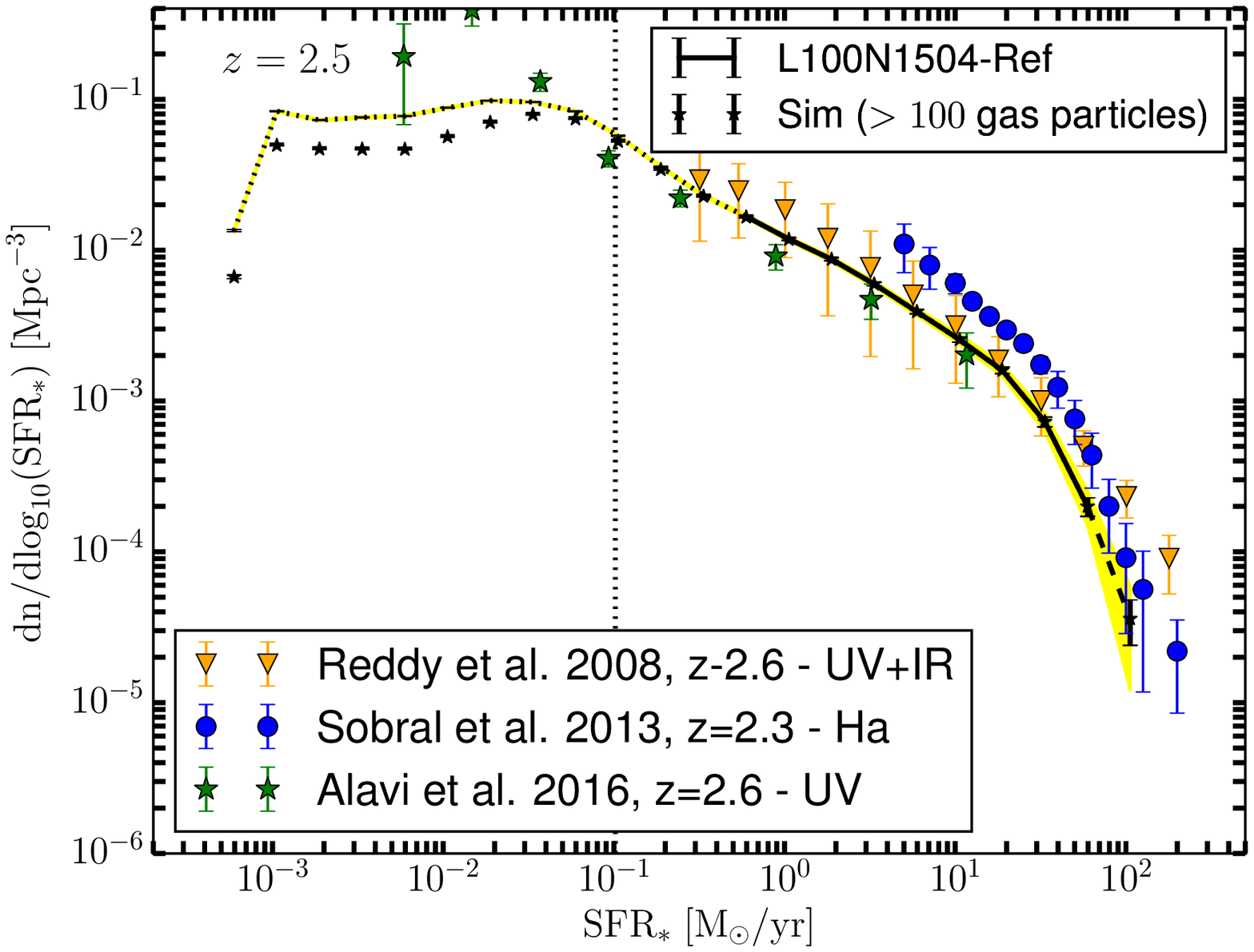}
\includegraphics[scale=0.40]{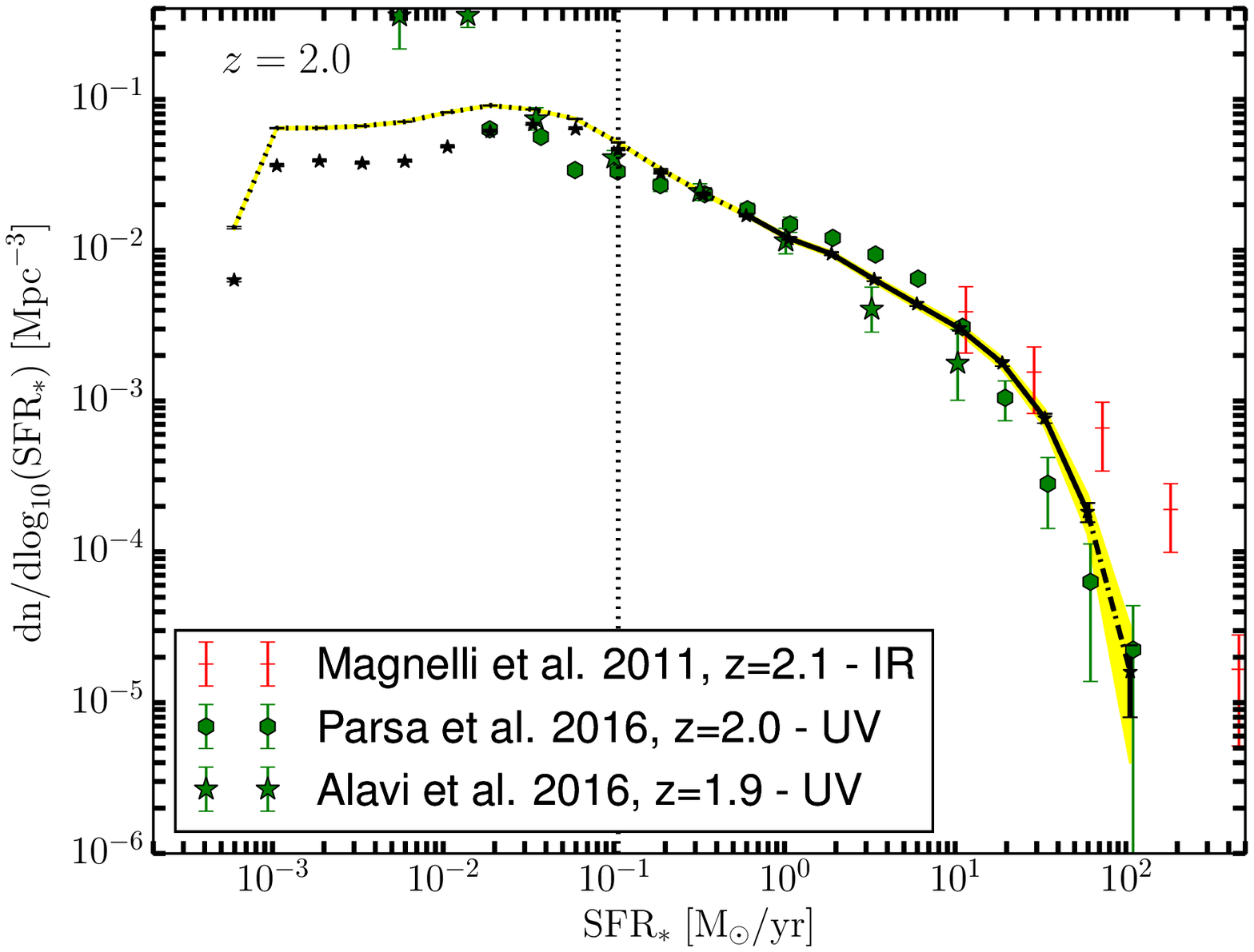}
\includegraphics[scale=0.40]{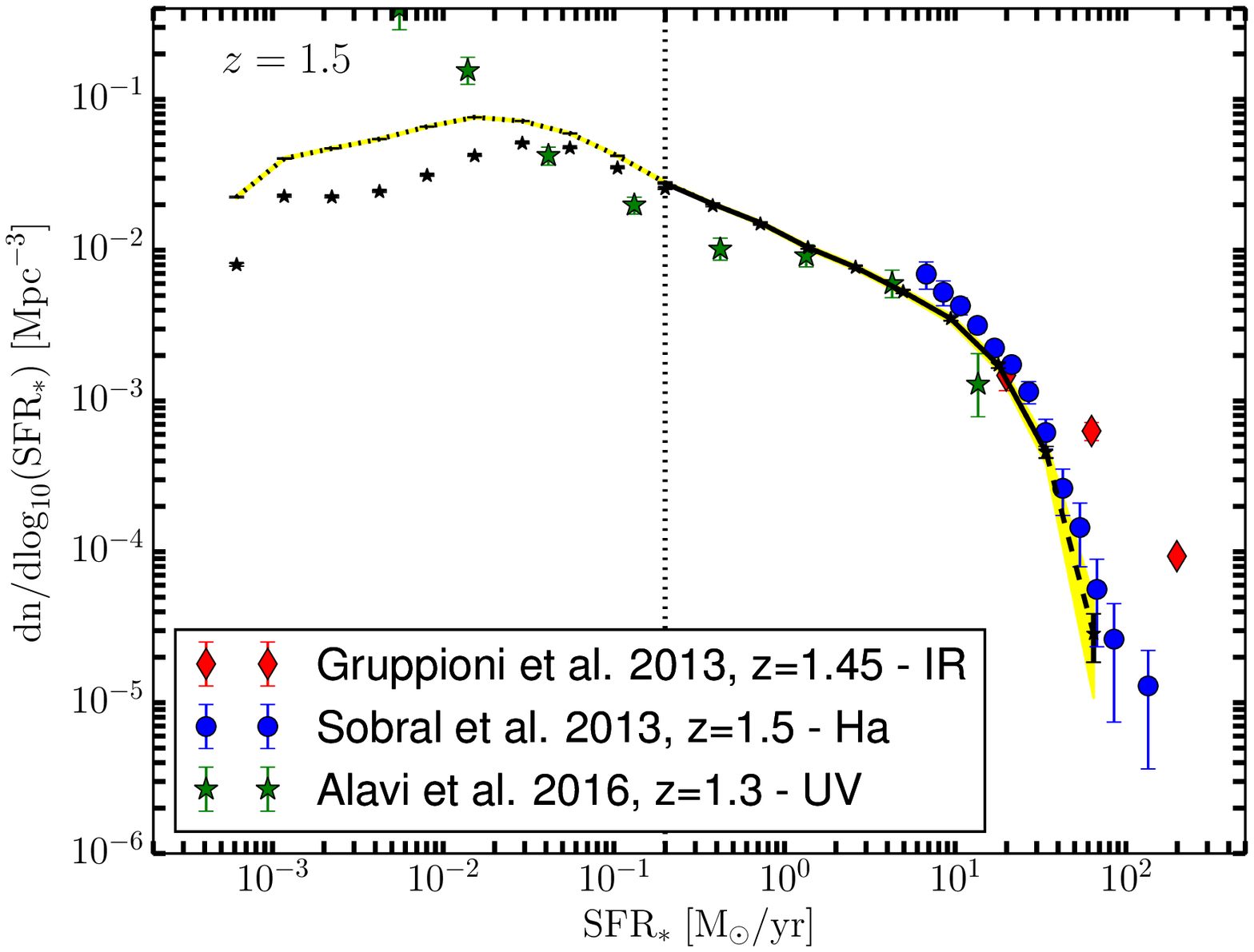}
\includegraphics[scale=0.40]{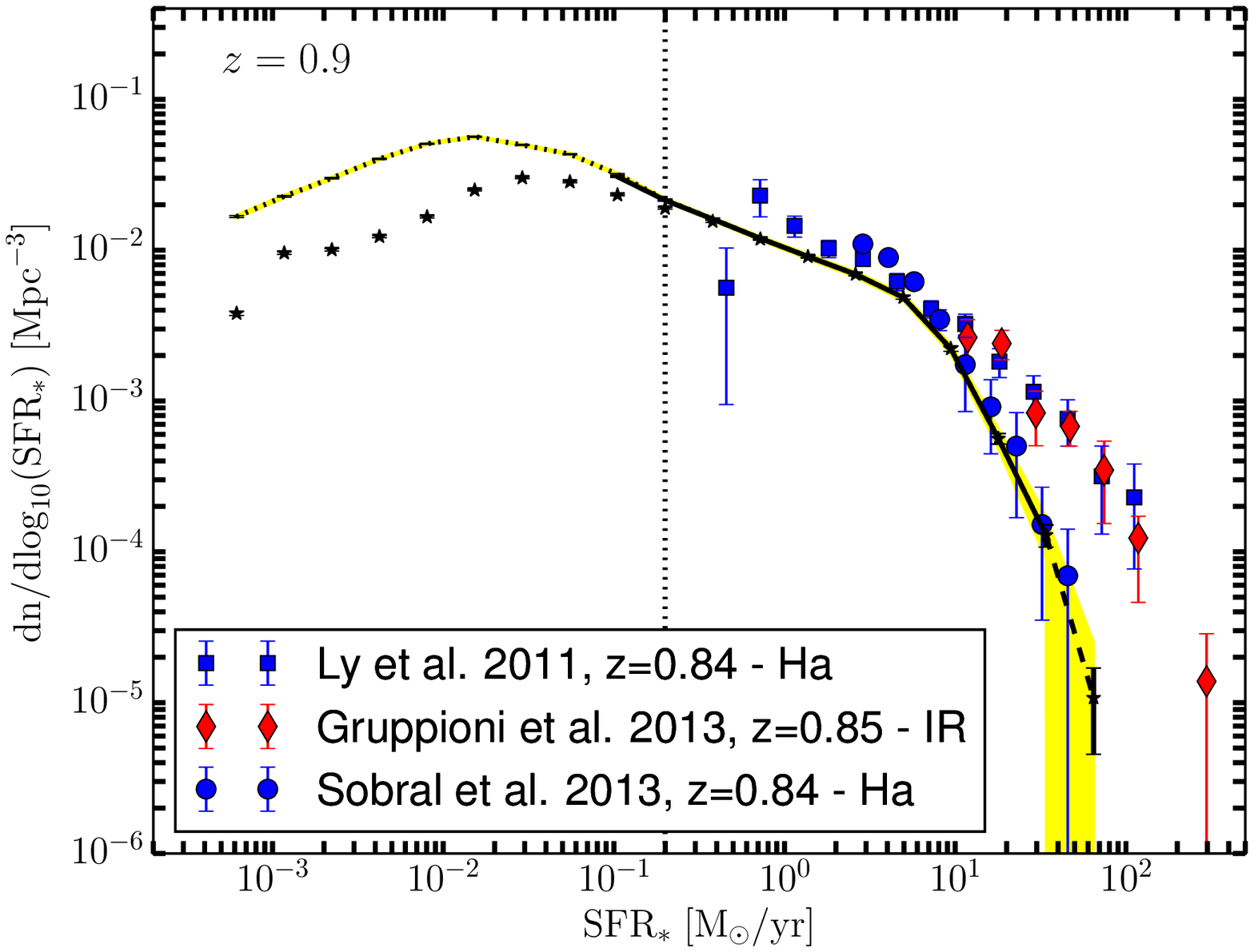}
\includegraphics[scale=0.40]{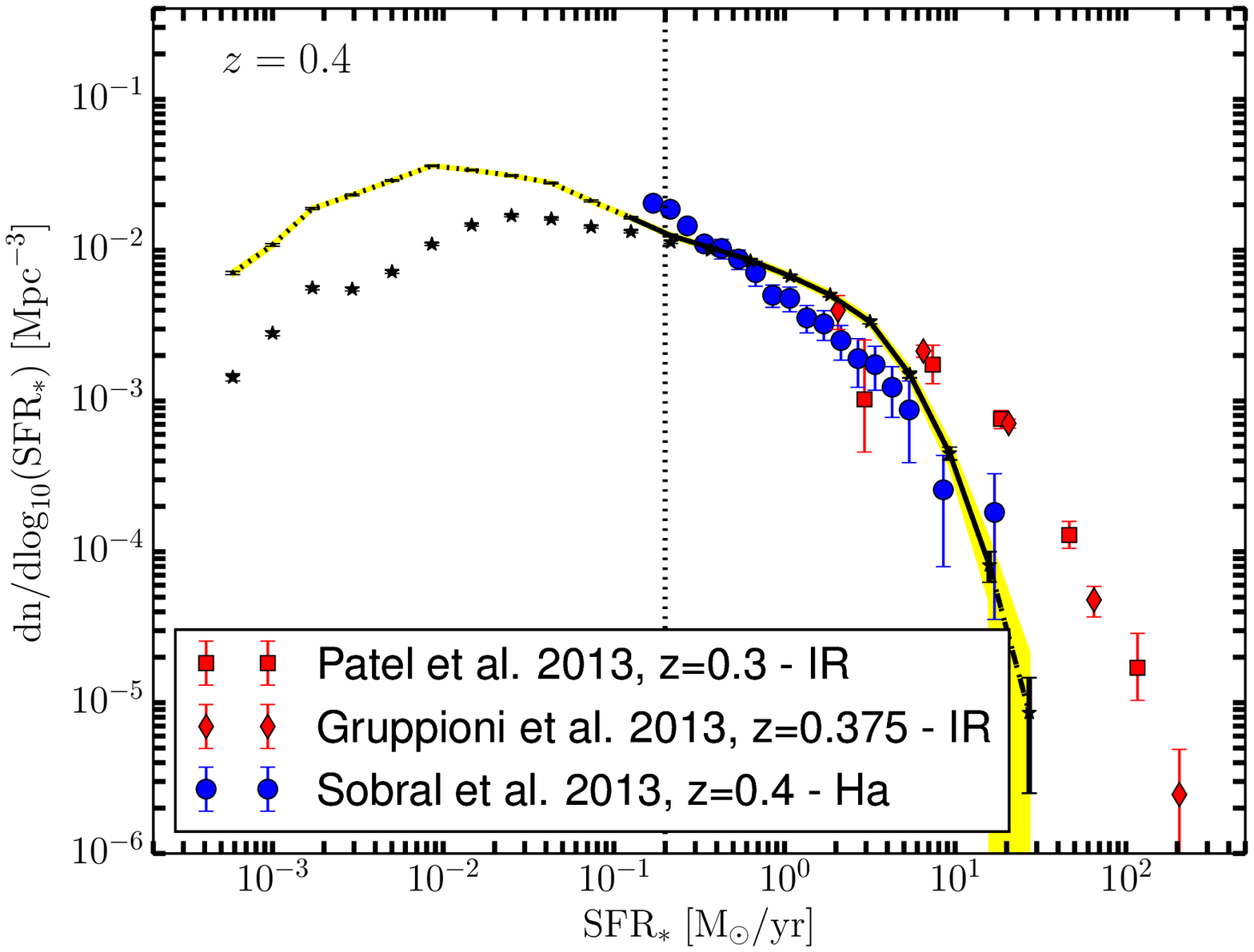}
\includegraphics[scale=0.40]{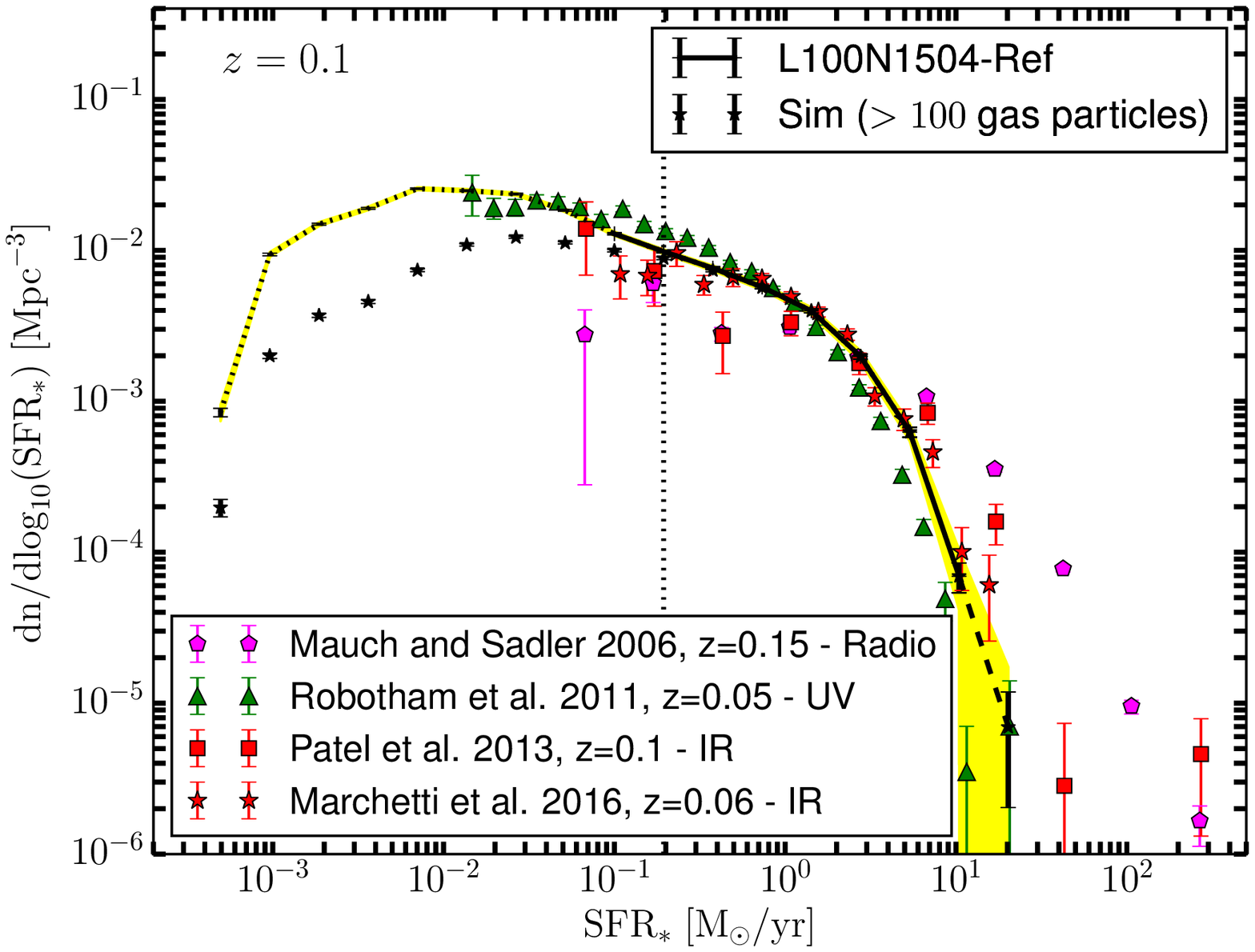}
\caption{Comparison between the L100N1504-Ref SFRF (black dashed line) and observations for redshifts $z \sim 2.5$ (top left panel), $z \sim 2.0$ (top right panel),  $z \sim 1.5$ (middle left panel), $z \sim 0.85$ (middle right panel), $z \sim 0.35$ (bottom left panel) and $z \sim 0.1$ (bottom right panel). In the Appendix \ref{table} we present detailed tables of the constraints we obtained using UV, IR and H$\alpha$ LFs (see also \citet{Katsianis2016}). The yellow area represents the 95$\%$ bootstrap confidence interval for 1000 re-samples of the EAGLE SFRs, while the black errorbars represent the poissonian errors. When a bin of the EAGLE SFRF contains objects with stellar masses below the mass limit of 100 baryonic particles curves are dotted, when there are fewer than 10 galaxies curves are dashed. To describe the limits due to poor sampling of gas particles we present the SFRF of objects which contain more than 100 gas particles (black stars+Vertical dotted line)}
\label{fig:EvolutionSFRF42}
\end{figure*}

Starting from redshift $z \sim8$ (top left panel of Fig. \ref{fig:EvolutionSFRF}) we present a comparison between the L100N1504-Ref and the constrains implied by the \citet[$z \sim8$, ][]{Bouwens2016} data. The yellow area represents the 95$\%$ bootstrap confidence interval for 1000 re-samples of the SFR of the simulated galaxies, while the black errorbars represent the poissonian errors. To describe the limits due to poor sampling of gas particles we present the SFRF of objects which contain more than 100 gas particles (black stars+Vertical dotted line). In addition, when a bin of the EAGLE SFRF contains objects with stellar masses below the mass limit of 100 (initial mass, $m_{GAS}$) baryonic particles curves are dotted since sampling effects associated with limited resolution become important below this regime \citep{Schaye2015}. The agreement is quite good despite the above problems and the fact that most of the simulated galaxies at this high redshift  suffer from resolution effects. At $z \sim 7.0$ (top right panel) we show that L00N1504-Ref is underproducing the number of objects with intermediate SFRs ($1 \, {\rm M_{\odot} \, yr^{-1}} < SFR < 10 \, {\rm M_{\odot} \, yr^{-1}}$) but this could be due to limits of resolution that still look prominent. The picture is similar at $z \sim 6.0$ (middle left panel of Fig. \ref{fig:EvolutionSFRF}) and at $z \sim 5.0$  (middle right panel of Fig. \ref{fig:EvolutionSFRF}), where the L100N1504-Ref run is able to describe a larger range of the SFRF free of resolution and boxsize effects. At these high redshifts we see a slight underproduction by 0.2 dex at the number density of objects with $1 \, {\rm M_{\odot} \, yr^{-1}} < SFR < 10 \, {\rm M_{\odot} \, yr^{-1}}$. Besides resolution effects, it is possible that the above is due to the strong SN feedback employed in the EAGLE reference model. We will see in detail the effect of both SN and AGN feedback prescriptions at various redshifts in section \ref{Feed}. At $z \sim4$ (bottom left panel of Fig. \ref{fig:EvolutionSFRF}) we see that UV studies \citep{smit12,Parsa2015} are able to probe successfully galaxies with low ($0.1 \, {\rm M_{\odot} \, yr^{-1}} < SFR < 1 \, {\rm M_{\odot} \, yr^{-1}}$), intermediate ($1 \, {\rm M_{\odot} \, yr^{-1}} < SFR < 10 \, {\rm M_{\odot} \, yr^{-1}}$) and high ($10 \, {\rm M_{\odot} \, yr^{-1}} < SFR < 100 \, {\rm M_{\odot} \, yr^{-1}}$) SFRs. On the other hand, IR studies \citep{Gruppionis13} are limited to constrain only high star-forming systems ($SFR > 100 \, {\rm M_{\odot} \, yr^{-1}}$) for this era. This can be due to the following reasons:
\begin{itemize}

\item different tracers possibly trace completely different population of galaxies and give only conditional SFRFs that do not represent the overall population,  
  
\item intermediate and low SFR objects do not have enough dust to reprocess a large number of UV photons into the IR, thus IR surveys are unable to probe the faint end slope of the distribution at early times,
  
\item Mid-FIR surveys have limitations of resolution and surface brightness,

\item different indicators of SFR correspond to different time averages (e.g. IR measurements represent less instantaneous measurements than UV and H$\alpha$ tracers).  
  
\end{itemize}
 At $z \sim 3.0$ (bottom right panel of Fig. \ref{fig:EvolutionSFRF}) we present a comparison between the reference model and the SFRFs obtained by \citet{Katsianis2016}, which rely on the LFs determinations of \citet[][Bolometric]{Reddy2008} and \citet[][UV-selected]{Parsa2015}. We see that the two distributions are in good agreement, except for objects with $SFR > 100 \, {\rm M_{\odot} \, yr^{-1}}$. With respect to the \citet[][bolometric]{Reddy2008} data, the EAGLE reference model slightly underpredicts the number density of high star-forming objects. On the other hand, the L100N1504-Ref run slightly overpredicts the number density of high star-forming galaxies with respect to the SFRF that relied on \citet[][UV]{Parsa2015} results.

To determine the SFRF at $z \sim 2.5$  (top left panel of Fig. \ref{fig:EvolutionSFRF42}), we employ the bolometric LF of \citet{Reddy2008}, the H$\alpha$ data of \citet{Sobral2013} and the UV-selected LF of \citet{Alavi2016}. \citet{Alavi2016} obtained near-UV imaging of 3 lensing clusters, to study the evolution of the faint end of the UV LF.  The L100N1504-Ref is broadly consistent with the constrains from observations. The different indicators provide SFRFs that are in agreement up to this redshift ($z \sim 2.5$). On the other hand, at redshift $z \sim2.0$ (top right panel of \ref{fig:EvolutionSFRF42}) we see that the SFRFs obtained from UV-selected samples \citep{Parsa2015,Alavi2016} are not consistent with the constraints from the IR data \citep{Magnelli2013} for galaxies with $SFR \ge 80 {\rm M_{\odot} \, yr^{-1}}$. This tension can be due to the following reasons:
\begin{itemize}
\item dust corrections for UV luminosities (and especially those implied by the  IRX-$\beta$ relation) are uncertain for highly star-forming systems and possibly underestimated, 
\item UV-LFs are usually incomplete at the bright end of the distribution since  bright objects have high dust contents and thus are invisible to UV surveys,
\item the SFR measured from IR light can be overestimated (see more details section \ref{thecode}).
\end{itemize}
The L100N1504-Ref SFRF at $z \sim2.0$ generally lay between UV and IR constrains, but typically are closer to the UV-SFRFs. However, we note that usually the highest star-forming bins contains less than 10 galaxies at most redshifts, thus the statistics are poor. Simulations with larger box size and resolution are needed to have more meaningful comparisons in the high and low star-forming ends of the SFRF.

At $z \sim1.5$ (middle left panel of Fig. \ref{fig:EvolutionSFRF42}) we employ the IR LF of \citet[][$z \sim1.45$]{Gruppionis13}, the H$\alpha$ data of \citet[][$z \sim1.5$]{Sobral2013} and the UV LF of \citet[][$z \sim1.3$]{Alavi2016}.  H$\alpha$ light is subject to dust attenuation effects. Thus, \cite{Sobral2013} applied a 1 mag correction across all bins of the observed H$\alpha$ distribution to estimate a dust free LF. \citet{Hopkins01} noted that a more sophisticated luminosity-dependent dust correction law produces similar to 1 magnitude correction but only for local galaxies. \citet{Hopkins01} suggest that at higher redshifts ($z > 0.3$), the 1 mag correction is possibly underestimating the dust corrections for objects with high luminosities. Thus, it is possible that the 1 mag correction to the H$\alpha$ luminosities employed by \citet{Sobral2013} is underestimated resulting artificially in low star formation rates at the high star-forming end. The L100N1504-Ref run is in agreement with the H$\alpha$ and UV SFRFs but is underproducing the number density of objects with respect to IR constraints. We will see in more detail in subsection \ref{AGNFeed} that the tension between the EAGLE reference model and the IR data could be due to the presence of the strong AGN feedback that is implemented. At $z \sim0.8$ (middle right panel of \ref{fig:EvolutionSFRF42}), we compare the reference model with the SFRFs reported by \citet{Katsianis2016} which rely on the H$\alpha$ LFs of \citet{Ly11} and \citet{Sobral2013} and the IR LF of \citet{Gruppionis13}. \citet{Ly11} obtained measurements of the H$\alpha$ luminosity function for galaxies at $z \sim0.8$, based on the NewH$\alpha$ Survey. In contrast to \citet{Sobral2013} who applied the 1 mag correction to their LF, \citet{Ly11} adopted the luminosity-dependent extinction relation of \citet{Hopkins01}. \citet{Sobral2013} found that their results are in excellent agreement with \citet{Ly11} if the authors assume the same dust corrections thus any differences between the H$\alpha$ SFRFs present in the top panel of Fig \ref{fig:EvolutionSFRF42} are possibly due to differences in the treatment of dust. At $z \sim0.4$ (bottom left panel of Fig. \ref{fig:EvolutionSFRF42}) we see again that SFRFs that rely on IR studies imply a higher number density of high SFRs than the H$\alpha$ LF of \citet{Sobral2013}. The L100N1504-Ref model is in agreement with the SFRFs obtained using \citet{Sobral2013} data but underproduces the number density of objects when compared to IR studies by an order of magnitude at $SFR>100 \, {\rm M_{\odot} \, yr^{-1}}$. Finally, the reference model at $z \sim0$ is in good agreement with the IR SFRF that rely on \citet{Marchetti2016}. However, it underpredicts the number density of high star-forming compared to constraints from the radio observations of \citet{Mauch2007}\footnote{We use the calibration given by \citet{Sullivan2001} which suggests that the relation between SFR and radio luminosity is  $SFR = \frac{L_{radio}}{8.85 \times 10^{20}}$. We convert the results for a Chabrier IMF by dividing them by 1.8.} and the IR data of \citet{Patel13}. On the other hand, the EAGLE SFRF slightly overpredicts the number of high star-forming objects with respect the SFRF that rely on the NUV luminosity function of \citet{Robotham2011}.

In conclusion, we find that the EAGLE SFRF at $z \sim0-8$ is consistent with the constraints from the observations, especially with those implied by UV and H$\alpha$ studies, at the regimes where simulations are considered free from resolution and volume effects. There is a slight underproduction of objects with $SFR \sim1-10 \,{\rm M_{\odot} \, yr^{-1}}$ in the L0100N1504-Ref run with respect the observations from all star formation indicators at $z \ge3$. In addition, the simulation underpredicts the number of objects at the high star-forming end ($SFR \sim10-100 \,{\rm M_{\odot} \, yr^{-1}}$) with respect to IR and Radio data for $ \le 2$. We will see in subsection \ref{CSFRD} that objects at these regimes dominate the cosmic star formation rate density, thus small disagreements between observed and simulated SFRFs for these galaxies may be responsible for disagreements between the EAGLE and observed CSFRDs.

\subsection{The evolution of the cosmic star formation rate density.}
\label{CSFRD}

\begin{figure}
\centering
\includegraphics[scale=0.40]{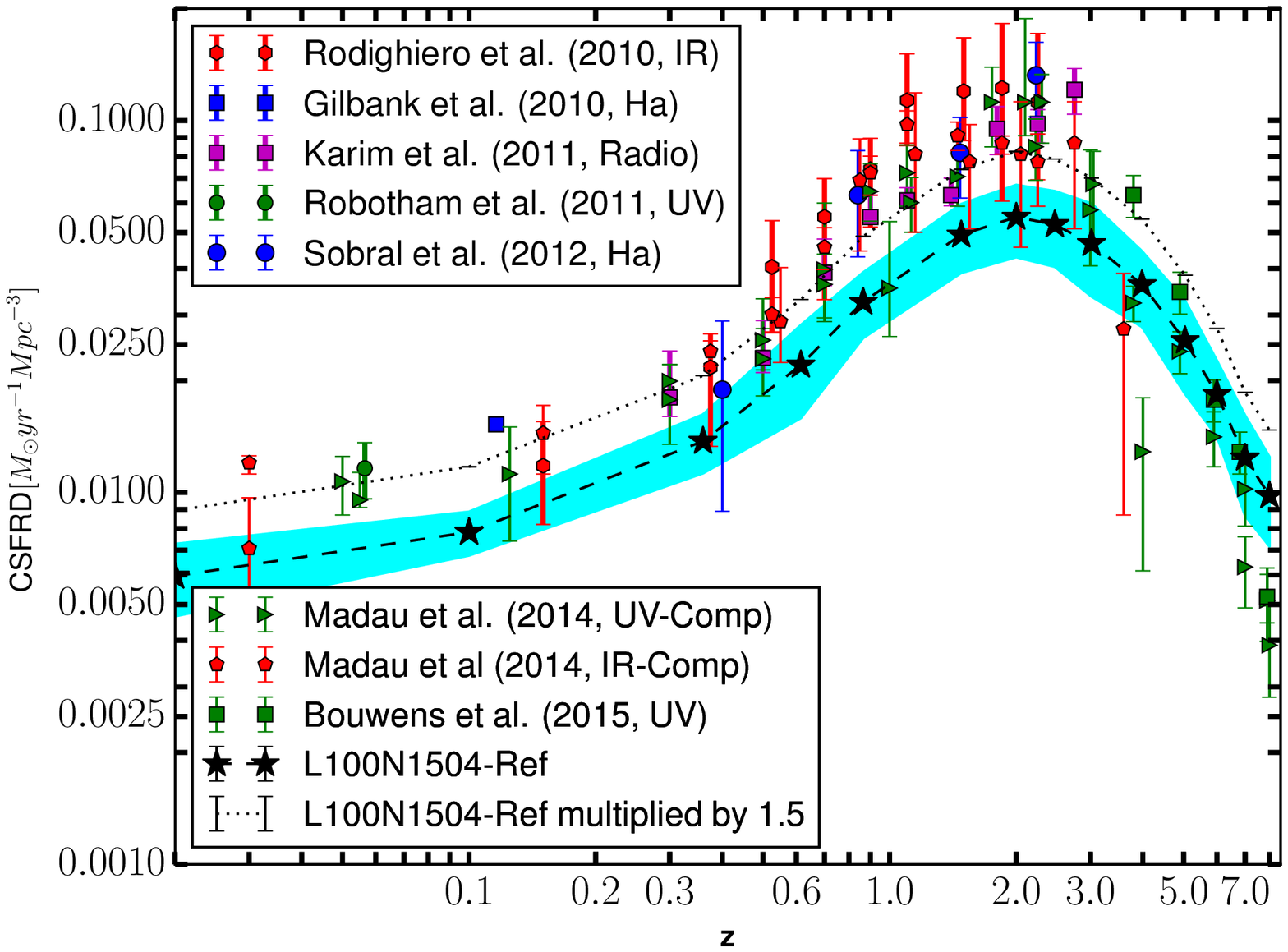}
\includegraphics[scale=0.40]{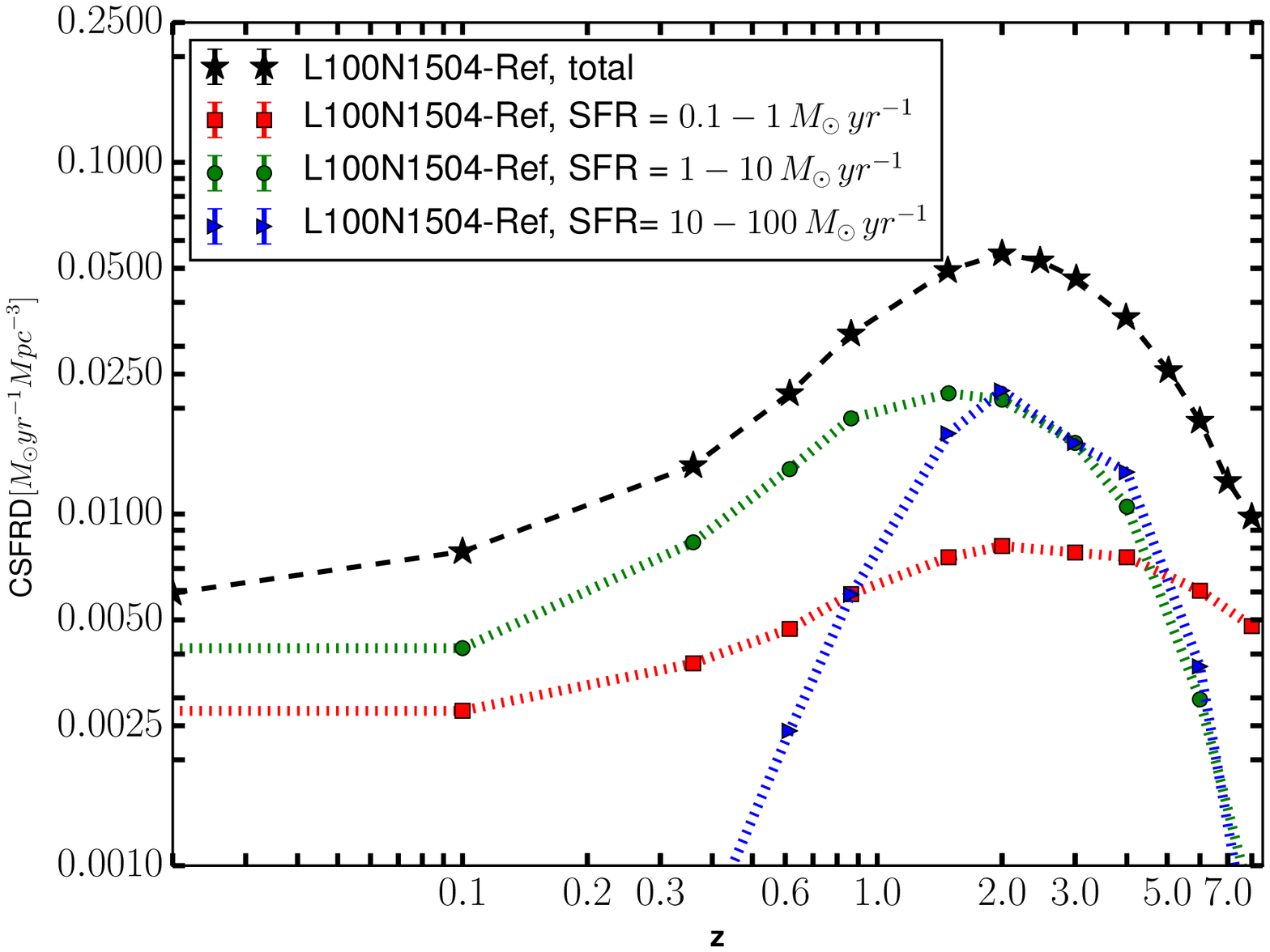}
\caption{Top panel: The evolution of the EAGLE L100N1504-Ref Cosmic Star Formation Rate Density (black dashed line) of the Universe alongside with the observations of \citet{Madau2014}. The cyan region represents the cosmic variance due to the limited boxsize of the simulation \citep{Driver2010}. Bottom panel: The contribution of galaxies with different SFRs in the EAGLE CSFRD. Galaxies with low SFRs ($SFR \sim 0.1-1 \, {\rm M}_{\rm \odot} \, {\rm yr^{-1}}$) dominate the CSFRD at $z > 6$. The contribution of $SFR \sim 10-100 \, {\rm M}_{\rm \odot} \, {\rm yr^{-1}}$ is significant down to $z \sim 2$, below which it drops quickly. Galaxies with intermediate $SFR \sim 1-10 \, {\rm M}_{\rm \odot} \, {\rm yr^{-1}}$ dominate the CSFRD for most of the history of the Universe.}
\label{fig:CSFRDz}
\end{figure}

A standard way to describe the evolution of the SFR of the universe as a whole is by providing estimates of the Cosmic Star Formation Rate Density (CSFRD) at various redshifts. These are obtained usually by integrating LFs or SFRFs \citep{Madau2014,Katsianis2016} and give useful constraints on the theory and simulations. In the top panel of Fig. \ref{fig:CSFRDz} we plot the cosmic star formation rate density as a function of redshift  for the L100N1504-Ref run alongside with the observations of \citet[][IR]{Rodighiero10}, \citet[][radio]{Karim2011}, \citet[][H$\alpha$]{Sobral2013}, \citet[][compilation]{Madau2014} and \citet[][UV]{Bouwens2016}. We use these measurements because combined they can trace the evolution of the CSFRD for various redshifts and at the same time they represent results from different indicators giving a sense of systematic scatter. The dark green triangles of Fig. \ref{fig:CSFRDz} and labeled as \citet[][UV-comp]{Madau2014} represent a compilation of UV observations \citep{Wyder2005,Schiminovich2005,Dahlen2007,Reddy2009,Robotham2011,cucciati12,bouwens2012}. The red pentagons of Fig. \ref{fig:CSFRDz} and labeled as \citet[][IR-comp]{Madau2014} is a compilation of IR observations \citep{Sanders2003,Takeuchi2003,Magenlli11,Magnelli2013,Gruppionis13}. The blue squares represent the results from the H$\alpha$ study of \citet{Guilbank2010}. The dark green circles were taken from \citet{Robotham2011}. We see that estimates of the CSFRD from different indicators have an offset of around 0.2 dex but are broadly consistent.

\citet{Katsianis2016} compared measurements of the CSFRD obtained from the integration of UV, IR and H$\alpha$ SFRFs at $z \sim1-4$. The authors reported that different SFR indicators produce results consistent for the CSFRD, despite  their differences (see also section \ref{SFRFEagle}). This is possibly due to the fact that all SFRFs regardless of SFR tracer agree well for objects with intermediate SFRs ($SFR \sim 1-10 \, {\rm M}_{\rm \odot} \, {\rm yr^{-1}}$), which dominate the CSFRD at most redshifts (bottom panel of Fig. \ref{fig:CSFRDz}). However, we note that the measurements from IR data are found to be typically 0.10-0.25 dex higher. This may be due to the following reasons:

\begin{itemize}
\item the faint-end slopes of the IR SFRFs/LFs are not directly constrained by individually detected sources and rely only on extrapolations which usually have artificially smaller negative slope $\alpha$.
\item the characteristic luminosity/SFR of IR LFs/SFRFs is typically higher than that of UV and H$\alpha$ studies, which are unable to trace dusty systems with high star formation rates. 
\end{itemize}

In the top panel of Fig. \ref{fig:CSFRDz} we see that the EAGLE CSFRD increases with time and peaks at $z \sim2$. That era is followed by a constant decrement so that the CSFRD is almost 10 times lower by $z \sim0$. As \citet{Furlong2014} pointed out this behavior agrees well with the observed one but the L100N1504-Ref run has typically a normalization $\sim 1.5$ times lower than the observations. In the bottom panel of Fig. \ref{fig:CSFRDz} we present the contribution of galaxies with different SFRs to the cosmic CSFRD in EAGLE. We see that galaxies with $SFR \sim 1-10 \, {\rm M}_{\rm \odot} \, {\rm yr^{-1}}$ have the largest contribution from redshift $z \sim 5$ and below. In addition, the high star-forming objects  ($SFR \sim 10-100 \, {\rm M}_{\rm \odot} \, {\rm yr^{-1}}$) also make a large contribution at $z \sim 1.5 -5$ but after that era they are suddenly quenched. In section \ref{SFRFEAGLE0}, where we presented the evolution of the SFRF, we showed that the number of the objects at both regimes are low with respect low star-forming objects at all redshifts. However, in this section we see that combined they dominate the CSFRD. In section \ref{SFRFEAGLE0} we also demonstrated that the EAGLE SFRF is typically lower by $\sim 0.1$ dex compared to the observations from various tracers for objects with $SFR \sim 1-10 \, {\rm M}_{\rm \odot} \, {\rm yr^{-1}}$ at $z \ge 3$. In addition, there is a tension of $0.1 -1 $ dex  with the IR data at the high star-forming end for $z \le 2.0$. Since the above objects (bottom panel of Fig. \ref{fig:CSFRDz}) dominate the CSFRD, any disagreements between the simulated and observed SFRFs at these regimes can lead to disagreements between the simulated and observed CSFRDs as well.

The feedback mechanisms employed in the L100N1504-Ref run are responsible for suppressing the star formation rates of these galaxies. This would have the effect of decreasing the simulated CSFRD and cause the tension between the model and the observations. However, we note that it is also possible that the observations of the CSFRD which mostly rely on the \citet{kennicutt1998} conversion laws are overestimated \citep{Katsianis2015}. New calibrations with lower normalizations have been suggested in the literature \citep{Oti2010,Kennicutt2012,Chang2015}. For example, \citet{Chang2015} added complementary data using Herschel observations to estimate SFRs and stellar masses. These measurements included dust emission and contamination from quiescent galaxies with the derived SFR-luminosity calibrations for $12 \mu m$ and $22 \mu m$ beeing $0.10-0.20$ dex lower than previous measurements. In addition, we have to keep in mind that selection effects and biases definitely affect the measurements, since usually the UV and H$\alpha$ LFs are incomplete above the characteristic SFR and IR measurements rely on uncertain extrapolations for low mass objects (see section \ref{SFRFEagle}). We demonstrate that the contribution of low SFR galaxies ($SFR \sim 0.1-1 \, {\rm M}_{\rm \odot} \, {\rm yr^{-1}}$) at most redshifts to the CSFRD is small. Any disagreements between our comparisons most likely arises from objects with higher SFR. We stress that the L100N1504-Ref run may suffer from resolution effects for low star-forming objects and their contribution could be higher in a simulation with higher resolution. Nevertheless, in section \ref{SFRFEAGLE0} we showed that even at this limit observations and simulations of the SFRF are in excellent agreement.

Finally, we investigate which objects drive the peak of the CSFRD. We see that besides objects with $SFR \sim 1-10 \, {\rm M}_{\rm \odot} \, {\rm yr^{-1}}$ the peak of the CSFRD at $z \sim 2$ in the L100N1504-Ref is driven by the strong contribution of rare high star-forming galaxies ($SFR \sim 10-100 \, {\rm M}_{\rm \odot} \, {\rm yr^{-1}}$). Their abundance and contribution decreases significantly and sharply after that epoch (bottom panel of Fig. \ref{fig:CSFRDz}). We note that the behavior described above is not in agreement with the evolution implied by IR studies \citep[e.g.][]{Magnelli2013}, at which the contribution of galaxies with high SFRs/Luminosities to the CSFRD/total IR luminosity decreases much flatter. This discrepancy reflects the tension between the SFRFs of the EAGLE reference model and IR constraints. The later imply larger number densities of objects with $SFR \sim 10-100 \, {\rm M}_{\rm \odot} \, {\rm yr^{-1}}$ (subsection \ref{SFRFEAGLE0}) at $z <2 $, and the disagreement with the L100N1504-Ref run increases with time. In section \ref{Feed} we will see the important role that feedback plays on keeping the number of these extremely high star-forming objects low at these redshifts.

\section{The contribution of halos with different masses to the SFRF and CSFRD.}
\label{SFRFhalos}

\begin{figure}
\centering
\includegraphics[scale=0.40]{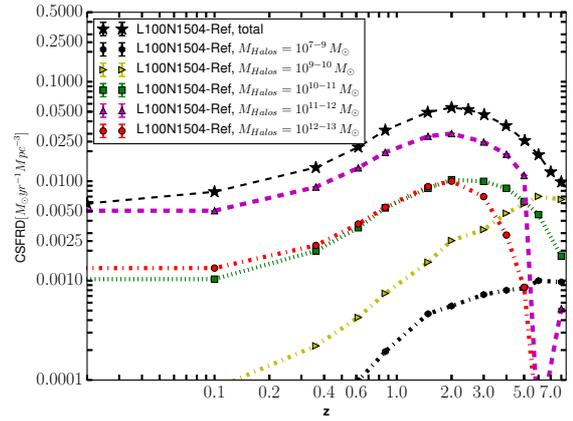}
\caption{CSFRD contribution of halos with different masses in EAGLE. Below redshift 5 objects with masses ${\rm M_{halo}} = 10^{11-12} \, {\rm M_{\odot}}$ dominate through their high efficiencies besides their low numbers (Fig. \ref{fig:CSFRDzHalos1}).}
\label{fig:CSFRDzHalos}
\end{figure}

\begin{figure*}
 \centering
\includegraphics[scale=0.40]{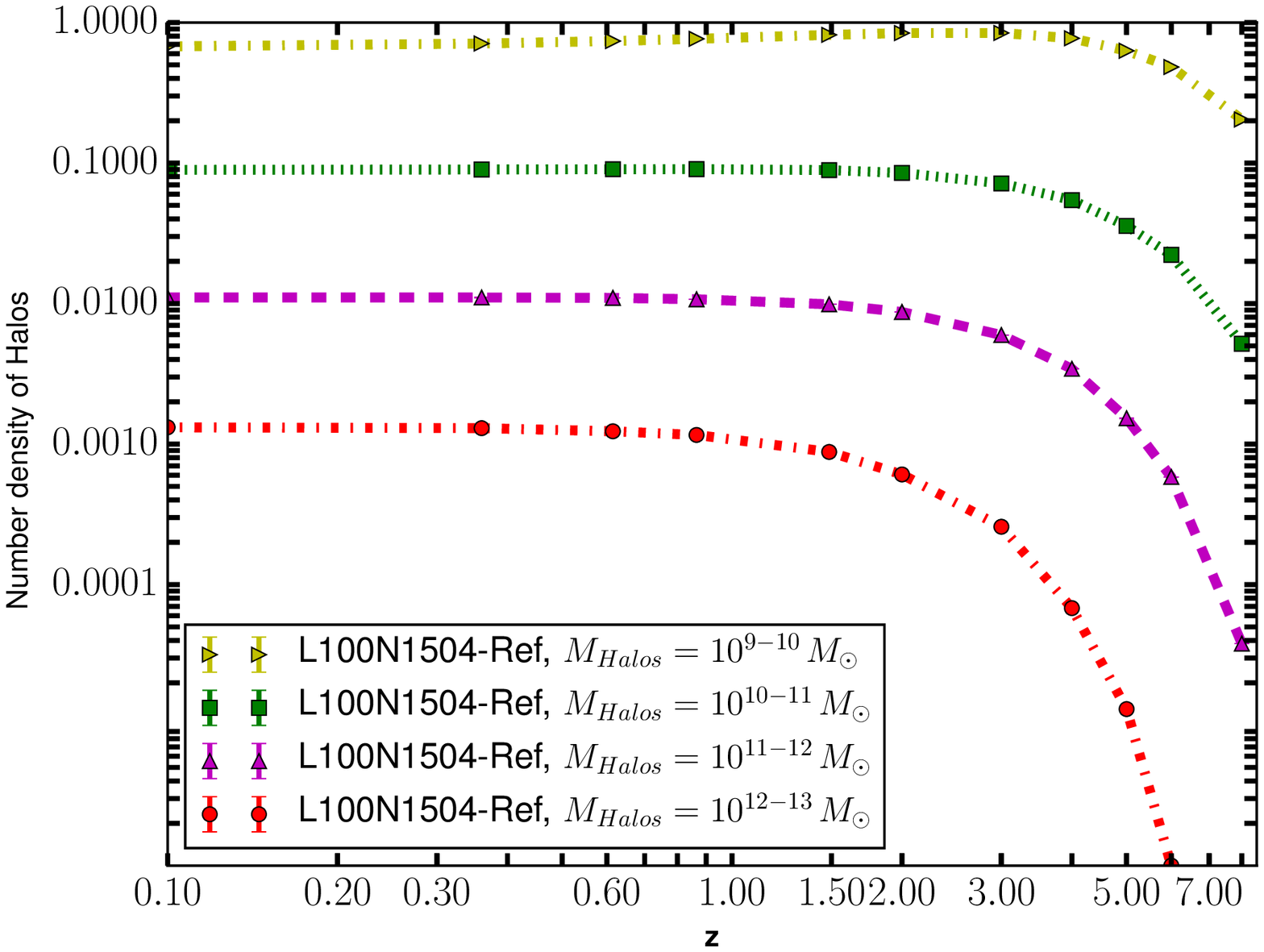}
\includegraphics[scale=0.40]{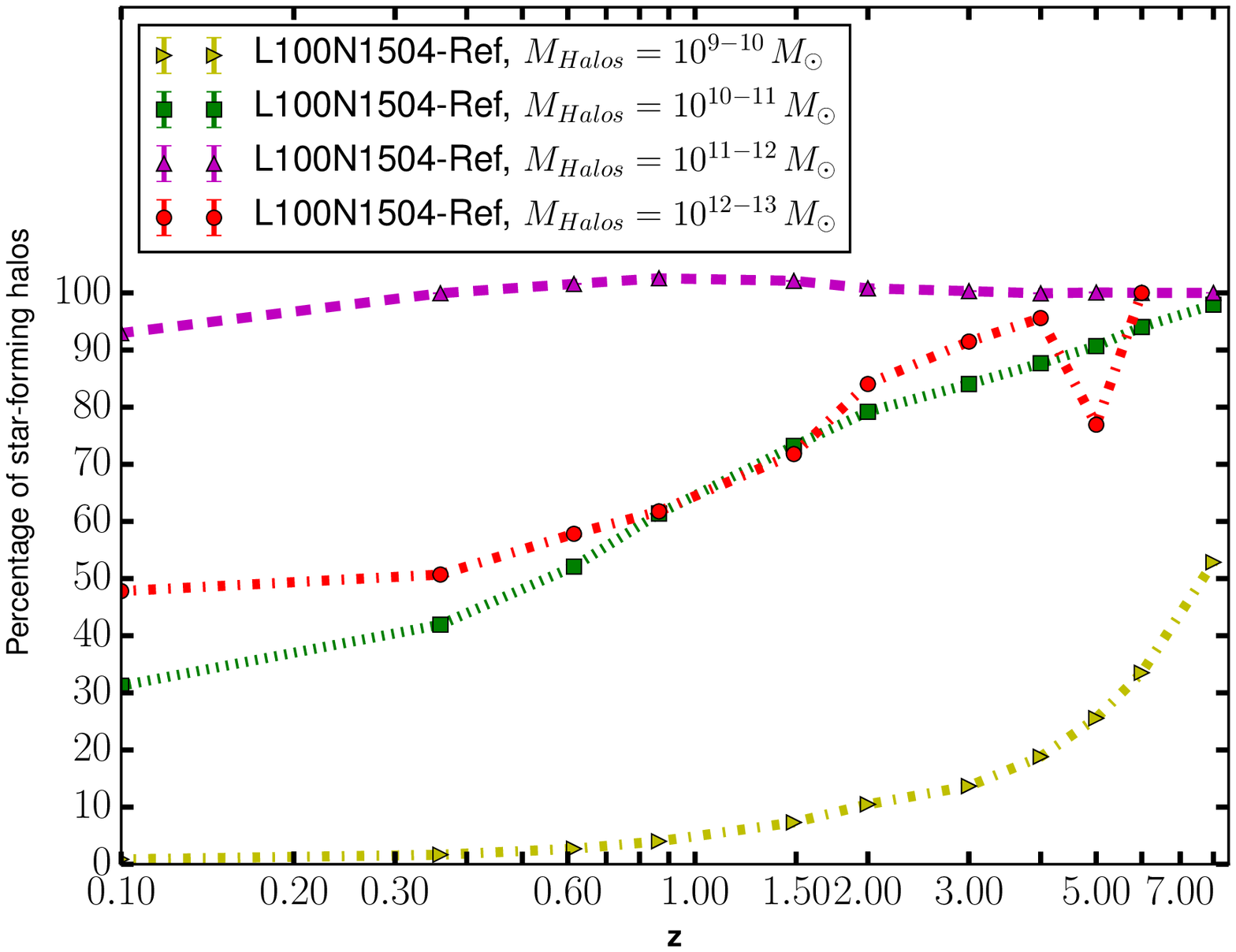}
\caption{Left panel: Evolution of the number of halos with different masses in the L100N1504-Ref. Halos with low masses are numerous even at high redshift while higher mass objects (e.g.  ${\rm M_{halo}} \sim 10^{11-13} \, {\rm M_{\odot}}$) form with high rate up to redshift $z \sim1.5$. Right panel: Percentage of star-forming halos at different masses. Halos with ${\rm M_{halo}} \sim 10^{11-12} \, {\rm M_{\odot}}$ are typically star-forming at all redshift.}
\label{fig:CSFRDzHalos1}
\end{figure*}

\begin{figure*}
\centering
\includegraphics[scale=0.40]{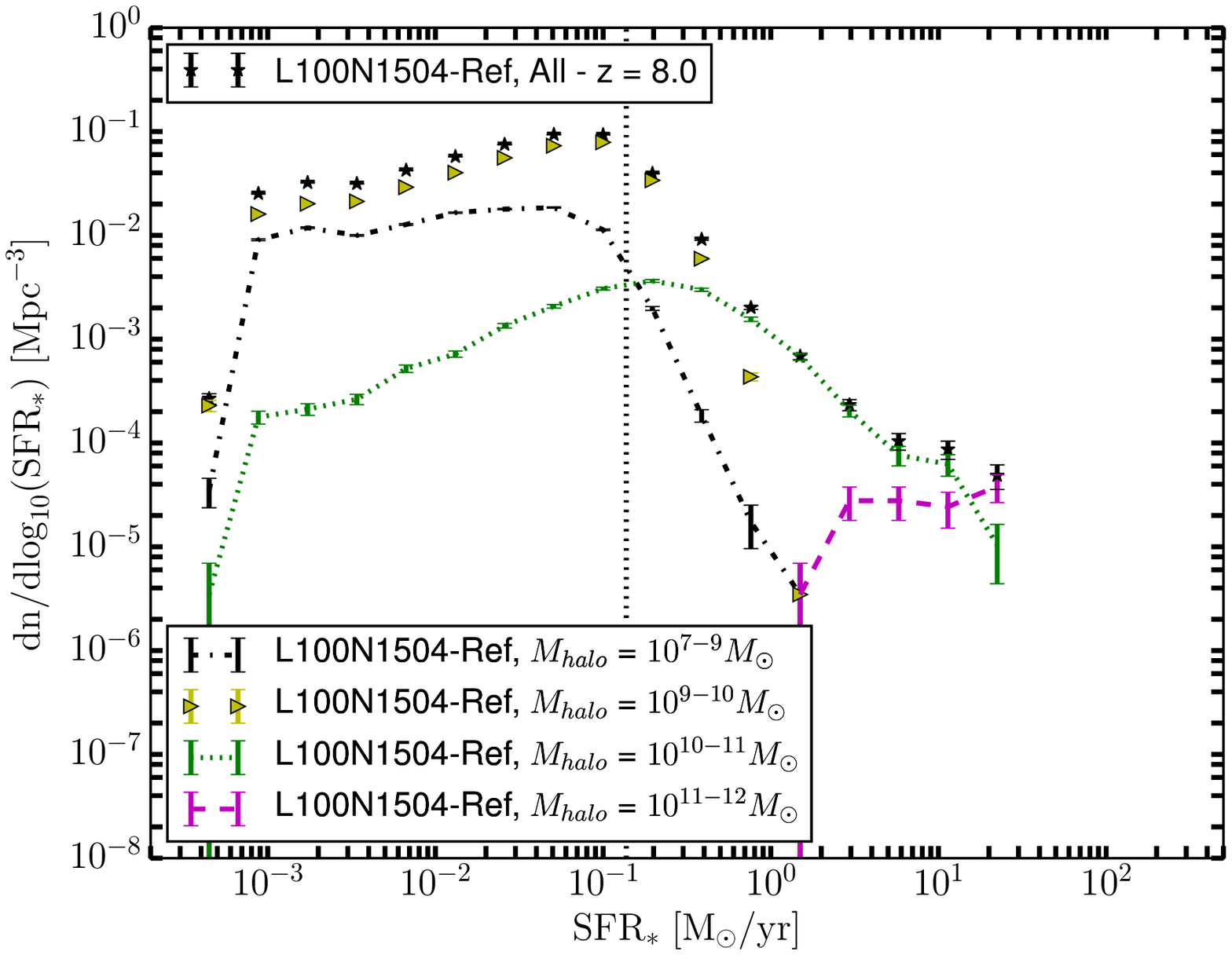}
\includegraphics[scale=0.40]{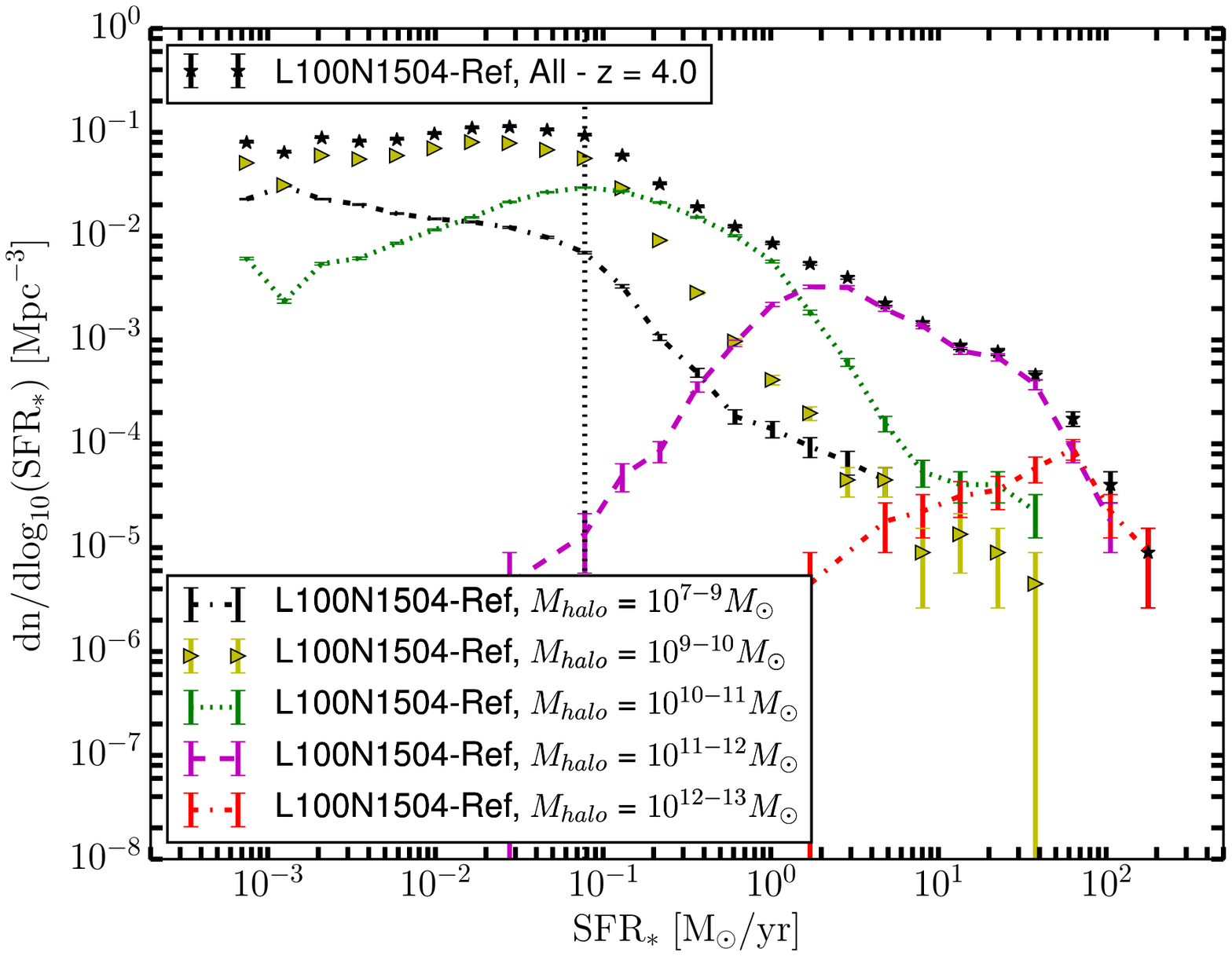}
\includegraphics[scale=0.40]{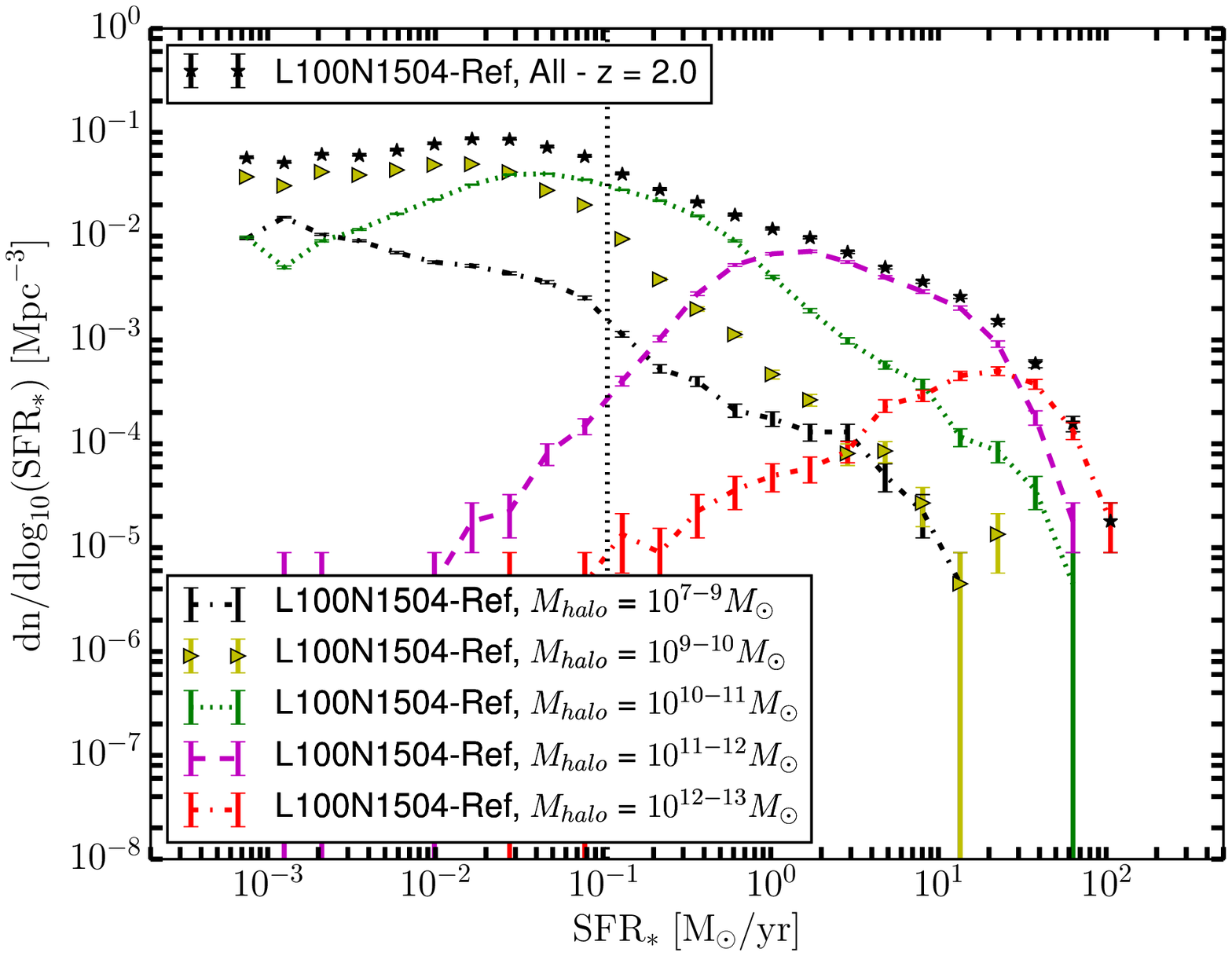}
\includegraphics[scale=0.40]{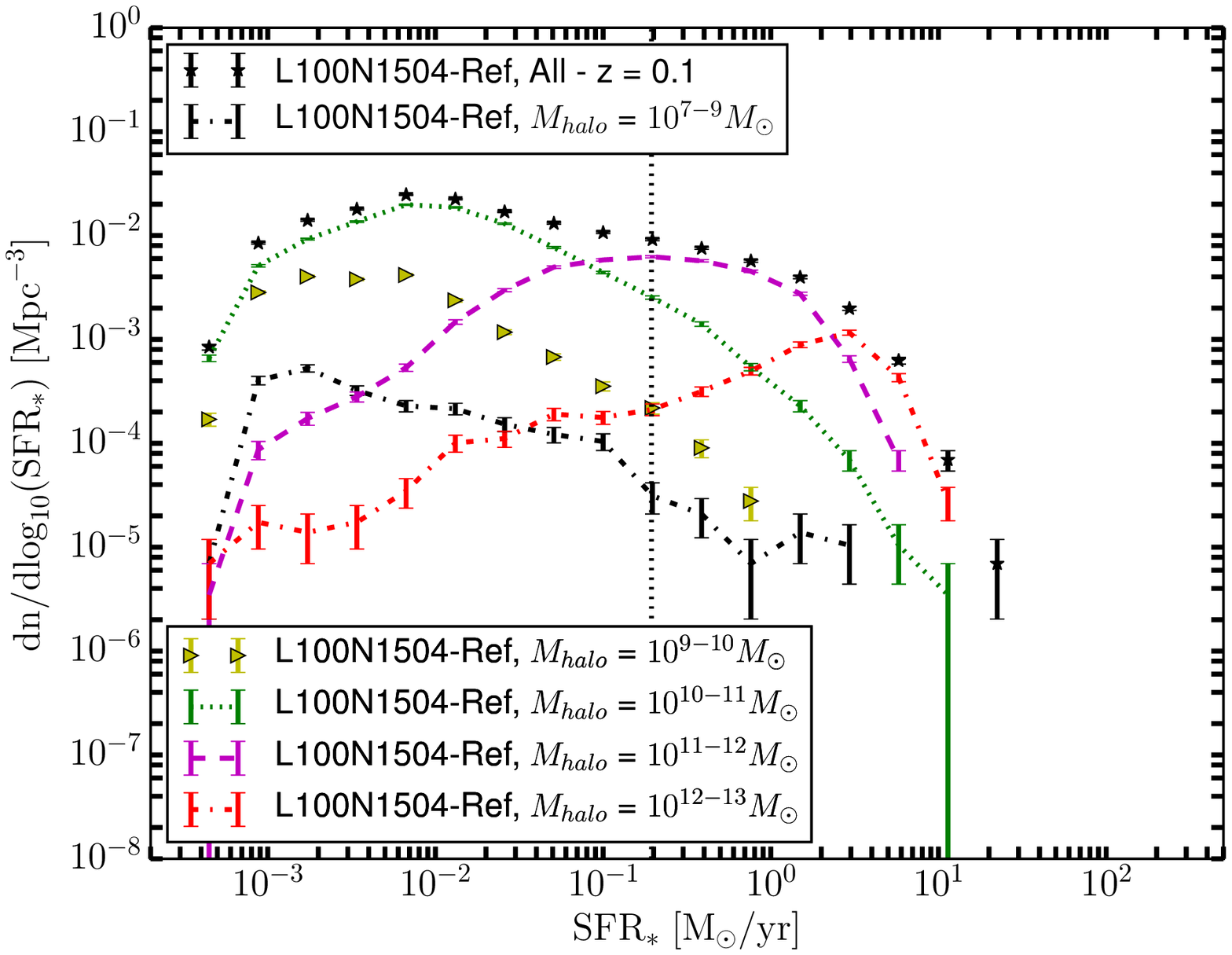}
\caption{The contribution of dark matter halos to the SFRF (black stars) for redshifts $z \sim 8.0$ (top left panel), $z \sim 4.0$ (top right panel), $z \sim 2.0$ (bottom left panel) and $z \sim 0.1$ (bottom right panel). Galaxies with $SFR \ge 1 \, {\rm M_{\odot} \, yr^{-1}}$ reside in large halos ${\rm M_{Halo}} \sim 10^{11-12} \, {\rm M_{\odot}}$ even at $z \sim4.0$. These halos represent only $ \sim 1\%$ of the total star-forming population at this era. In the local Universe the number of massive halos ${\rm M_{halo}} \ge 10^{11} \, {\rm M_{\odot}}$ is significantly higher than that of $z \sim 4$. They dominate almost entirely the SFRF at $SFR \ge 0.1 \, {\rm M_{\odot} \, yr^{-1}}$.}
\label{fig:halosSFRF}
\end{figure*}

In this section we present the contribution of halos with various masses to the CSFRD (Fig. \ref{fig:CSFRDzHalos}), the number density of halos of different masses at different redshifts (Fig. \ref{fig:CSFRDzHalos1}) and their contribution on the star formation rate function (Fig. \ref{fig:halosSFRF}) in the EAGLE L100N1504-Ref run.

\citet{Behroozi2013} and \citet{Moster2013} argued that halos near $10^{12} \, {\rm M_{\odot}}$ are the most efficient at forming stars at all redshifts with the baryon conversion efficiency  ($M_{\star}/M_{baryon}$) droping rapidly at both higher and lower masses. In addition, the stellar to halo mass ratio also peaks at $10^{12} \, {\rm M_{\odot}}$ \citep{Behroozi2013}. However, in the paradigm of hierarchical structure formation, small halos form earlier than larger ones, which need some time to emerge. In Fig. \ref{fig:CSFRDzHalos} we see that at $z \ge 5$ the cosmic star formation rate density is dominated by relatively small halos of mass ${\rm M_{halo}} = 10^{9-10} \, {\rm M_{\odot}}$\footnote{Halos with masses ${\rm M_{halo}} \le 10^{9} \, {\rm M_{\odot}}$ contain less than 100 dark matter particles. In this regime sampling effects associated with the limited resolution become important \citep{Katsianis2014} so we do not focus on objects that have masses less than ${\rm M_{halo}} \le 10^{9} \, {\rm M_{\odot}}$. We note that the exclusions of objects with masses ${\rm M_{halo}} = 10^{7-9} \, {\rm M_{\odot}}$ from our discussion would not have an effect in our conclusions as the CSFRD and SFRF is dominated by more massive halos (Fig. \ref{fig:CSFRDzHalos} and \ref{fig:halosSFRF}).}. The left panel of Fig. \ref{fig:CSFRDzHalos1} reveals that the number of objects of this mass interval is high. In addition, the right panel of the same figure, where we present the percentage of star-forming halos at different masses (i.e. $n_{Halos, SFR>0}/n_{Halos, total}$), shows that a large percentage ($ > 50\%$) of those are able to form stars. At $z \sim 8$ the total $\dot{M}_{\star}/M_{gas}$ ratio for halos with ${\rm M_{halo}} = 10^{9-10} \, {\rm M_{\odot}}$ is high ($\dot{M}_{\star}/M_{gas} \sim 1.3 \times 10^{-10} yr^{-1}$), while combined they contain $\sim 40\%$ of the total gas present in halos in the simulation. Objects with masses around the Milky way ${\rm M_{halo}} = 10^{11-12} \, {\rm M_{\odot}}$ are more efficient by almost an order of magnitude but have significantly lower counts. Thus, low mass halos through their large numbers and low star formation rates (Fig. \ref{fig:halosSFRF}) dominate the cosmic budget of star formation at this era.  Up to $z \sim 5$ the number of low mass halos is slowly increasing but a lot of them are merging to form larger structures. However, at this epoch only 20$\%$ of them are able to form stars, and their efficiency ($\dot{M}_{\star}/M_{gas} \sim 4.2 \times 10^{-11} \, yr^{-1}$) is significantly lower than in the past. In the following paragraphs we will see that at this era higher mass halos ($M_{halo} = 10^{11-12} \, M_{\odot}$) dominate the CSFRD. From $z \sim 5$ to $z \sim 1$ the total number of halos with masses $M_{halo} = 10^{9-10} \, M_{\odot}$ slightly increases, while from $z \sim 1$ to $ z\sim 0$ it is kept almost constant \citep{Warren2006,Lukic2007}. At $z \sim0$ only $0.6 \%$ of them are able to form stars, with their  efficiency being only $\dot{M}_{\star}/M_{gas} \sim 6.6 \times 10^{-12} \, yr^{-1}$. Their contribution to the cosmic budget is almost negligible (Fig. \ref{fig:CSFRDzHalos}).

In Fig. \ref{fig:CSFRDzHalos} we see that at $z \ge 5$ the contribution of large halos (${\rm M_{halo}} \ge 10^{12} \, {\rm M_{\odot}}$) to the total star formation rate of the universe is negligible. The left panel of Fig. \ref{fig:CSFRDzHalos1} reveals that the number density of these objects is small compared to low mass halos, which at this era dominate the Universe. However, halos grow rapidly at high redshifts \footnote{\citet{Correa2015} proposed that the total mass growth rate of a halo can be approximated by $ \dot{M}_{tot} = 71.6 \, M_{\odot} yr^{-1} \, (\frac{M(z)}{10^{12} \, M_{\odot}}) (\frac{h}{0.7}) \times f(M_{o})[(1+z)-\alpha][\Omega_{m}(1+z)^3]$}. Mergers play an important role in the creation of large halos in overdense regions, while diffuse accretion dominates the growth in voids. \citet{Fakhouri2008} and \citet{Fakhouri2010} using the Millenium simulations demonstrated that the rate of mergers increases with mass as $\propto M^{1.1}$ and with redshift as $\propto (1+z)^{2.5}$. In addition \citet{Qu2017} demonstrated that the merger fraction (the fraction of massive galaxies that are merging with a less massive companion) in the EAGLE simulation is large at higher redshifts, a behavior that is broadly consistent with observations \citep{Man2016}. The left panel of Fig. \ref{fig:CSFRDzHalos1} reveals that the number density of halos with masses $ {\rm M_{halo}} = 10^{11-12} \, {\rm M_{\odot}}$ has been increasing at high rate at $z \sim5-8$ in the L100N1504-Ref run. To indicate this we note that in only $0.5 \, {\rm Gyrs}$,  the number of halos in this mass regime has increased by more than $\sim 50$ times. Starting from redshift $z \sim5$ there is a sudden and immense increase in the contribution of ${\rm M_{halo}} \ge 10^{11} \, {\rm M_{\odot}}$ to the CSFRD. We note that the objects with ${\rm M_{halo}} = 10^{11-12} \, {\rm M_{\odot}}$ represent less than $\sim 1 \%$ of the total star-forming population at $ z \sim5$ (Fig. \ref{fig:CSFRDzHalos1}), they contain only $\sim 8\%$ of the total gas present in halos, yet they are so successful at forming stars ($\dot{M}_{\star}/M_{gas} \sim 4.2 \times 10^{-10} \, yr^{-1}$) with respect to other halos that they dominate the total CSFRD below $z < 5$. At $z \sim2-4$ the number of halos with ${\rm M_{halo}} = 10^{11-12} \, {\rm M_{\odot}}$ keeps increasing but at a relatively slower rate. By redshift $ \sim1$ most of the high mass halos (${\rm M_{halo}} = 10^{11-13} \, {\rm M_{\odot}}$) have already been formed in the EAGLE L100N1504-Ref something that is in agreement with the prediction from N-body simulations \citep{Diemand2007}. From redshift $z \sim 2$ to $z \sim 0$ the numbers of ${\rm M_{halo}} = 10^{11-12} \, {\rm M_{\odot}}$ and ${\rm M_{halo}} = 10^{12-13} \, {\rm M_{\odot}}$ halos remain almost constant. At $z \sim0$ objects with ${\rm M_{halo}} = 10^{11-12} \, {\rm M_{\odot}}$ represent almost $\sim 20 \%$ of the total star-forming population and contain only $\sim 6\%$ of the total gas present in halos in the simulation. Halos of this mass interval have been dominating the CSFRD in the EAGLE simulation for more than 11 billion years through their high efficiencies.

It is important to see the contribution of halos with different masses to the SFRF since different mechanisms that affect the SFRs of galaxies are related to the masses of their host halos. In Fig. \ref{fig:halosSFRF} we present the contribution of dark matter halos to the SFRF (black stars) for redshifts $z \sim 8.0$ (top left panel), $z \sim 4.0$ (top right panel), $z \sim 2.0$ (bottom left panel) and $z \sim 0.1$ (bottom right panel). We see that typically at all times star-forming efficient, high mass halos around $10^{12} \, {\rm M_{\odot}}$ completely dominate the SFRF at $SFR \ge 1.0 {\rm \, M_{\odot} \, yr^{-1}}$. As discussed in the previous paragraphs the growth of halos is rapid at high redshifts. At $z \sim4.0$ we see that galaxies with $SFR \ge 1 \,{\rm M_{\odot} \, yr^{-1}}$ reside entirely in large halos at the mass regime of ${\rm M_{Halo}} \sim 10^{11-13} \, {\rm M_{\odot}}$. Going to lower redshifts these objects make their dominance more absolute not only at the high star-forming rate end but to the whole distribution of SFRs. At $z \sim2$ we see that the high star-forming galaxies $SFR \sim 10-100 \, {\rm M}_{\rm \odot} \, {\rm yr^{-1}}$ which contribute significantly to the peak of the CSFRD (see section \ref{CSFRD}) mostly reside in halos of mass ${\rm M_{Halo}} \sim 10^{12-13} \, {\rm M_{\odot}}$, where AGN feedback can play a central role. We will see in detail in section \ref{AGNFeed} the effect of the AGN feedback implementation used by EAGLE for these objects. Finally, at $ z \sim0$ we see that any galaxy with $SFR \ge 0.1 \, {\rm M}_{\rm \odot} \, {\rm yr^{-1}}$ resides entirely in large halos with ${\rm M_{Halo}} \sim 10^{11-13} \, {\rm M_{\odot}}$.



\section{The effect of feedback prescriptions on the star formation rate function.}
\label{Feed}

\subsection{The effect of the SN feedback on the star formation rate function}
\label{SNe}

\begin{figure*}
\centering
\includegraphics[scale=0.40]{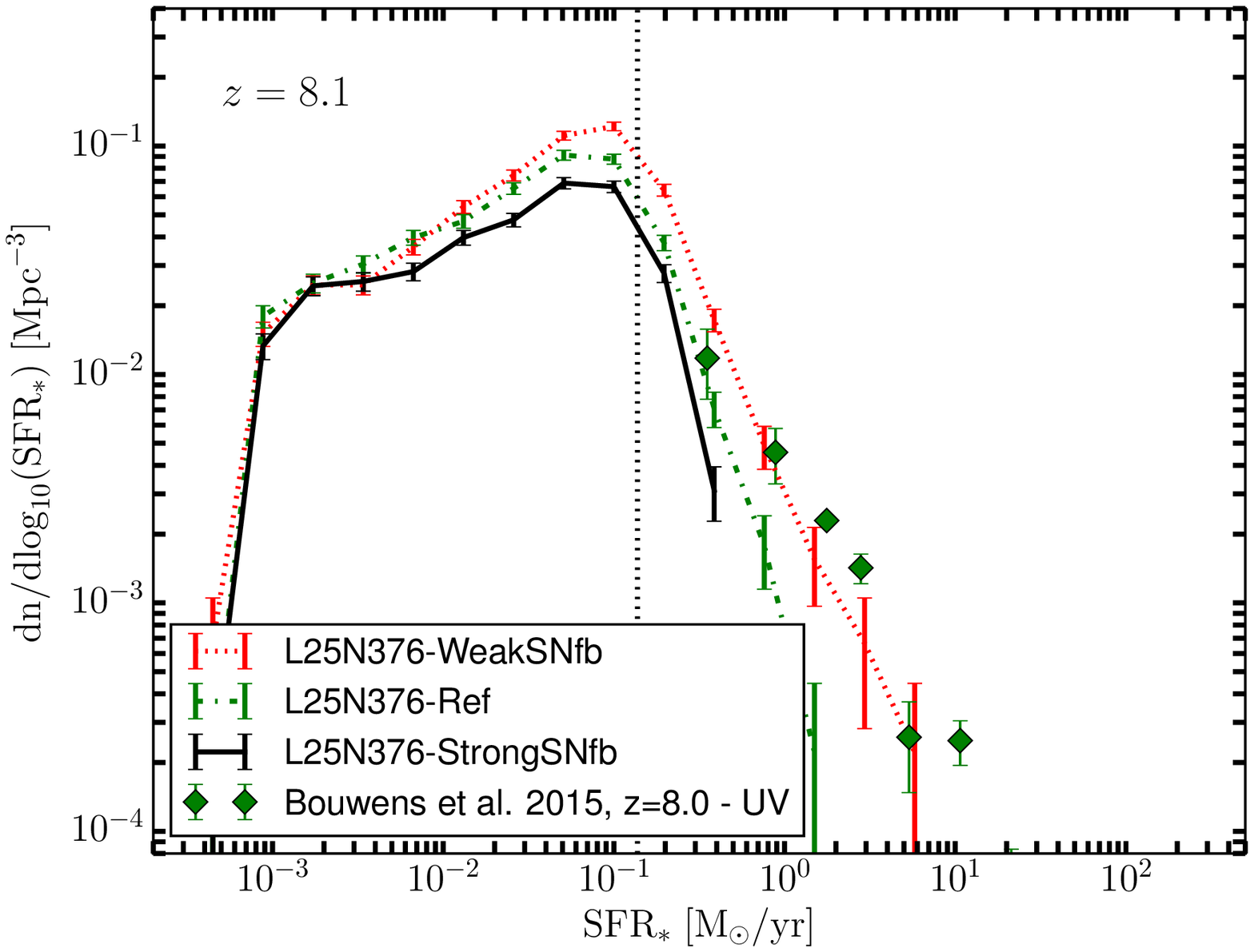}
\includegraphics[scale=0.40]{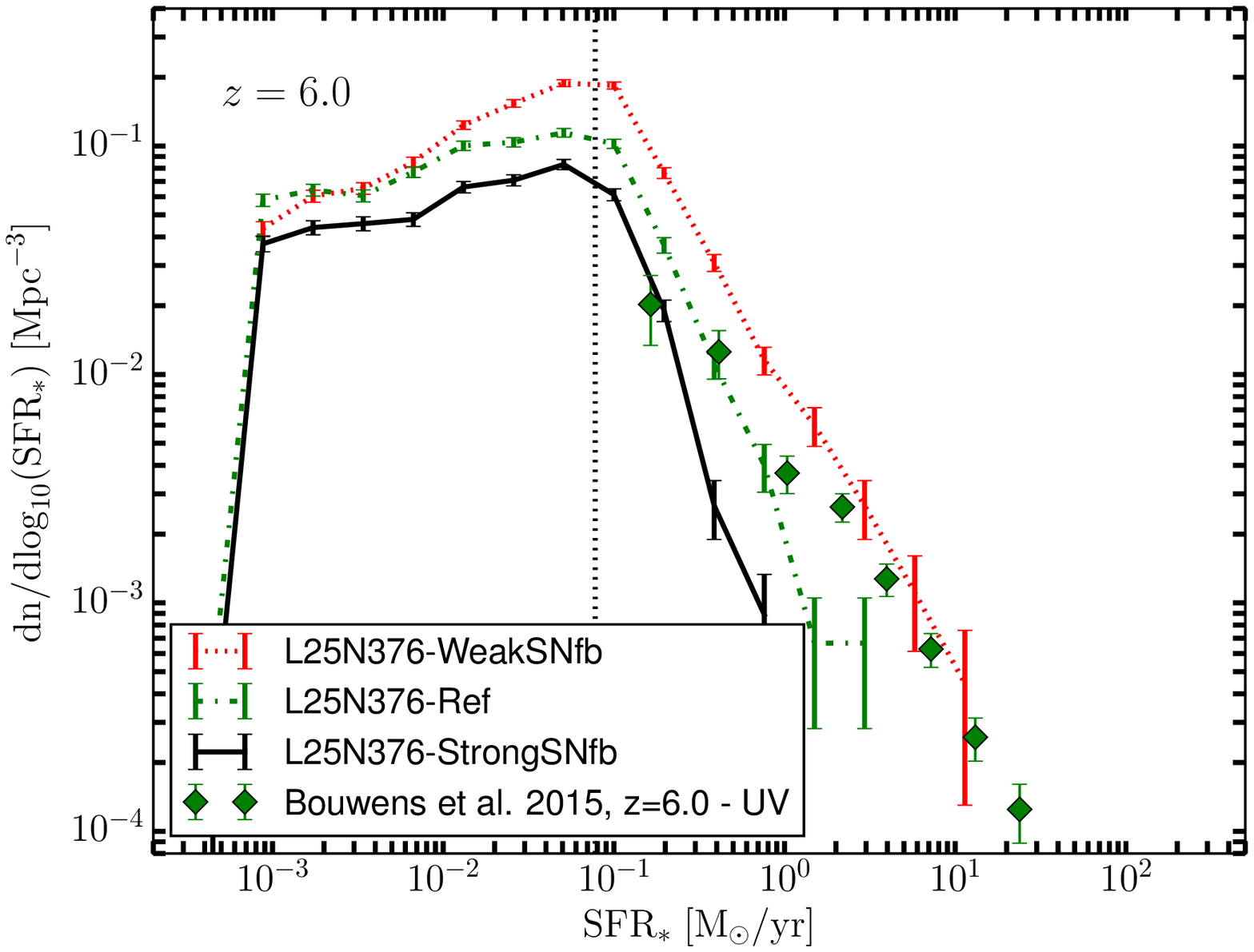}
\includegraphics[scale=0.40]{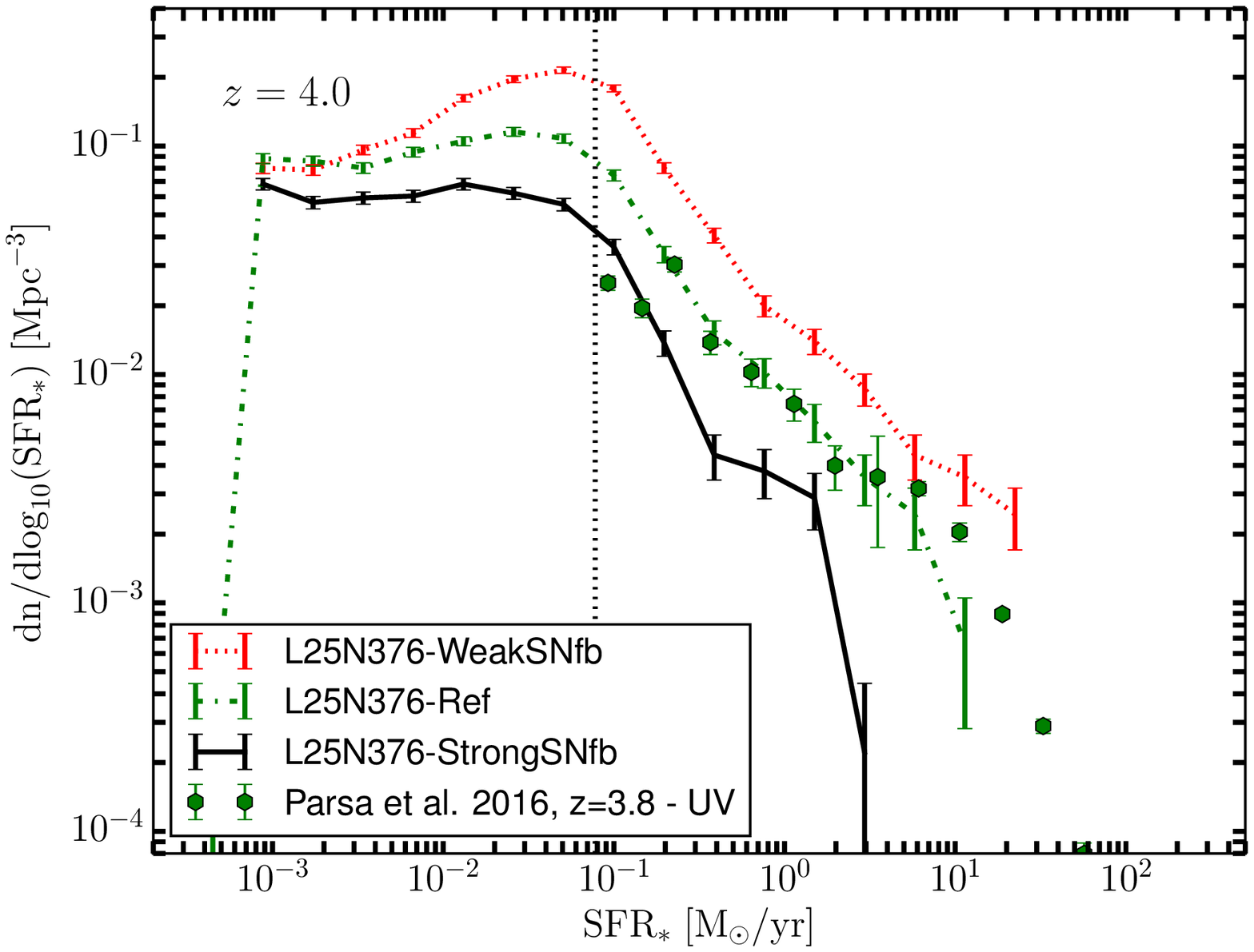}
\includegraphics[scale=0.40]{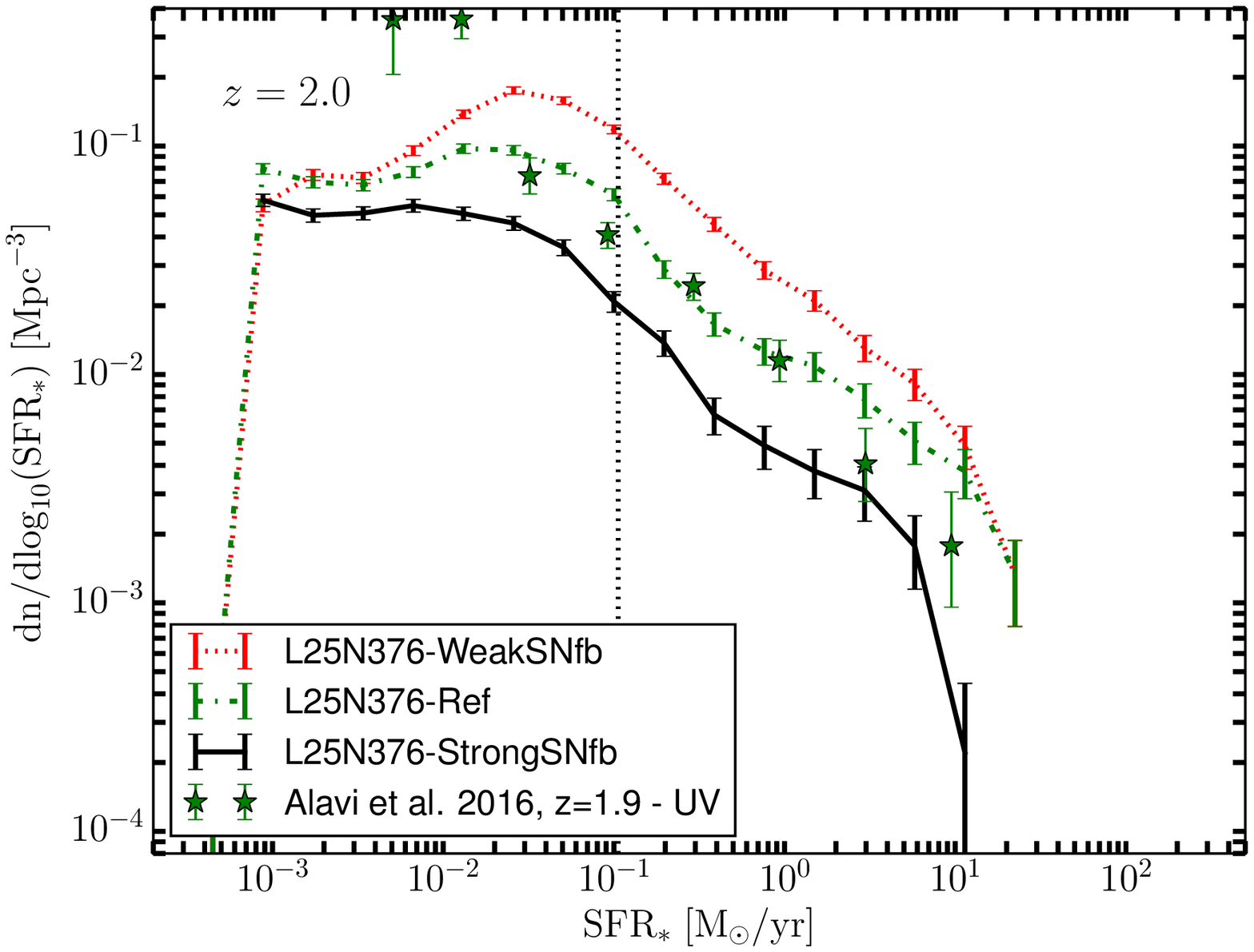}
\includegraphics[scale=0.40]{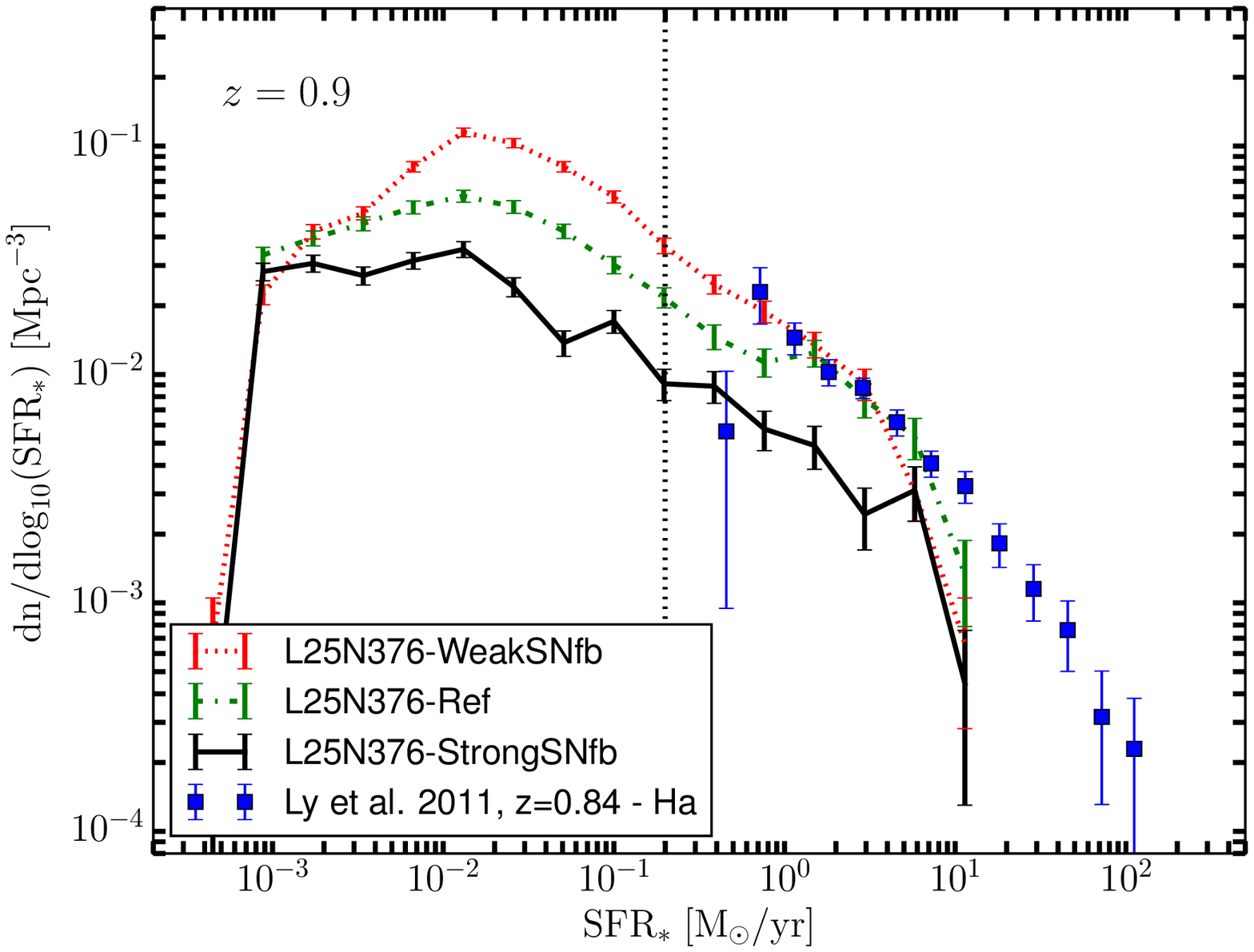}
\includegraphics[scale=0.40]{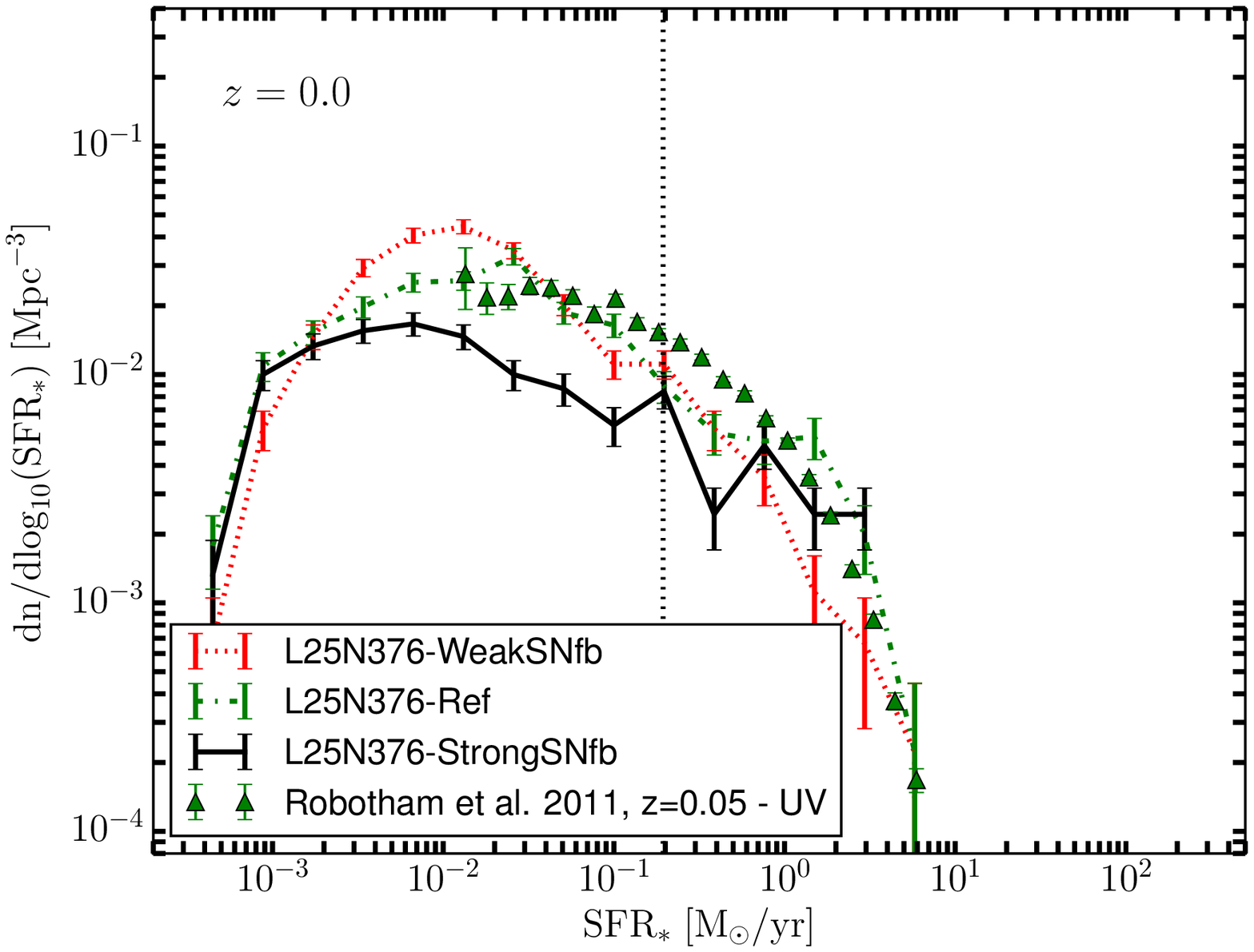}
\caption{We present the effect of the SN feedback prescription used in the EAGLE-Ref simulation on the SFRF at $z \sim 0-8$. At redshifts $z \ge 2.0$ SN feedback decrease effectively the number of intermediate and high star-forming galaxies. At redshift $z \sim0$ the SN feedback is less efficient at decreasing the SFR of highly star-forming systems. The agreement between the three configurations for objects $SFR \le 0.1 \, M_{\odot} \, {\rm yr^{-1}}$ is probably due to the fact that feedback is not resolved at these small galaxies.}
\label{fig:SFRFFeed0}
\end{figure*}

In $\Lambda$CDM simulations there is a tendency for the gas in galaxies to transform into stars too efficiently and too early. As a result, the stellar mass fractions of simulated objects are larger with older stellar populations than those implied by observations \citep{White1991,Guo2010,Moster2013}. There has been a large effort in the last decades to find and understand mechanisms that can decrease this discrepancy. One such mechanism is stellar feedback.

Currently, simulations of large scales lack the resolution necessary to model the self-consistent development of outflows from feedback and rely on subgrid models. The most widely used types of prescriptions in the literature are:
\begin{itemize}
\item injecting energy in kinetic form \citep[i.e.][]{springel2003,dvecchia&schaye08},
\item decoupling wind particles from hydrodynamical forces \citep[i.e.][]{oppe06,TescariKaW2013},
\item turning off radiative cooling and decoupling different thermal phases \citep[i.e.][]{Scannapieco2006},
\item and thermal feedback \citep[i.e.][]{Stinson2006}. 
\end{itemize}

The EAGLE simulations adopt the stochastic thermal feedback scheme described in \citet{DVecchia2012}. In addition to the effect of re-heating interstellar gas from star formation, which  is  already  accounted in by the equation of state, galactic winds produced by Type II Supernovae are also considered following the implementation described below \footnote{The feedback prescription used in EAGLE does not distinguish type II from type Ib/c Supernovae. The physics of the later are not well understood so any event of core-collapse Supernovae is considered to follow the physics of type II incidents.}. When a stellar particle has reached the age $3 \times 10^7 \, {\rm yr}$,  which corresponds to the maximum lifetime of a star that explodes as core collapse supernovae, it injects thermal energy to its neighbouring elements, increasing their internal energy and giving them a temperature jump $\Delta T$. The total available energy per unit stellar mass provided by SNII, ${\rm \epsilon_{SNII} = n_{SNII}E_{SNII}}$, is described by 
\begin{eqnarray}
\label{eSN}
\epsilon_{SNII} = 8.73 \times 10^{15} \, erg \, g^{-1} \left( \frac{n_{SNII}}{1.736 \times 10^{-2} M_{\rm\odot}^{-1}} \right) \times E_{51},
\end{eqnarray}
where $E_{SNII} $ is the available energy from a single SNII, $E_{51}$ a value that is related to the energy released by a single type II SNe and $n_{SNII}$ is their number. Therefore, the amount of energy available in a SSP particle is $m_{\star} \epsilon_{SNII}$, where $m_{\star}$ is the mass of the star particle. To obtain the $E_{51}$ and $n_{SNII}$ in the simulation, the feedback from type II SNe is subject to the two following assumptions:
\begin{enumerate}
\item $6-100$ M$_{\rm\odot}$ stars are the progenitors of type II SNe,
\item each SN releases $10^{51}$ erg (i.e $E_{SNII} = 10^{51} erg$ and $E_{51} = 1$).

\end{enumerate}

In the feedback scheme employed by EAGLE a fraction of the energy given by eq. \ref{eSN}, $f_{th}$, is used. Implementing the density and metallicity requirements described in \citet{Schaye2015} to the functional form of $f_{th}$, the efficiency adopted in the simulations is written as:
\begin{eqnarray}
\label{fth}
f_{th} = f_{th, min} + \frac{f_{th, max}-f_{th, min}}{1+(\frac{Z}{0.1 \, Z_{\rm\odot}})^{n_Z} \, (\frac{n_{H, birth}}{n_{H,0}})^{-n_n}} ),
\end{eqnarray}
where $n_{H,birth}$ is the density of a gas particle at the instant it is converted into a star particle, $n_{H,0} = 0.67 cm^{-3}$ set after comparing test simulations to the observed present-day GSMF and galaxy sizes, $Z_{\odot} = 0.0127$ the solar metallicity, $n_z=n_n = 2/ln10$, $f_{th, min} = 0.3$ and $f_{th, max} = 3$. The maximum value $f_{th, max}$ is achieved at low metallicities and high densities and vice versa \footnote{Values of $f_{th}$ greater than  unity  are  physically motivated by appealing to other sources of energy than supernovae, e.g.  stellar  winds,  radiation  pressure, cosmic rays,  or  if supernovae  yield  more  energy  per  unit mass than initially assumed. An other important motivation is the need to deal with the finite numerical resolution of the simulations.}.

Fig. \ref{fig:SFRFFeed0} shows the effect of scaling the $f_{th}$ function and thus the feedback efficency adopted by the Ref model (L25N376-Ref, dashed green line) by factors of 0.5 (L25N376-WeakSNfb, red dotted line) and 2 (L25N376-StrongSNfb, black solid line) on the SFRF at redshifts $z = 8.0$ (top left panel), $z = 6.0$ (top right panel), $z = 4.0$ (middle left panel), $z = 2.0$ (middle right panel), $z = 0.85$ (bottom left panel) and $z = 0$ (bottom right panel)\footnote{The runs L25N376-Ref, L25N376-WeakSNfb and L25N376-StrongSNfb were performed in a small volume of $25$ MPC and an intermediate resolution same with L100N1504-Ref was employed. This volume is small to sample high star-forming objects. However, we demonstrate in Appendix \ref{ResBox} that if the resolution is kept the same, simulations with the same physics but different volumes produce SFRFs with the same shape, even for relatively high star-forming objects.}. The asymptotic efficiencies of these models are $f^{min}_{th} = (0.15, 0.6)$ and $f^{max}_{th} = (1.5, 6.0)$, respectively. Galaxies in the EAGLE simulations start with high gas fractions and initially form stars too efficiently like in any other $\Lambda$CDM scenario so a feedback prescription is required to be efficient at these epochs. This can be achieved by employing a high $f_{th, max}$ for the subgrid SNe feedback. In addition, since metallicity decreases with increasing redshift at a fixed stellar mass, the metallicity dependence on $f_{th}$ described by eq. \ref{fth} implies a feedback prescription that is relatively more efficient at high redshifts \citep{Schaye2015}. However, at $z \sim8.0$ the effect of the feedback prescription on the SFRF is not strong enough, possibly due to the fact that most of the objects are not resolved properly (see section \ref{SFRFEAGLE0} for more details). On the other hand, at redshift 6 we see that the difference by a factor of 2 in $f_{th}$ affect significantly the SFRF. The number of galaxies of intemediate star-forming objects ($SFR \sim 1-10 \, {\rm M}_{\rm \odot} \, {\rm yr^{-1}}$) is affected almost by the same factor. We note that the run with weak feedback is closer to the observations but the three different configurations we study have issues of resolution that play a major role at these high redshifts at the faint end of the distribution (the resolution employed is the same as the L100N1504-Ref). At redshifts $z \sim4.0$ and $z \sim2.0$ the reference model is in good agreement with observations while the simulation with 2 times lower $f_{th}$ has an abundance of galaxies with intermediate ($SFR \sim 1-10 \, {\rm M}_{\rm \odot} \, {\rm yr^{-1}}$) and low SFRs ($SFR \sim 1-10 \, {\rm M}_{\rm \odot} \, {\rm yr^{-1}}$) larger by a factor of $ \sim 3$. Thus, a small variation in the efficiency of feedback can affect the star formation rate by a large factor at epochs close to the peak of the CSFRD. However, going to lower redshfits we see that the change of $f_{th}$ by the factor of 2 affect the SFRF only by $ \sim 1.5$.  \citet{Crain2015} presented the galaxy stellar mass function at $z \sim0.1$ and demonstrated that a lower (higher) star formation feedback efficiency corresponds to a greater (smaller) abundance of galaxies with masses below the characteristic mass of the \citet{schecter1976} form (i.e. the SNe feedback prescription affects the low mass end of the distribution) almost by a factor of 3. The SFRF at $z \sim0$ is not affected significantly from changes to the $f_{th}$, thus the difference in the stellar masses between the three different configurations reported by \citet{Crain2015} for present day galaxies can be attributed mostly to the effects of SNe feedback at higher redshifts.

We find that SNe feedback is importanct at all redshifts and plays a major part for replicating the observed SFRFs at all epochs. However, we note that this prescription plays an increasing role at decreasing the star formation rates of galaxies with redshift and especially at epochs close to the peak of the CSFRD. In addition, we demonstrated that the SN feedback mechanism employed by EAGLE affects the simulated objects similarly over the entire SFR range. This is in accordance with the findings of \citet{Katsianis2016} where the authors demonstrated that SN feedback prescriptions in cosmological hydrodynamic simulations need to play a major role at changing the abundances of low, intermediate and high star-forming galaxies at $z \sim1-4$ and not only the low star-forming objects to match observations.

\subsection{The effect of AGN feedback on the star formation rate function.}
\label{AGNFeed}

\begin{figure*}
\centering
\includegraphics[scale=0.40]{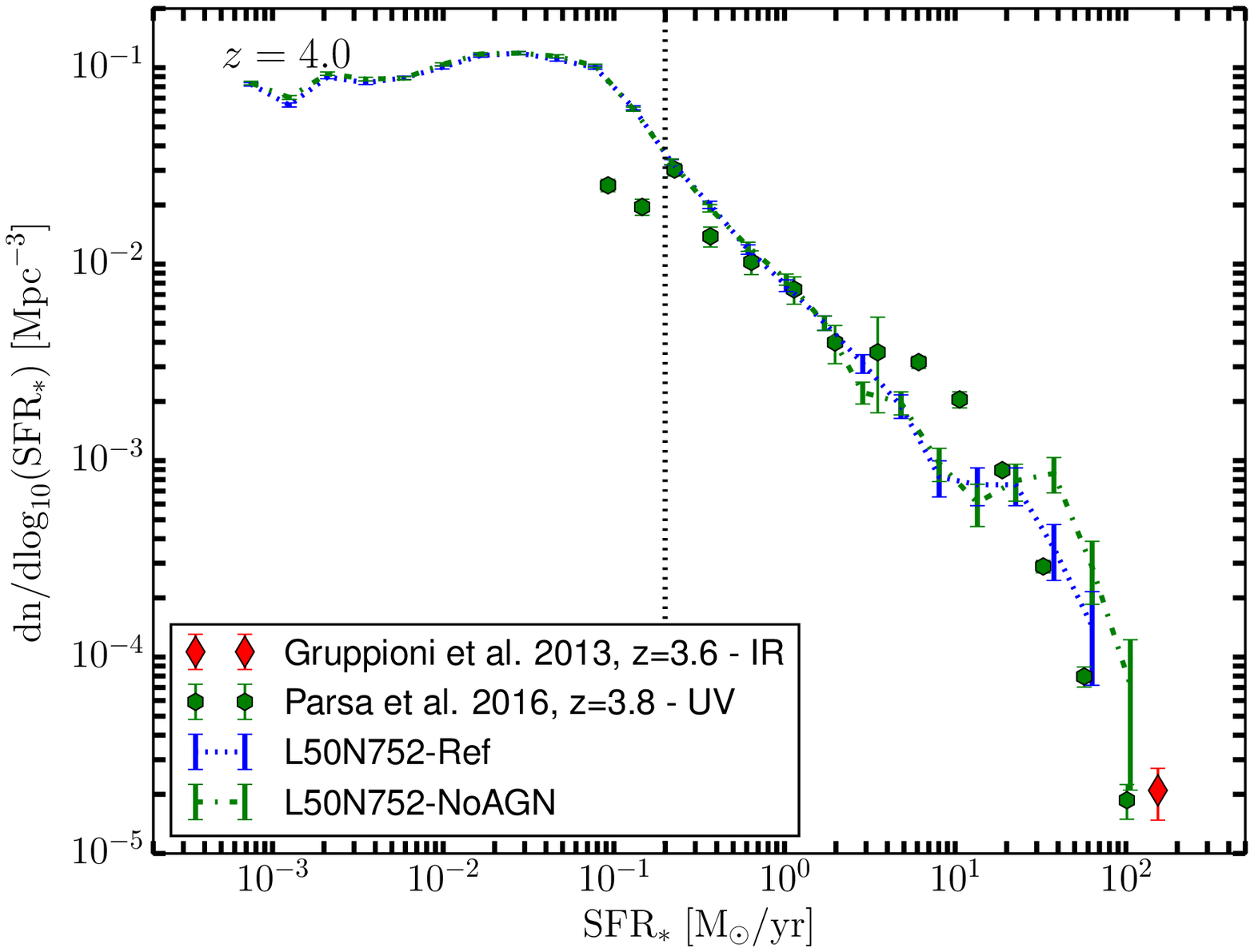}
\includegraphics[scale=0.40]{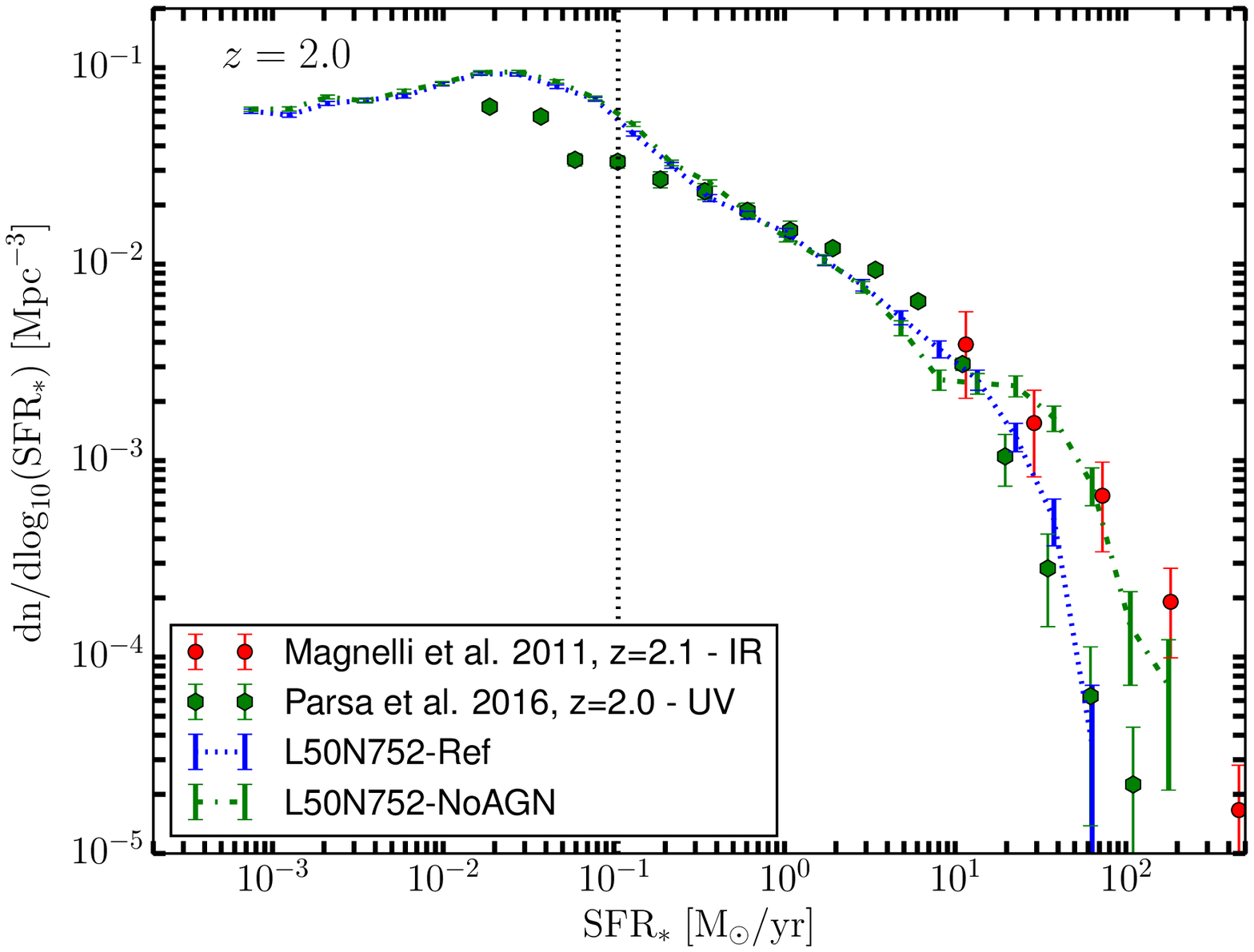}
\includegraphics[scale=0.40]{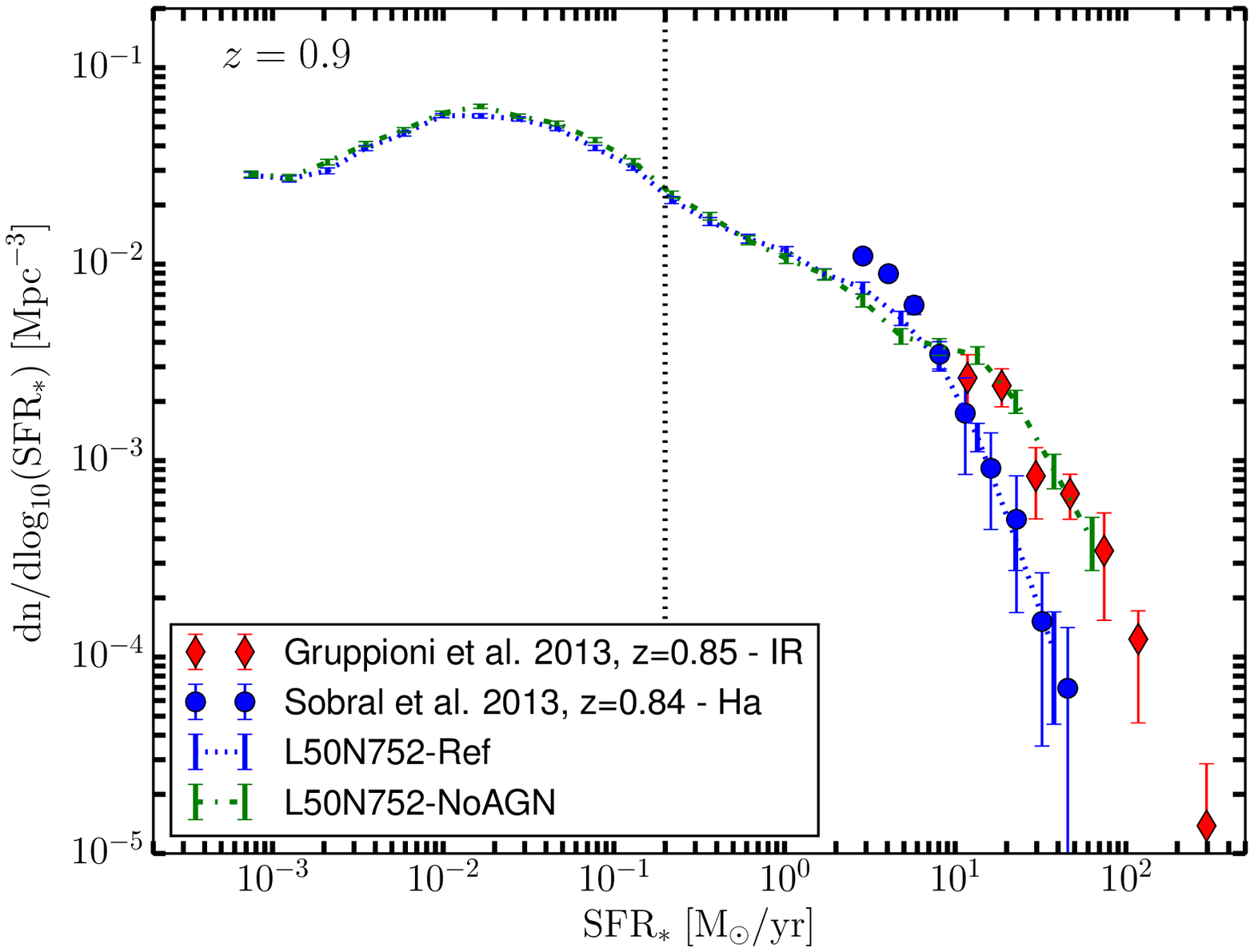}
\includegraphics[scale=0.40]{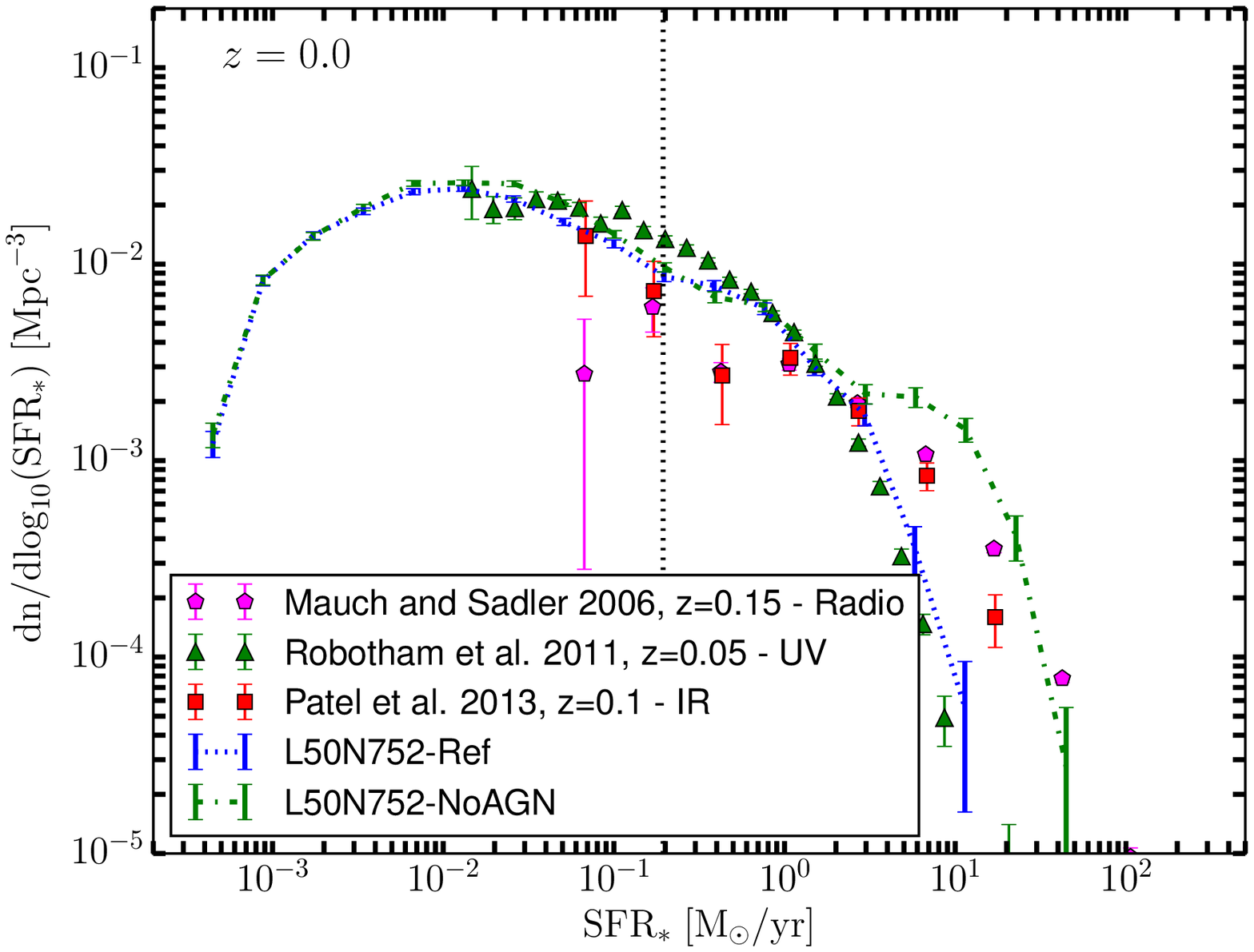}
\caption{We present the effect of the AGN feedback prescription used in the EAGLE-Ref simulation on the SFRF at $z \sim 0-4$. Starting from intermediate redshifts ($z \sim 2$) the AGN feedback mechanism decreases the number of objects with high star formation rates ($SFR \ge 10 \, {\rm M_{\odot} \, yr^{-1}}$). The effect of the prescription becomes significant from intemediate redshifts and shape the high star-forming end at $z \le 2$.}
\label{fig:SFRFFeed0AGN}
\end{figure*}

Since AGN feedback quenches star formation in massive galaxies and regulates the growth of BHs Implementing a feedback prescription associated with Supermasive Black Holes (BHs) in cosmological simulations is essential to reproduce a range of observables like the high mass end of the stellar mass function \citep{Furlong2014} and black hole masses \citep{Rosas2016}. The method that EAGLE employs to seed galaxies with BHs is described by \citet{Springel2005}, where seed BHs are placed at the centre of every halo more massive than $10^{10}$ M$_{\rm\odot}/h$ that does not already contain a BH. When a seed is needed to be implemented at a halo, its highest density gas particle is converted into a collisionless BH particle inheriting the particle mass. These BHs grow by accretion of nearby gas particles or through mergers. The gas accretion obeys the Bondi-Hoyle-Lyttleton formula:
\begin{eqnarray}
  \label{Bondi}
 \dot{m}_{Bondi} = \frac{4 \pi \, G^2 \, m^2_{BH} \, \rho }{(c^2_{s}+u^2)^{3/2}},
\end{eqnarray}
where u the relative velocity of the BH and the gas, $c_{s}$ the sound speed and $\rho$ the density of the gas. EAGLE takes into account gas circulation to calculate a revised Bondi rate $\dot{m}_{bondi, circ}$, which can be written as:
\begin{eqnarray}
  \label{bonnn}
 \dot{m}_{Bondi, circ} = \dot{m}_{Bondi} \times min (C_{visc}^{-1}(c_s/V_{\phi})^3, 1),
\end{eqnarray}
where $\dot{m}_{Bondi}$ is the \citet{Bondi1994} rate applicable to spherically symmetric accretion (eq. \ref{Bondi}), $V_{\Phi}$ is the circulation speed of the gas around the BH computed using equation 16 of \citet{Rosas-Guevara2015} and $C_{visc}$ is a free parameter related to the viscosity of a notional subgrid accretion disc. The accretion rate also cannot exceed the Eddington limit:
\begin{eqnarray}
 \dot{m}_{Edd} = \frac{4 \pi G \, m_{BH} \, m_{p}}{\epsilon_{r} \sigma_T \, c},
\end{eqnarray}
where $m_{p}$ is the proton mass, $\sigma_T$ the Thomson cross section, $\epsilon_{r}$ the radiative efficiency of the accretion disc and c the speed of light. Thus the final accretion rate can be written as:
\begin{eqnarray}
  \label{bonnn}
\dot{m}_{accr} = min(\dot{m}_{Bondi, circ},\dot{m}_{Edd}).
\end{eqnarray}
If the radiation of the accretion disc is taken into account, the growth of the BH can be written as:
\begin{eqnarray}
\dot{m}_{BH} = (1-\epsilon_r) \, \dot{m}_{accr}.
\end{eqnarray}
EAGLE simulations assume a radiative efficiency of $\epsilon_r = 0.1$. Apart from accretion, BHs can grow via merging.

In the EAGLE simulations a single  mode  of  AGN  feedback  is  adopted in which energy is injected thermally and stochastically, in a similar way to energy feedback from star formation described in Section \ref{Feed}. The energy injection rate is specified as:
\begin{eqnarray}
\label{EAGN}
\dot{E}_{AGN} = \epsilon_f \, \epsilon_r \, \dot{m}_{accr} \, c^2,
\end{eqnarray}
where $\epsilon_f$ is the fraction of the radiated energy that couples with the ISM. Like in the case of feedback associated with star formation where $f_{th}$ was specified, the value of $\epsilon_f$ must be chosen by calibrating the simulations to the observations. The parameter $\epsilon_f$ was calibrated by ensuring the normalization of the observed relation between BH mass and stellar mass is reproduced at z = 0 and set to $\epsilon_{f} = 0.15$ as in OWLS simulations. This implies that a fraction of $\epsilon_{f} \, \epsilon_{r} = 0.015$ of the accreted rest mass energy is returned to the local ISM.

\begin{figure*}
\centering
\includegraphics[scale=0.40]{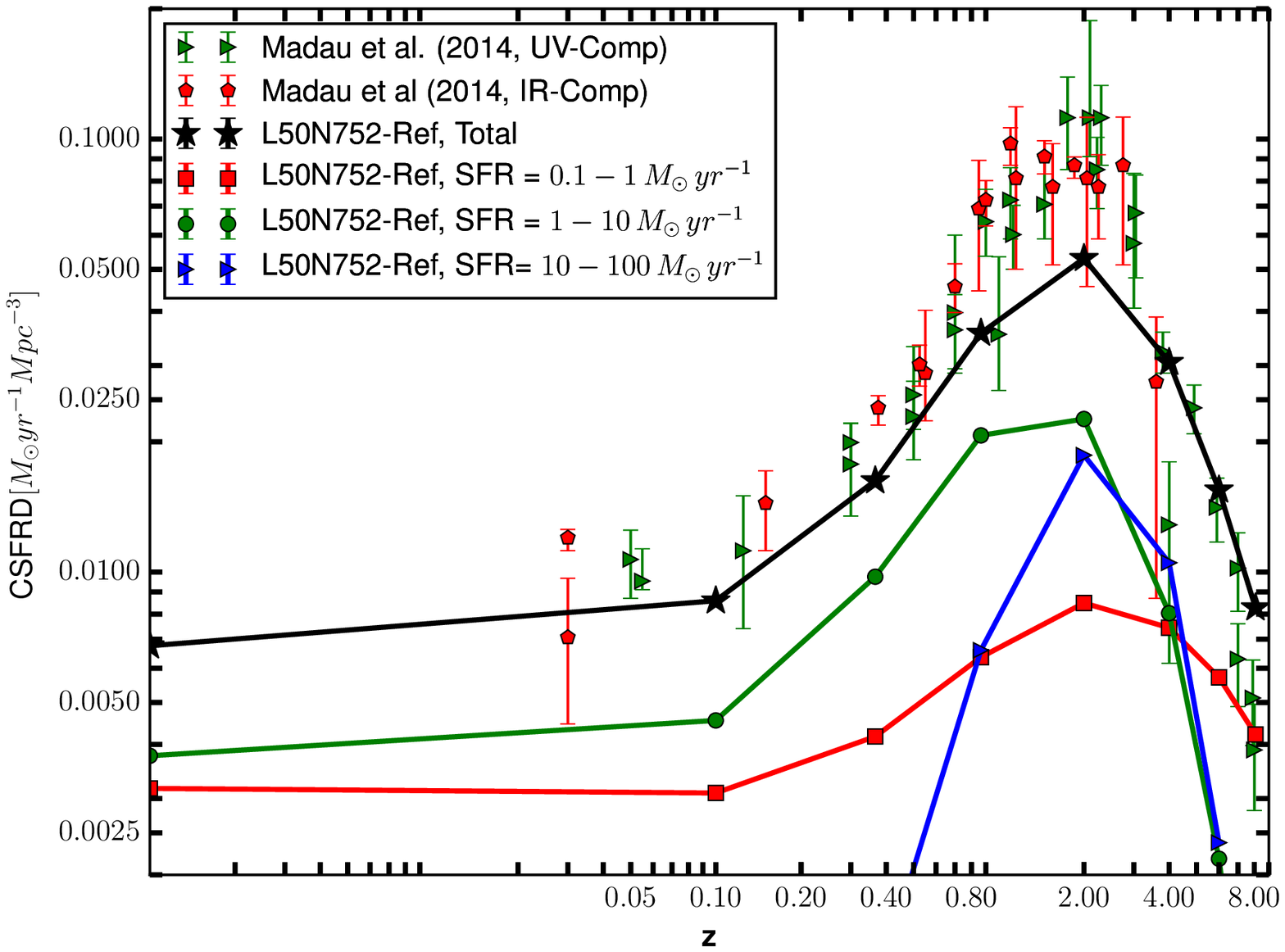}
\includegraphics[scale=0.40]{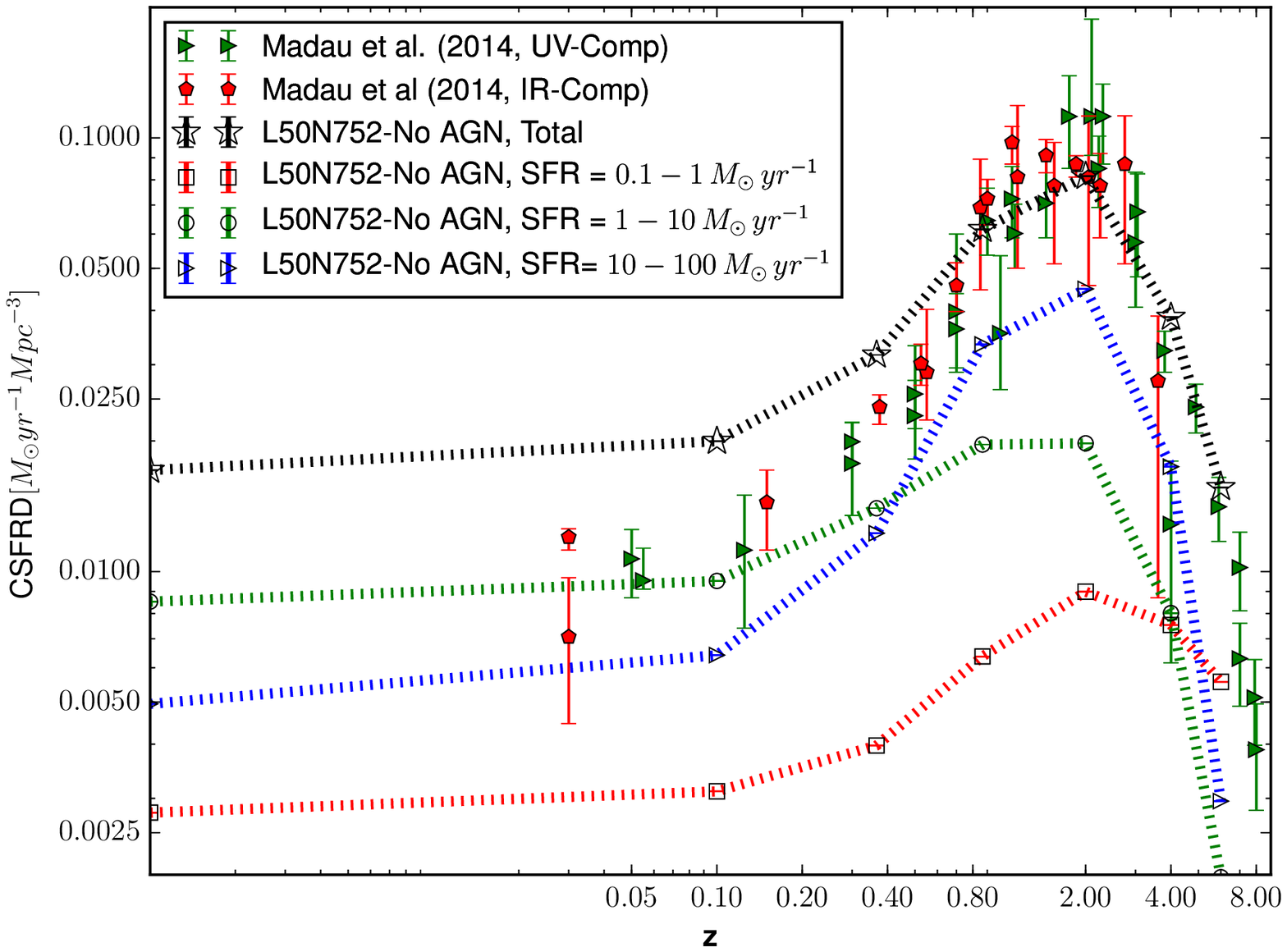}
\caption{We present the effect of the EAGLE AGN feedback prescription on the CSFRD. We compare a run implemented with the AGN feedback mechanism (L50N752-Ref/left panel) and a simulation without (L50N752-NoAGN/right panel). AGN feedback makes its presence at $z < 3$ since it is proportional to the mass of the SMBHs in the simulation which increase with time. In the reference model the contribution of objects with SFR of $10-100 \, {\rm M_{\odot} \, yr^{-1}}$ is increasing at $z \sim 8-2 $ but these galaxies are quenched by the AGN feedback. In the configuration without AGN feedback these objects continue to contribute significantly at $z < 2$ resulting in a CSFRD which is significantly higher than the constraints from observations.}
\label{fig:CSFRDznoAGN}
\end{figure*}

In Fig. \ref{fig:SFRFFeed0AGN} we present the effect of the EAGLE AGN feedback prescription on the SFRF by comparing a run with AGN feedback (L50N752-Ref) and a simulation without (L50N752-NoAGN). Starting from redshift $z \sim4$ we see that the two configurations have almost identical SFRFs in agreement with UV and IR constraints. The energy injection rate from a supermassive black hole described by equation \ref{EAGN} shows a dependency with its accretion rate. The accretion rate is proportional to the mass of the black hole and thus the AGN feedback prescription employed by EAGLE is dependent on the masses of the BHs. At high redshifts (e.g. $z\sim4$) these do not have enough time to grow via mergers or accretion and hence the effect of the AGN feedback on the SFR of galaxies is negligible. However, going to $z \sim2.0$ we see that the AGN feedback decreases the number density of highly star-forming objects ($SFR \sim 10-100 \, {\rm M}_{\rm \odot} \, {\rm yr^{-1}}$). We see that the reference model is in agreement with the UV constraints but the configuration without AGN feedback is actually in better agreement with the IR observations. The effect of the mechanism on the SFRF becomes more prominent at lower redshift and eventually at $z \sim0$ the difference between the L50N752-Ref and L50N752-NoAGN runs can be even by a factor of 10 at the high end of the SFR range.

In Fig. \ref{fig:CSFRDznoAGN} we present the evolution of the CSFRD of L50N752-Ref and L50N752-NoAGN runs at $z \sim0-8$. At high redshifts $z \ge 4.0$ the two configurations are in agreement since AGN feedback is not effective yet. Going to lower redshifts we start seeing that the AGN feedback is responsible for decreasing the CSFRD significantly.  In section \ref{CSFRD} we demonstrated that galaxies with $SFR \sim 10-100 \, {\rm M}_{\rm \odot} \, {\rm yr^{-1}}$ contribute significantly at the CSFRD at its peak era. In the reference model, these are subject to the AGN feedback employed by EAGLE and from redshift $z \le 2$ they significantly suppress any contribution to the cosmic budget. However, in the run where AGN feedback is not implemented (L50N752-NoAGN) galaxies with $SFR \sim 10-100 \, {\rm M}_{\rm \odot} \, {\rm yr^{-1}}$ are not quenched and they keep contributing significantly to the CSFRD at low and intermediate redshifts. We see that the peak of the CSFRD for the L50N752-NoAGN is 1.5 times higher than that of L50N752-Ref. The case without feedback is actually in better agreement with the compilation of UV, H$\alpha$ and IR observations at $z\sim2$. However, going to lower redshifts we see that the configuration without AGN feedback has values $\sim2.5$ larger, which are significantly higher than observational constraints. The prescription is required as well to reproduce a range of observables, like the GSMF at $z \sim0$. We have to note though, that AGN feedback is just a candidate for being the quenching mechanism that is necessary to decrease the mass and SFRs of high mass galaxies at lower redshifts. It is possible that other physical mechanisms are involved to the quenching of high star-forming systems and maybe these could give a better match between observations and simulations at all redshifts for the CSFRD.  

In conclusion, AGN feedback is crucial in EAGLE simulations for decreasing the number of highly star-forming systems below ${\rm z = 2}$. The mechanism becomes more important with time since its effects are proportinal to the masses of supermassive black holes which at high redshifts are small. We note that UV SFRF constraints are in agreement with the reference model but the IR observations are closer to the configuration without AGN feedback implementation. The prescription plays a major role for the peak value of the simulated CSFRD and regulates the cosmic budget of star formation rate at lower redshifts since it affects significantly the numbers of galaxies with  ${\rm SFR \sim 10-100} \, {\rm M}_{\rm \odot} \, {\rm yr^{-1}}$ which have a large contribution if not quenched. This is in agreement with the results of \citet{Vandevoort2011}.

\section{Conclusions and discussion}
\label{concl6}

In this paper, we  investigated the evolution of the galaxy Star Formation Rate Function (SFRF) in the EAGLE simulations comparing the results to a compilation of UV, IR and H$\alpha$ observations at ${\rm z \sim0-8}$. We present the constraints from various star-formation tracers which can be used to constrain models and theory (for detailed tables see the Appendix \ref{table} and \citep{Katsianis2016}). In addition, using cosmological hydrodynamic simulations we explored which halos and what kind of objects dominate the Cosmic Star Formation Rate Density (CSFRD) alongside with the importance of SNe and AGN feedback prescriptions. In the following we summarize the main results and conclusions of our analysis:
\begin{itemize}

\item  There is a tension between the SFRFs of different indicators for high star-forming objects (${\rm SFR \sim 10-100} \, {\rm M}_{\rm \odot} \, {\rm yr^{-1}}$) at $ {\rm 0 \le z \le 2}$ (subsection \ref{SFRFEAGLE0}). This discrepancy is more prominent with time and has its roots possibly in selection biases and the limitations of the different tracers. UV studies are possibly inclomplete for the high star-forming systems, while IR data can give information only for dusty massive galaxies and are limited especially at high redshifts. UV and H$\alpha$ light are subjects to dust attenuation effects and dust correction laws (e.g. IRX-$\beta$, 1 mag correction for the H$\alpha$) suggested in the literature may underestimate these. Thus, both tracers possibly underestimate the SFRF for high star-forming systems with high dust contents. On the other hand, IR light may overestimate the SFRs of galaxies due to various factors (e.g. the dust that it originates from can be heated by older stars, not related to newly born stars or AGN).  

 \item The SFRF of the EAGLE reference model is in good agreement with the constraints from UV and H$\alpha$ observations at ${\rm z \sim0-8}$. IR and Radio data typically suggest a higher number density of high star-forming systems compared to the above at ${\rm z \le2}$  (subsection \ref{SFRFEAGLE0}). There is a slight underproduction of objects between $1-10 \, {\rm M_{\odot} \, {\rm yr^{-1}}}$ in the reference model of EAGLE with respect the observed estimates, which may be due to the feedback implementations used. This small difference can be the origin of the offset between observed and simualted CSFRDs.

\item The cosmic star formation rate density is dominated by galaxies with SFRs of $1-10 \, {\rm M_{\odot} \, yr^{-1}}$ at ${\rm z \le5}$. Objects with lower star formation rates do not contribute significantly to the cosmic budget at these redshifts, despite the fact that they are abundant (subsection \ref{CSFRD}). The peak of the simulated CSFRD at ${\rm z \sim 2}$ is partially driven by a large contribution of rare high star-forming galaxies (${\rm SFR \sim 10-100} \, {\rm M}_{\rm \odot} \, {\rm yr^{-1}}$). They decrease significantly and suddenly at ${\rm z \le0.8}$ due to the presence of AGN feedback.

\item At ${\rm z \ge 5}$ the CSFRD is mostly dominated by a large number of low mass halos (${\rm M_{halo}} = 10^{9-10} \, {\rm M_{\odot}}$) while the contribution from larger objects (${\rm M_{halo}} \ge 10^{11} \, {\rm M_{\odot}}$) is negligible (section \ref{SFRFhalos}). However, halos grow rapidly at high redshifts due to mergers and accretion. There is a sudden and significant increase in the numbers and contribution of halos with masses ${\rm M_{halo}} = 10^{11-12} \, {\rm M_{\odot}}$ to the CSFRD at ${\rm z \sim5}$. We note that galaxies which reside in halos in this mass range represent less than the $\sim 1 \%$ of the total population at ${\rm z \sim5}$ but still, these halos are so efficient at forming stars that they dominate the total budget of cosmic star formation (section \ref{SFRFhalos}). The above objects keep increasing in numbers at lower redshifts but at a relatively slower rate and their numbers are kept almost constant after ${\rm z \sim2}$, when the peak of the CSFRD is finally achieved. Halos in the mass interval of ${\rm M_{halo}} = 10^{11-12} \, {\rm M_{\odot}}$ have been dominating the CSFRD for most of the history of the Universe (i.e. most of the stars in the Universe, including ours, were born in Milky way-like halos). 
  
\item We find that SNe feedback is of great importance at all redshifts and plays a major part in replicating the observations at all epochs. We note that the prescription used by EAGLE plays an increasing role at decreasing the star formation rates of galaxies at higher redshifts and especially at epochs close to the peak of the CSFRD (subsection \ref{SNe}). We demonstrated that the mechanism affects the simulated objects similarly at all SFR regimes for ${\rm z \ge 1}$ and not only at the faint end of the distribution.

\item AGN feedback is crucial for decreasing the number of high star-forming systems (${\rm SFR \sim 10-100} \, {\rm M}_{\rm \odot} \, {\rm yr^{-1}}$) and a thoughtful tuning is required to bring observations and simulations in agreement. The mechanism becomes prominent with time and plays a major role for the peak value of the simulated CSFRD since it affects the high star-forming systems which, if not quenched, can rise the cosmic budget of star formation rate at lower redshifts extensively (subsection \ref{AGNFeed}).   

\item We require higher resolution simulations to make meaningful comparisons between the observed and simulated SFRFs at ${\rm z \ge 4}$ for low star-forming objects. Even the EAGLE reference model L100N1504-Ref which is one of the state-of-the-art simulations in terms of resolution and volume is limited at high redshifts. In appendix \ref{ResBox}, we perform resolution and boxsize tests and demonstrate that simulations with representative cosmological volumes (e.g. 25 or 50 Mpc) produce similar populations of galaxies in terms of star formation rate with configurations which employ significantly larger box-sizes (e.g. 100 Mpc), provided that the resolution and subgrid physics are the same. Thus, for the study of SFRs of low and intermediate star-forming galaxies we suggest higher resolution cosmological hydrodynamic simulations run in representative volumes and a larger focus on subgrid physics. Larger volumes can be useful for the study of high star-forming systems and possibly can unravel the reasons for the tension between different SFR indicators.  

\end{itemize}

\section*{Acknowledgments}

We would like to thank Joop Schaye, Edoardo Tescari, Stuart Wyithe, Kristian Finlator, Lee Spitler, Marko Stalevski and the anonymous referee for their suggestions and insightful discussions on the subject. We would also like to thank Carlotta Gruppioni, Kenneth Duncan, Naveen Reddy, Richard Bouwens, Renske Smit, David Sobral, Shaghayegh Parsa, Chun Ly, Harsit Patel, Aaron Robotham, Lucia Marchetti and Tom Mauch for making their results publicly available. This work used the DiRAC Data Centric system at  Durham  University,  operated  by  the  Institute for  Computational  Cosmology  on  behalf  of  the  STFC  DiRAC  HPC  Facility  (www.dirac.ac.uk). A.K is supported by the {\it CONICYT/FONDECYT fellowship, project number: 3160049}. G.B. is supported by {\it CONICYT/FONDECYT, Programa de Iniciacion, Folio 11150220}. N.T. acknowledges support from {\it CONICYT PAI/82140055}. V.G. was supported by {\it CONICYT/FONDECYT iniciation grant number 11160832}. S.L has been supported by {\it CONICYT/FONDECYT, grant number 1140838}.

\bibliographystyle{mn2e}	
\bibliography{Katsianis_mnrasRev.bbl}

\appendix

\section{The observed UV, IR and H$\alpha$ star formation rate function.}
\label{table}

In this Appendix we present the observed SFRF at ${\rm z \sim0-8}$, used for this work. The following tables are complimentary to those reported by \citet{smit12}, \citet{Duncan2014} and \citet{Katsianis2016}. Alltogether, they give a description of the evolution of UV, IR and H$\alpha$ SFRF functions for most of the history of the Universe. For the following estimations a \citet{chabrier03} IMF and $\Lambda$CDM cosmology same with EAGLE was assumed. We note that an uncertainty of $50 \%$ in the Keniccut calibrations could lead to uncertainties for the estimates of the observed SFR by $\sim 0.3$ dex.

\begin{table}
  \centering
\resizebox{0.45\textwidth}{!}{%
  \begin{tabular}{cccc}
    \hline \\
    & {\large $\frac{{\rm SFR}}{{\rm M}_{\odot}\ {\rm yr}^{-1}}$}  &
    {\large $ dn/dlog_{10} (\phi_{\rm SFR})\ \left({\rm Mpc}^{-3}\ \right) \times 10^{-2}$ } \\ \\
    \hline \hline
    & $z\sim8.0$  & UV-dust corrected \\ 
    \hline
    & 43.269 & 0.0010$\pm$0.0006 \\
    & 21.704 & 0.0026$\pm$0.0010 \\
    & 10.891 & 0.0116$\pm$0.0030 \\
    & 5.469 & 0.0120$\pm$0.0050 \\
    & 2.850 & 0.0662$\pm$0.0208 \\
    & 1.803 & 0.1066$\pm$0.0452 \\
    & 0.902 & 0.2120$\pm$0.0680 \\
    & 0.359 & 0.5480$\pm$0.2080 \\
    \hline 
    & $z\sim7.0$  & UV-dust corrected \\ 
    \hline
    & 73.533 & 0.0002$\pm$0.0004 \\
    & 41.186 & 0.0062$\pm$0.0017 \\
    & 23.070 & 0.0090$\pm$0.0028 \\
    & 12.921 & 0.0362$\pm$0.0064 \\
    & 7.239 & 0.0578$\pm$0.0114 \\
    & 4.235 & 0.1224$\pm$0.0187 \\
    & 2.674 & 0.1697$\pm$0.0331 \\
    & 1.687 & 0.3212$\pm$0.0894 \\
    & 0.534 & 1.0925$\pm$0.2731 \\
    & 0.171 & 1.5901$\pm$0.5499 \\
    \hline
    & $z\sim6.0$  & UV-dust corrected \\ 
    \hline
    & 141.748 & 0.0004$\pm$0.0004 \\
    & 77.951 & 0.0028$\pm$0.0012 \\
    & 42.862 & 0.0100$\pm$0.0024 \\
    & 23.585 & 0.0330$\pm$0.0047 \\
    & 12.974 & 0.0598$\pm$0.0077 \\
    & 7.132 & 0.1305$\pm$0.0015 \\
    & 3.921 & 0.2330$\pm$0.0026 \\
    & 1.860 & 0.3554$\pm$0.0598 \\
    & 0.742 & 1.2496$\pm$0.2581 \\ 
    & 0.309 & 2.5517$\pm$0.7857 \\
    \hline
    & $z\sim5.0$  & UV-dust corrected \\ 
    \hline
    & 382.081 & 0.0004$\pm$0.0004 \\
    & 208.215 & 0.0012$\pm$0.0006 \\
    & 113.462 & 0.0063$\pm$0.0015 \\
    & 61.828 & 0.0189$\pm$0.0026 \\
    & 33.695 & 0.0495$\pm$0.0047 \\
    & 18.369 & 0.1270$\pm$0.0086 \\
    & 10.001 & 0.1925$\pm$0.0125 \\
    & 5.452 & 0.2486$\pm$0.0175 \\    
    & 2.974 & 0.3900$\pm$0.0319 \\
    & 1.280 & 0.8343$\pm$0.0101 \\
    & 0.512 & 1.6080$\pm$0.0331 \\
    & 0.203 & 4.5640$\pm$0.0133 \\    
\hline \hline
  \end{tabular}%
}
  \caption{Stepwise SFR functions at $z\sim5-8$ using the luminosity functions from \citet[][ UV]{Bouwens2016}, equation \ref{eq:SFRpara3} and the dust corrections described in section \ref{dustcorrectionlaws}.}
  \label{tab_stepsfrf3}
\end{table}

\begin{table}
  \centering
\resizebox{0.45\textwidth}{!}{%
  \begin{tabular}{cccc}
    \hline \\
    & {\large $\frac{{\rm SFR}}{{\rm M}_{\odot}\ {\rm yr}^{-1}}$}  &
    {\large $ dn/dlog_{10} (\phi_{\rm SFR})\ \left({\rm Mpc}^{-3}\ \right) \times 10^{-2}$ } \\ \\
    \hline \hline
    & $z \sim2.6$  & UV-dust corrected \\ 
    \hline
    & 11.662 & 0.2020$\pm^{0.1216}_{0.0804}$ \\
    & 3.214 & 0.4714$\pm^{0.1627}_{0.1235}$ \\ 
    & 0.881 & 0.9121$\pm^{0.2113}_{0.1739}$ \\
    & 0.259 & 2.2186$\pm^{0.2843}_{0.2843}$ \\
    & 0.938 & 4.0594$\pm^{0.4957}_{0.4957}$ \\
    & 0.371 & 13.1310$\pm^{1.7996}_{1.7996}$ \\
    & 0.015 & 39.1263$\pm^{10.5453}_{8.4968}$ \\
    & 0.006 & 19.2142$\pm^{25.3427}_{12.4122}$ \\
    \hline 
    & $z \sim1.9$  & UV-dust corrected \\ 
    \hline
    & 10.337 & 0.1777$\pm^{0.1197}_{0.0767}$ \\
    & 3.259 & 0.4059$\pm^{0.1627}_{0.1197}$ \\
    & 1.024 & 1.1150$\pm^{0.2451}_{0.2057}$ \\
    & 0.325 & 2.4450$\pm^{0.3105}_{0.3105}$ \\
    & 0.102 & 4.0894$\pm^{0.4920}_{0.4920}$ \\
    & 0.035 & 7.4155$\pm^{1.3675}_{1.1673}$ \\
    & 0.014 & 35.9610$\pm^{7.1612}_{6.0518}$ \\
    & 0.006 & 35.7384$\pm^{21.3488}_{14.1764}$ \\      
    \hline
    & $z \sim1.3$  & UV-dust corrected \\ 
    \hline
    & 13.599 & 0.1291$\pm^{0.0767}_{0.0505}$ \\
    & 4.266 & 0.5986$\pm^{0.1484}_{0.1141}$ \\
    & 1.339 & 0.9166$\pm^{0.1664}_{0.1421}$ \\
    & 0.420 & 1.0158$\pm^{0.1852}_{0.1571}$ \\
    & 0.132 & 1.1990$\pm^{0.2563}_{0.2562}$ \\
    & 0.041 & 4.2391$\pm^{0.5761}_{0.5761}$ \\
    & 0.014 & 15.4802$\pm^{3.5113}_{2.9711}$ \\
    & 0.006 & 39.8072$\pm^{14.3915}_{10.8988}$ \\
\hline \hline
  \end{tabular}%
}
  \caption{Stepwise SFR functions at $z\sim1.3-2.6$ using the luminosity functions from \citet[][ UV]{Alavi2016}, equation \ref{eq:SFRpara3} and the dust corrections described in section \ref{dustcorrectionlaws}.}
  \label{tab_stepsfrf3}
\end{table}

\begin{table}
  \centering
\resizebox{0.45\textwidth}{!}{%
  \begin{tabular}{cccc}
    \hline \\
    & {\large $\frac{{\rm SFR}}{{\rm M}_{\odot}\ {\rm yr}^{-1}}$}  &
    {\large $ dn/dlog_{10} (\phi_{\rm SFR})\ \left({\rm Mpc}^{-3}\ \right) \times 10^{-2}$ } \\ \\
    \hline \hline
    & $z \sim0.4$  & H$\alpha$-dust corrected (1 mag) \\ 
    \hline
    & 16.998 & 0.0182$\pm{0.0157}$ \\
    & 8.513 & 0.0257$\pm{0.0190}$ \\
    & 5.374 & 0.0873$\pm{0.0516}$ \\
    & 4.271 & 0.1233$\pm{0.0486}$ \\
    & 3.397 & 0.1742$\pm{0.0603}$ \\
    & 2.695 & 0.1910$\pm{0.0723}$ \\
    & 2.144 & 0.2518$\pm{0.0696}$ \\
    & 1.709 & 0.3243$\pm{0.0776}$ \\
    & 1.353 & 0.3556$\pm{0.0782}$ \\
    & 1.071 & 0.4797$\pm{0.0960}$ \\
    & 0.852 & 0.5023$\pm{0.0903}$ \\
    & 0.684 & 0.7095$\pm{0.1420}$ \\
    & 0.542 & 0.8729$\pm{0.1389}$ \\
    & 0.435 & 1.0986$\pm{0.1632}$ \\
    & 0.343 & 1.4487$\pm{0.1277}$ \\
    & 0.276 & 1.4487$\pm{0.1363}$ \\
    & 0.217 & 1.8663$\pm{0.1756}$ \\
    & 0.171 & 2.0463$\pm{0.1924}$ \\ 
\hline \hline
  \end{tabular}%
}
  \caption{Stepwise SFR functions at $z\sim0.4$ using the luminosity functions from \citet[][ H$\alpha$]{Sobral2013} and equation \ref{eq:SFRpara1}.}
  \label{tab_stepsfrf3}
\end{table}

\begin{table}
  \centering
\resizebox{0.45\textwidth}{!}{%
  \begin{tabular}{cccc}
    \hline \\
    & {\large $\frac{{\rm SFR}}{{\rm M}_{\odot}\ {\rm yr}^{-1}}$}  &
    {\large $ dn/dlog_{10} (\phi_{\rm SFR})\ \left({\rm Mpc}^{-3}\ \right) \times 10^{-2}$ } \\ \\
    \hline \hline
    & $z \sim0.3$  & IR \\ 
    \hline
    & 116.953 & 0.0017$\pm^{0.0007}_{0.0018}$ \\
    & 46.564 & 0.0129$\pm^{0.0024}_{0.0030}$ \\
    & 18.541 & 0.0760$\pm^{0.0105}_{0.0105}$ \\
    & 7.382 & 0.1742$\pm^{0.0441}_{0.0602}$ \\
    & 2.948 & 0.1025$\pm^{0.0567}_{0.1512}$ \\
    \hline \hline
    & $z \sim0.375$  & IR \\ 
    \hline
    & 205.554 & 0.0002$\pm$0.0002 \\
    & 65.002 & 0.0048$\pm$0.0011 \\
    & 20.557 & 0.0709$\pm$0.0049 \\
    & 6.504 & 0.2143$\pm$0.0197 \\
    & 2.061 & 0.3990$\pm$0.1011 \\
\hline \hline
  \end{tabular}%
}
  \caption{Stepwise SFR functions at $z \sim 0.3$ and $z \sim 0.375$ using the luminosity functions from \citet[][ IR]{Patel13} and \citet[][ IR]{Gruppionis13}, respectively and equation \ref{eq:SFRpara2}.}
  \label{tab_stepsfrf3}
\end{table}

\begin{table}
  \centering
\resizebox{0.45\textwidth}{!}{%
  \begin{tabular}{cccc}
    \hline \\
    & {\large $\frac{{\rm SFR}}{{\rm M}_{\odot}\ {\rm yr}^{-1}}$}  &
    {\large $ dn/dlog_{10} (\phi_{\rm SFR})\ \left({\rm Mpc}^{-3}\ \right) \times 10^{-2}$ } \\ \\
    \hline \hline
    & $z \sim0.15$  & Radio \\ 
    \hline
    & 267.179 & 0.0002$\pm_{0.0001}^{0.0001}$ \\
    & 106.363 & 0.0009$\pm_{0.0001}^{0.0001}$ \\
    & 42.345 & 0.0077$\pm_{0.0004}^{0.0004}$ \\
    & 16.866 & 0.0356$\pm_{0.0016}^{0.0016}$ \\
    & 6.711 & 0.1073$\pm_{0.0025}^{0.0025}$ \\
    & 2.671 & 0.1954$\pm_{0.0090}^{0.0090}$ \\
    & 1.063 & 0.3096$\pm_{0.0214}^{0.0214}$ \\
    & 0.423 & 0.2824$\pm_{0.0325}^{0.0260}$ \\
    & 0.169 & 0.6038$\pm_{0.1530}^{0.1252}$ \\
    & 0.067 & 0.2759$\pm_{0.2470}^{0.1272}$ \\       
\hline \hline
  \end{tabular}%
}
  \caption{Stepwise SFR functions at $z\sim0.15$ using the luminosity function from \citet[][ Radio]{Mauch2007} and the radio-SFR conversion law given by \citet{Sullivan2001}.}
  \label{tab_stepsfrf3}
\end{table}

\begin{table}
  \centering
\resizebox{0.45\textwidth}{!}{%
  \begin{tabular}{cccc}
    \hline \\
    & {\large $\frac{{\rm SFR}}{{\rm M}_{\odot}\ {\rm yr}^{-1}}$}  &
    {\large $ dn/dlog_{10} (\phi_{\rm SFR})\ \left({\rm Mpc}^{-3}\ \right) \times 10^{-2}$ } \\ \\
    \hline \hline
    & $z \sim0.1$  & IR \\ 
    \hline
    & 271.705 & 0.0005$\pm$0.0003  \\
    & 43.062 & 0.0003$\pm$0.0005 \\
    & 17.143 & 0.0159$\pm$0.0047 \\
    & 6.825 & 0.0837$\pm$0.0135 \\
    & 2.717 & 0.1790$\pm$0.0289 \\
    & 1.082 & 0.3334$\pm$0.0614 \\
    & 0.431 & 0.2710$\pm$0.1186 \\
    & 0.171 & 0.7293$\pm$0.3025 \\    
    & 0.068 & 1.3897$\pm$0.7044 \\
\hline \hline
  \end{tabular}%
}
  \caption{Stepwise SFR functions at $z \sim 0.1$ using the luminosity functions from \citet[][ IR]{Patel13} and equation \ref{eq:SFRpara2}.}
  \label{tab_stepsfrf3}
\end{table}

\begin{table}
  \centering
\resizebox{0.45\textwidth}{!}{%
  \begin{tabular}{cccc}
    \hline \\
    & {\large $\frac{{\rm SFR}}{{\rm M}_{\odot}\ {\rm yr}^{-1}}$}  &
    {\large $ dn/dlog_{10} (\phi_{\rm SFR})\ \left({\rm Mpc}^{-3}\ \right) \times 10^{-2}$ } \\ \\
    \hline \hline
    & $z \sim0.06$  & IR \\ 
    \hline
    & 15.634 & 0.0060$\pm$0.0035 \\
    & 10.817 & 0.0100$\pm$0.0045 \\
    & 7.313 & 0.0460$\pm$0.0096 \\
    & 4.944 & 0.0764$\pm$0.0123 \\
    & 3.343 & 0.1078$\pm$0.0149 \\
    & 2.312 & 0.2773$\pm$0.0242 \\
    & 1.563 & 0.3917$\pm$0.0297 \\
    & 1.082 & 0.4931$\pm$0.0392 \\
    & 0.731 & 0.6500$\pm$0.0504 \\
    & 0.494 & 0.6652$\pm$0.0897 \\
    & 0.334 & 0.5928$\pm$0.0876 \\
    & 0.231 & 0.9145$\pm$0.1790 \\
    & 0.156 & 0.6807$\pm$0.1790 \\
    & 0.108 & 0.6965$\pm$0.2202 \\
\hline \hline
  \end{tabular}%
}
  \caption{Stepwise SFR functions at $z \sim 0.1$ using the luminosity functions from \citet[][ IR]{Marchetti2016} and equation \ref{eq:SFRpara2}.}
  \label{tab_stepsfrf3}
\end{table}

\begin{table}
  \centering
\resizebox{0.45\textwidth}{!}{%
  \begin{tabular}{cccc}
    \hline \\
    & {\large $\frac{{\rm SFR}}{{\rm M}_{\odot}\ {\rm yr}^{-1}}$}  &
    {\large $ dn/dlog_{10} (\phi_{\rm SFR})\ \left({\rm Mpc}^{-3}\ \right) \times 10^{-2}$ } \\ \\
    \hline \hline
    & $z \sim0.05$  & UV-Dust corrected \\ 
    \hline
    & 20.611 & 0.0007$\pm$0.0007 \\
    & 11.544 & 0.0004$\pm$0.0004 \\
    & 8.640 & 0.0049$\pm$0.0014 \\
    & 6.466 & 0.0147$\pm$0.0017 \\
    & 4.839 & 0.0326$\pm$0.0028 \\
    & 3.622 & 0.0740$\pm$0.0042 \\
    & 2.711 & 0.1235$\pm$0.0052 \\
    & 2.028 & 0.2118$\pm$0.0070 \\
    & 1.518 & 0.3107$\pm$0.0087 \\
    & 1.113 & 0.4507$\pm$0.0112 \\
    & 0.850 & 0.5640$\pm$0.0140 \\
    & 0.636 & 0.7246$\pm$0.0186 \\
    & 0.476 & 0.8337$\pm$0.0235 \\
    & 0.356 & 1.0438$\pm$0.0312 \\
    & 0.267 & 1.2122$\pm$0.0403 \\
    & 0.199 & 1.3437$\pm$0.0484 \\
    & 0.149 & 1.4899$\pm$0.0624 \\
    & 0.112 & 1.8835$\pm$0.0863 \\
    & 0.084 & 1.6113$\pm$0.1199 \\
    & 0.063 & 1.9389$\pm$0.1210 \\
    & 0.047 & 2.1058$\pm$0.1554 \\
    & 0.035 & 2.1441$\pm$0.1884 \\
    & 0.026 & 1.9287$\pm$0.2448 \\
    & 0.020 & 1.0907$\pm$0.2999 \\
    & 0.015 & 2.4152$\pm$0.7257 \\
\hline \hline
  \end{tabular}%
}
  \caption{Stepwise SFR functions at $z \sim 0.05$ using the luminosity functions from \citet[][ UV]{Robotham2011}, equation \ref{eq:SFRpara3} and the dust corrections described in section \ref{dustcorrectionlaws}.}
  \label{tab_stepsfrf3}
\end{table}

\section{Resolution and boxsize effects on the EAGLE star formation rate function.}
\label{ResBox}

\begin{figure}
\centering
\includegraphics[scale=0.40]{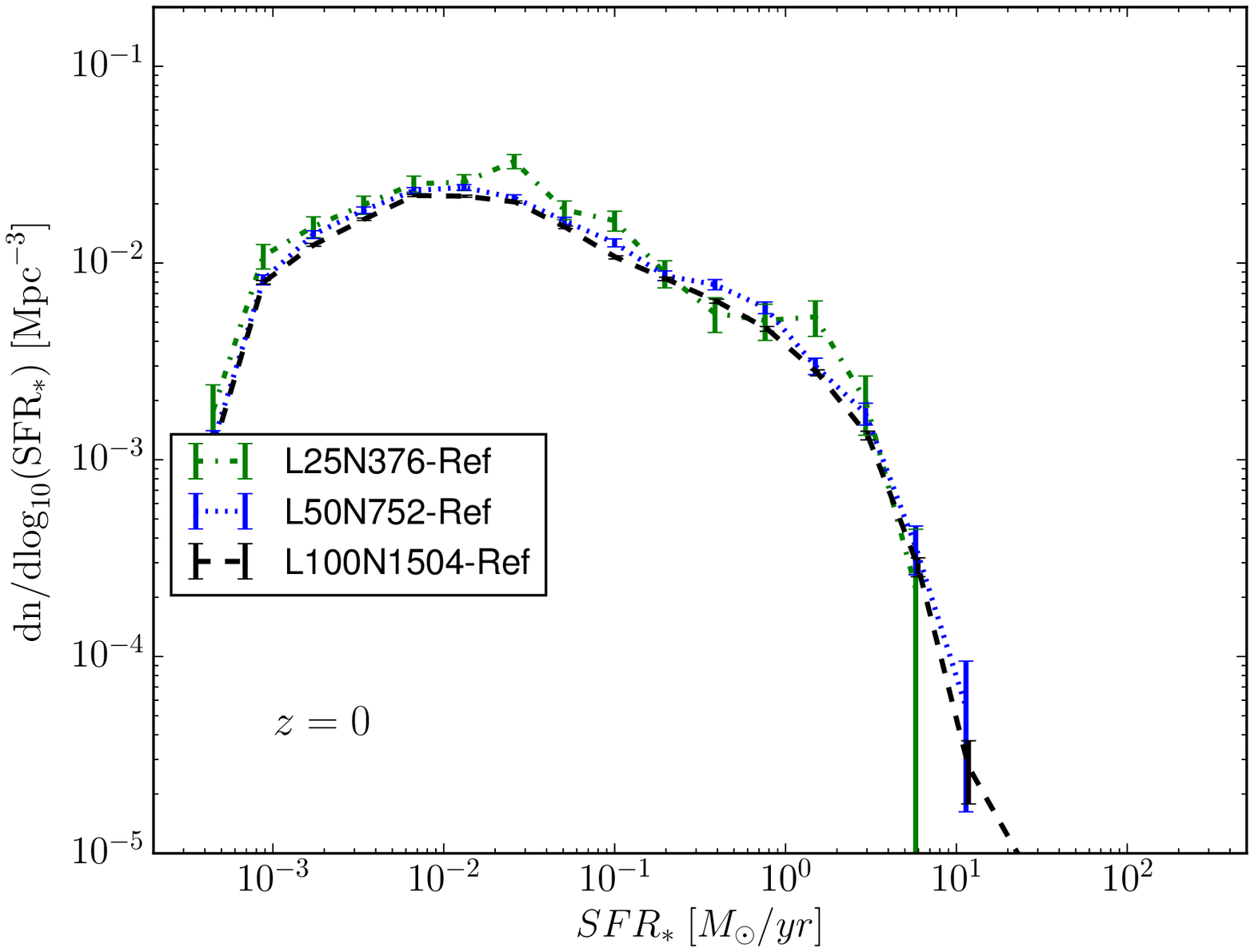}
\includegraphics[scale=0.40]{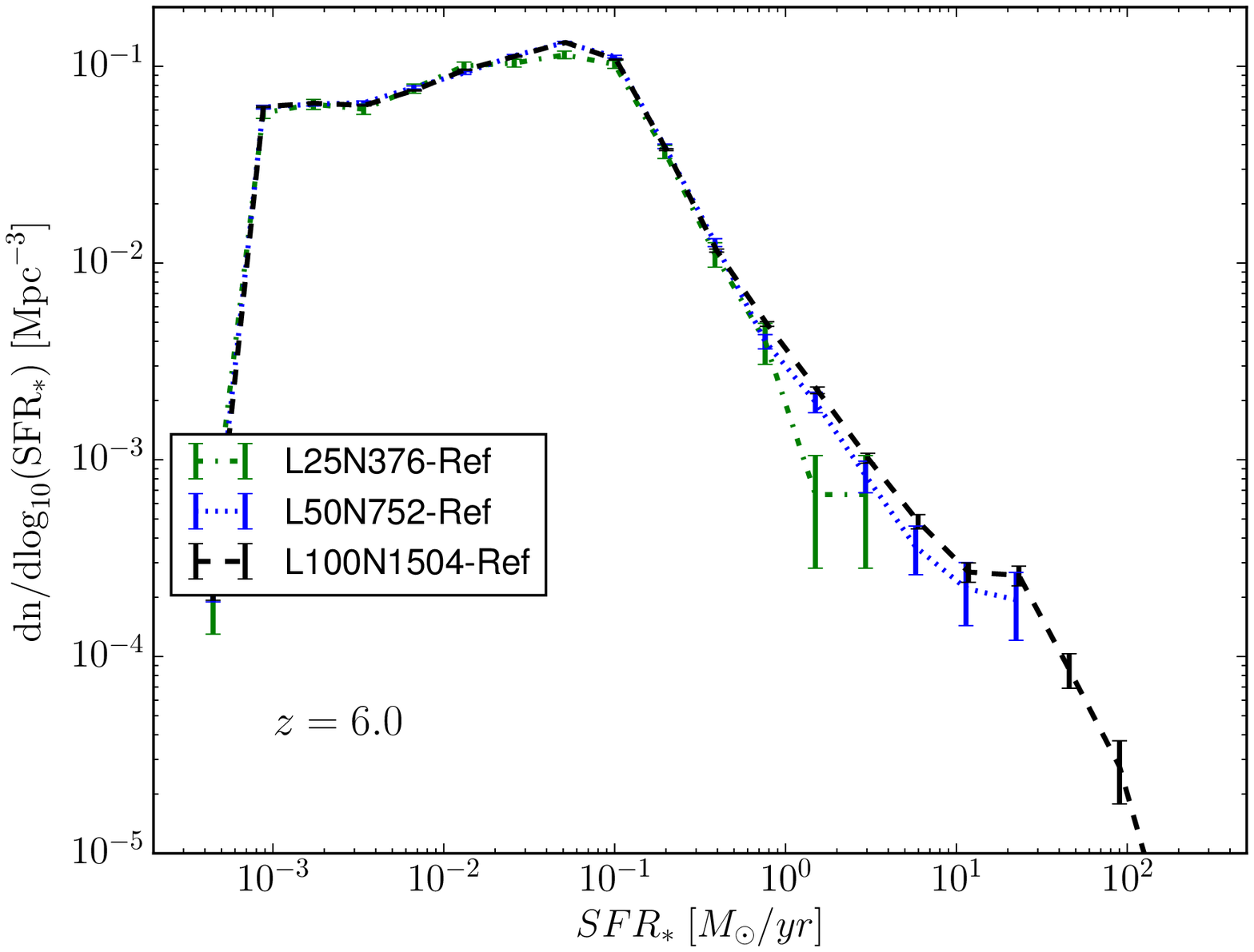}
\caption{To illustrate convergence as the simulation volume is varied, the Reference model at intermediate resolution is shown in volumes of $L = 25$, $50$ and $100$ Mpc boxes for redshifts z $\sim0$ (top) and z $\sim6$ (bottom). The 2 configurations with the largest volumes give better statistics at the high star-forming end, while the run with the smallest box size is unable to sample active galaxies with $SFR \ge 10 \, {\rm M}_{\rm \odot} \, {\rm yr^{-1}}$. Otherwise there are no differences besides the huge differences in box size.}
\label{fig:SFRFRes}
\end{figure}

In this Appendix we present the resolution and boxsize effects on the star formation rate function. In Fig. \ref{fig:SFRFRes} we compare the SFRFs of the L25N376-Ref, L50N752-Ref and L100N1504-Ref simulations. The three different runs employ the same subgrid parameters, feedback prescriptions and identical resolution. However, the boxsize is changed by a factor of 8 and 64 with respect the L25N376-Ref run. We see that the different configurations are in excellent agreement with each other and illustrate convergence as the simulation volume is varied both at redshifts $z=0$ and $z=6$. The largest volumes can give better statistics at the high star-forming end, while the simulation with the smallest box size considered is unable to sample active galaxies with $SFR \ge 10 \, {\rm M}_{\rm \odot} \, {\rm yr^{-1}}$. Otherwise there are no differences in the three distributions besides the huge differences in box size. The above, point to the direction that simulations with representative cosmological volumes (e.g. 25 Mpc) can produce similar populations of galaxies with state-of-the-art cosmologcial hydrodynamic simulations which employ significantly larger box-sizes (e.g. 100 Mpc), provided that the resolutions and subgrid physics are similar \citep{Katsianis2015}.

\begin{figure}
\centering
\includegraphics[scale=0.40]{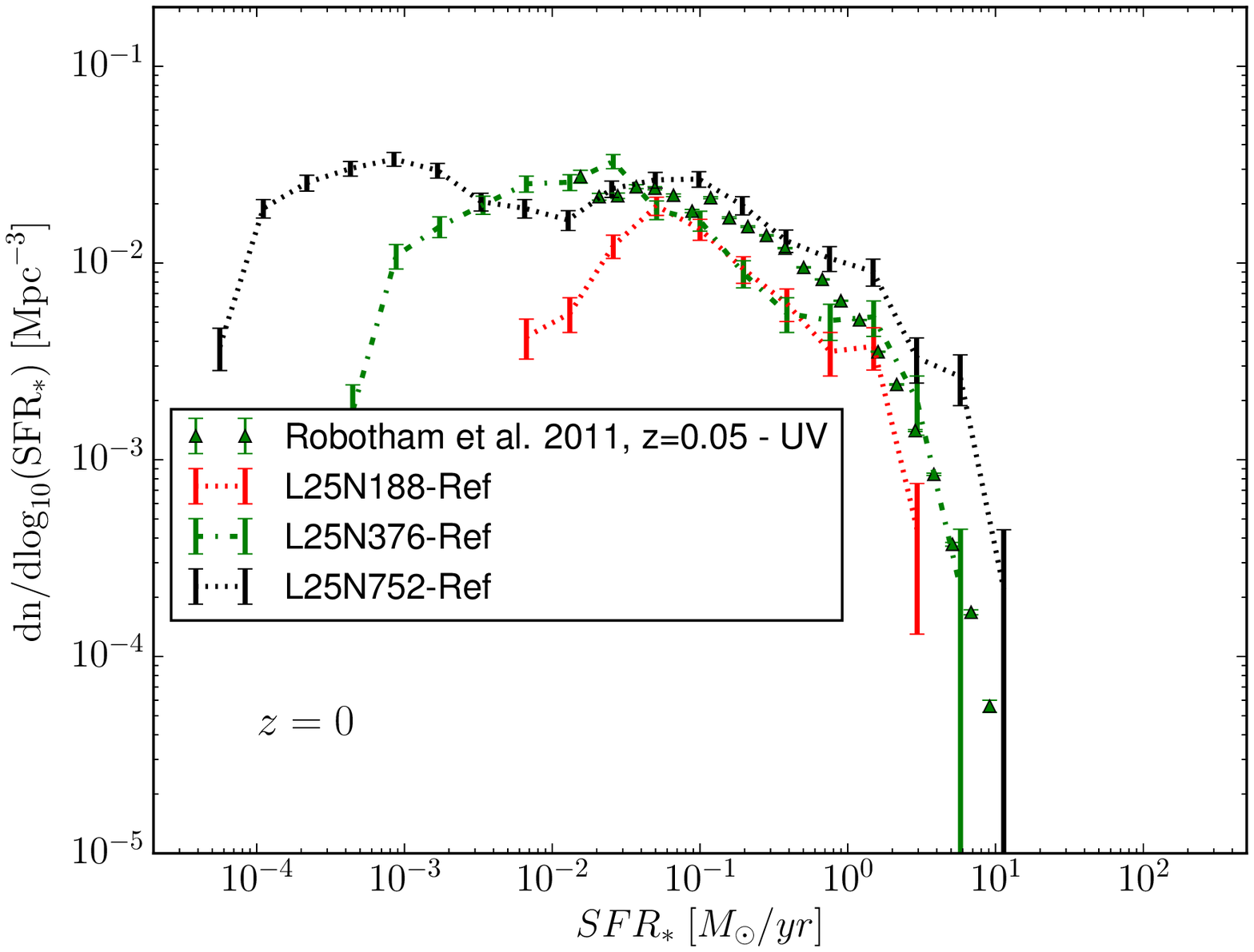}
\includegraphics[scale=0.40]{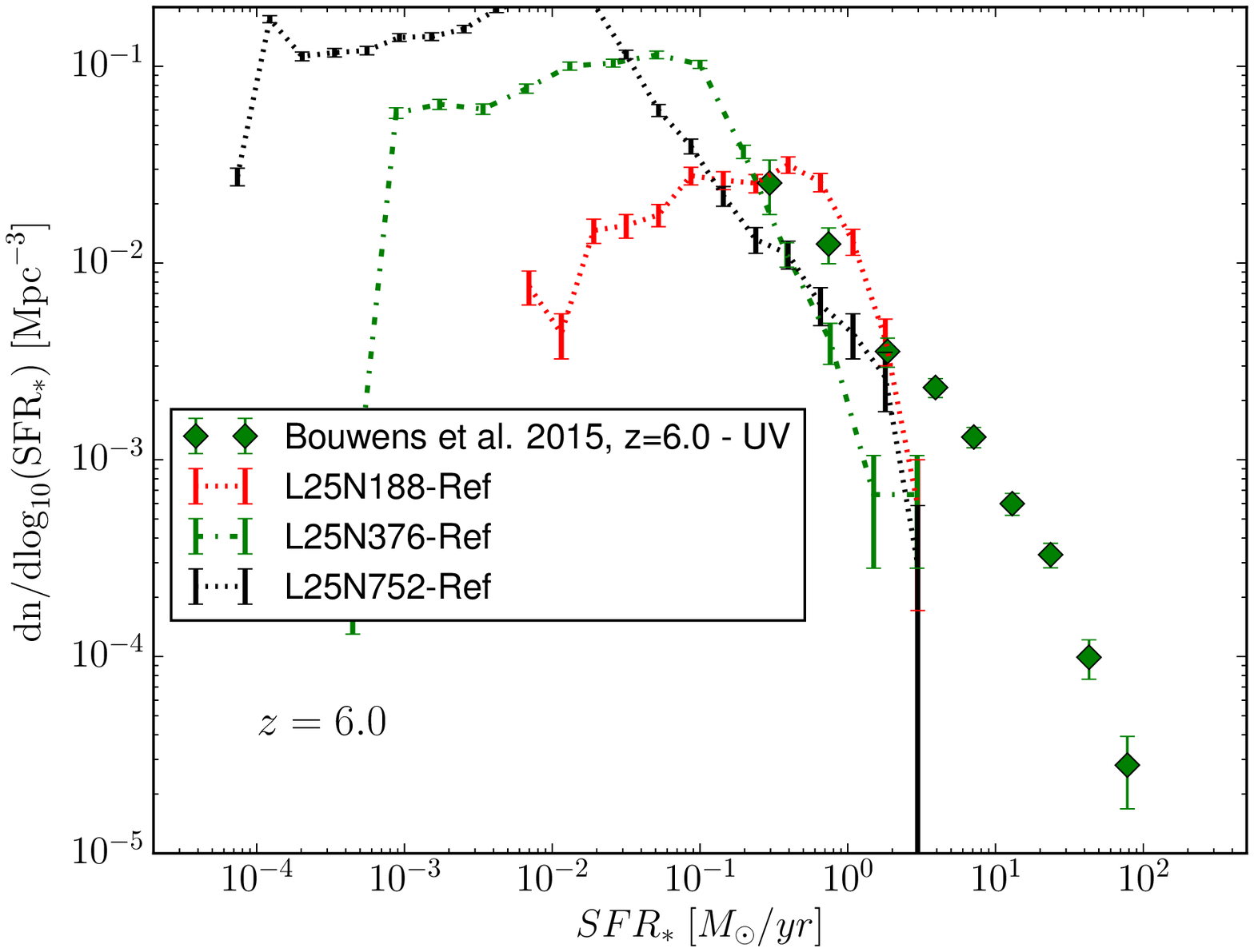}
\caption{Resolution tests for redshifts z $\sim0$ (top) and z $\sim6$ (bottom) where the Box size is kept fixed while the number of particles is varied. The mass and spatial resolution differ by factors of 8 and 64, respectively for the L25N376 and L25N752, with regard the lowest resolution run L25N188-Ref. The simulated SFRFs do not converge, since the parameters for subgrid feedback are kept the same ('strong convergence test') besides the large differences in resolution. We perform a weak convervence test in Fig. \ref{fig:SFRFRes2}.}
\label{fig:SFRFRes1}
\end{figure}

In Fig. \ref{fig:SFRFRes1} we present a comparison between the L25N188-Ref, L25N376-Ref and L25N752-Ref runs to investigate the effect of changing the resolution in the EAGLE reference model. In the last two configurations the mass and spatial resolution differ by factors of 8 and 64, respectively, with regard the lowest resolution run L25N188-Ref.  \citet{Schaye2015} and \citet{Crain2015} argued that hydrodynamical simulations such as EAGLE should recalibrate the efficiency of the subgrid feedback when the resolution is changed substantially since keeping the parameters the same does not guarantee that the physical  models  remain unchanged.  In section \ref{Feed} we showed that the efficiencies of SNe and AGN feedback are not predicted from first principles and are actually constrained from observations (e.g. GSMF at $z \sim0)$. Thus, in the case that the resolution changes substantially a recalibration is desired and that is the reason that L25N188-Ref, L25N376-Ref and L25N752-Ref runs may not converge.

\begin{figure*}
\centering
\includegraphics[scale=0.40]{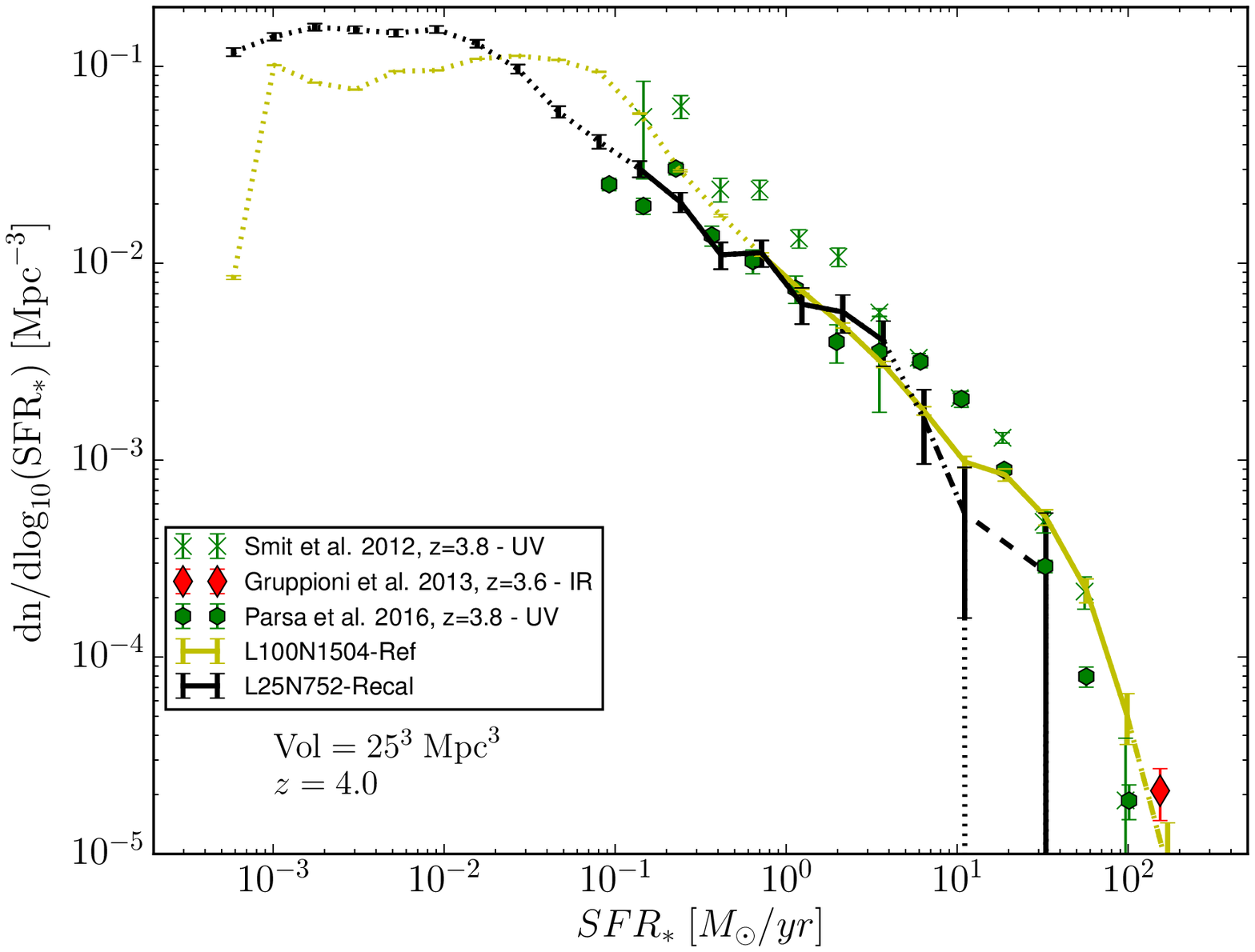}
\includegraphics[scale=0.40]{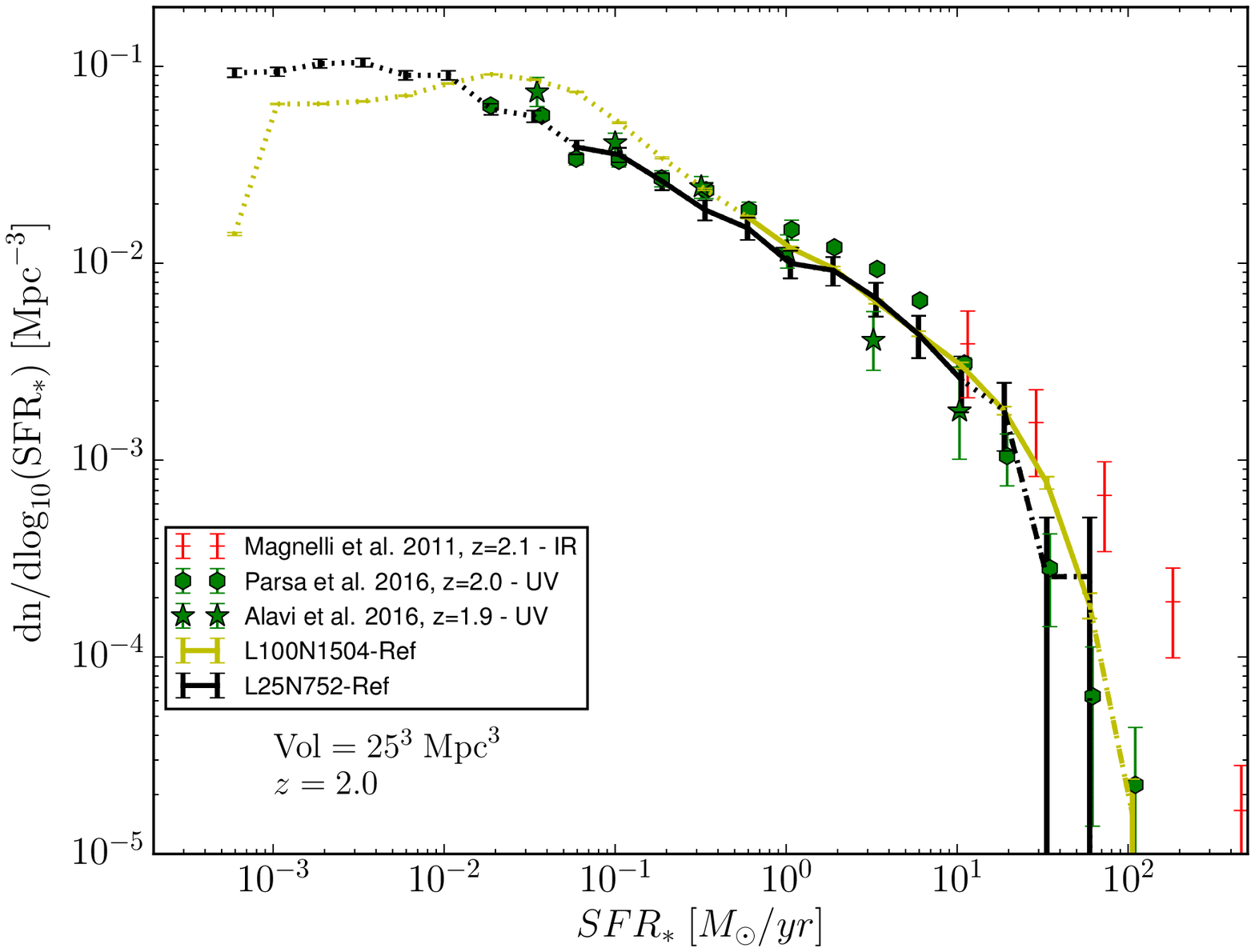}
\includegraphics[scale=0.40]{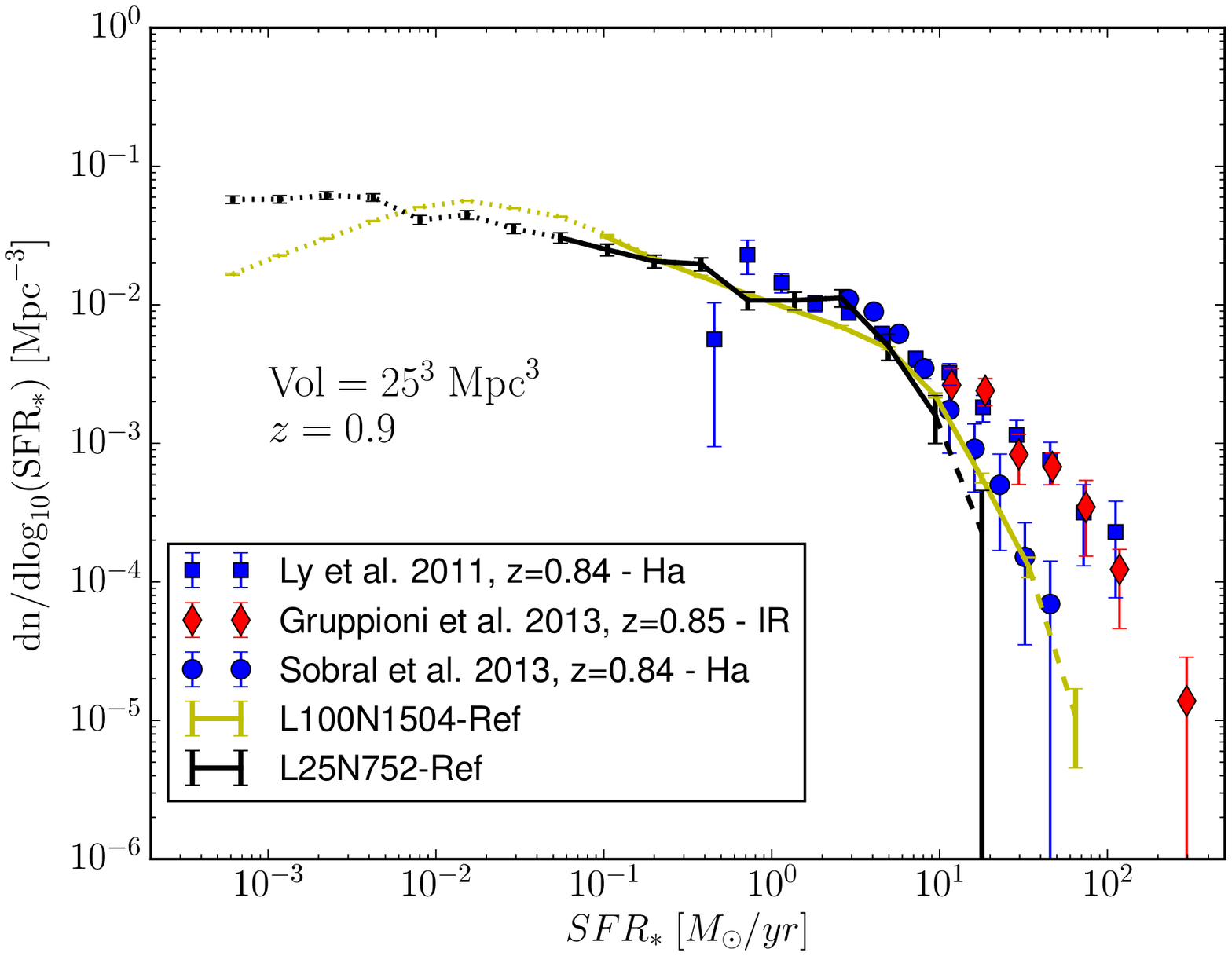}
\includegraphics[scale=0.40]{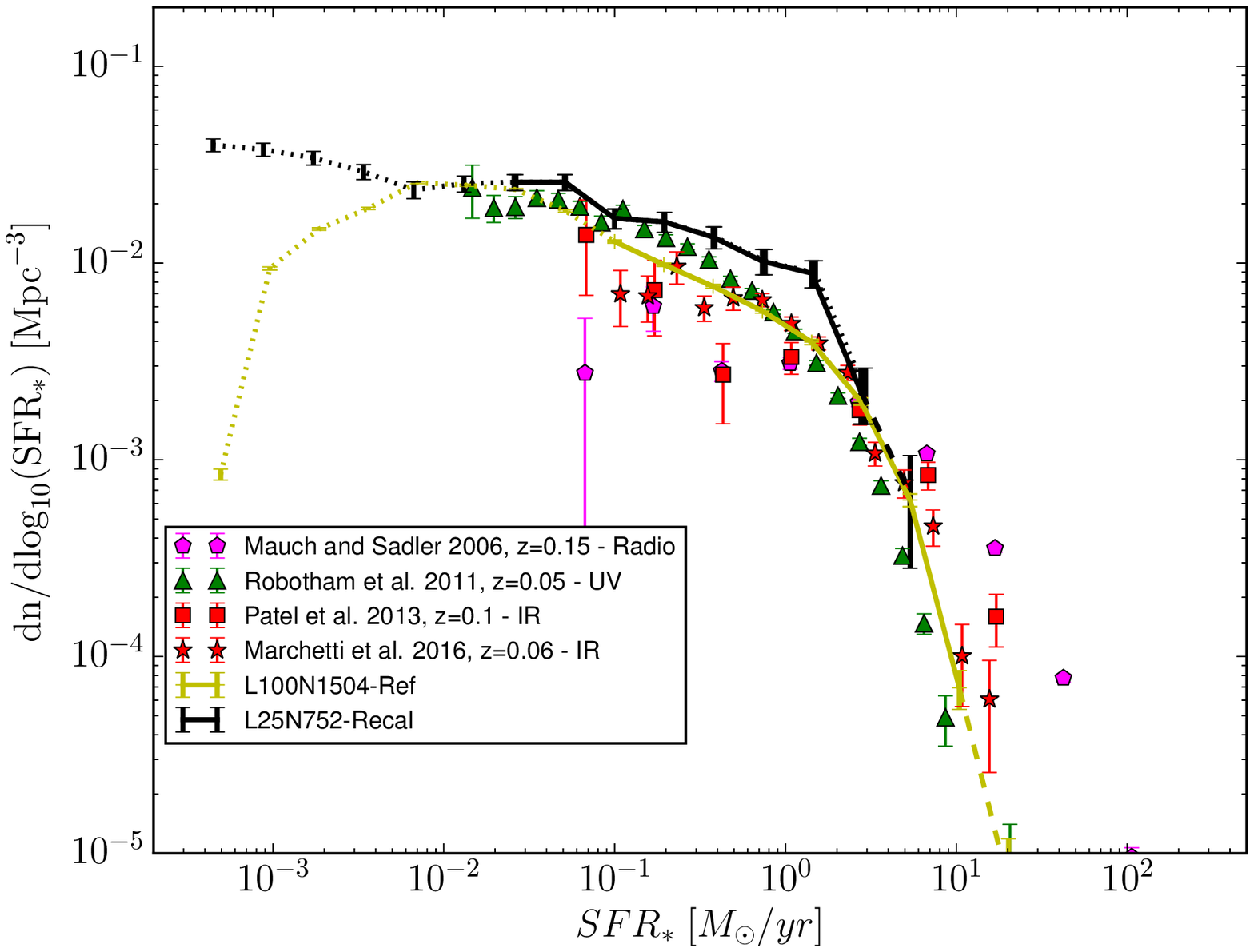}
\caption{We present the weak convergence of the EAGLE simulated SFRF by comparing the L100N1504-Ref and L25N752-Recal runs at redshifts $z \sim 4$, $z \sim 2$, $z \sim0.85$ and $z \sim 0$. The  L25N752-Recal configuration has 8 times higher resolution than the reference model so a recalibration of its subgrid physics is required to replicate the observed stellar mass function. When a bin of the EAGLE SFRF contains objects with stellar masses below the mass limit of 100 baryonic particles curves are dotted, when there are fewer than 10 galaxies curves are dashed.}
\label{fig:SFRFRes2}
\end{figure*}

\citet{Schaye2015}  introduced  the  terminology of 'weak convergence' to  describe the consistency of simulation outcomes in the case at which subgrid parameters are recalibrated when the resolution is changed, as opposed to the 'strong convergence' at which we hold the parameters fixed. To test if the simulations can fulfil the weak convergence test the L25N752-Recal simulation was run. In this simulation the following parameters for the feedback were changed with respect the reference model:
\begin{itemize}
\item  Temperature increment from the AGN ${\rm \Delta T_{AGN} =  10^{9}}$ K instead of ${\rm \Delta T_{AGN} =  10^{8.5}}$, \footnote{Increasing the temperature increment helps to suppress the increase in the cooling losses that would otherwise occur due to the higher gas densities that are resolved in the higher resolution model. Without this change the AGN feedback would be insufficient effective.}
\item subgrid BH viscosity parameter ${\rm C_{visc} = 2 \pi \times 10^3}$ instead of ${\rm C_{visc} = 2 \pi}$, \footnote{L25N752-Recal uses a different value for the parameter that controls the importance of angular momentum in suppressing accretion onto BHs, making the accretion rate more sensitive to the angular momentum of  the  accreting  gas.  Without  this  change,  AGN  feedback would  become  important  at  too  low  masses.}
\item ${\rm n_{H,0}=0.25 \, cm^{-3}}$ instead of ${\rm n_{H,0}=0.67 \, cm^{-3}}$ (eq. \ref{fth}), 
\item power-law exponent for the density term ${\rm n_n = 1/ln10}$ instead of ${\rm n_n = 2/ln10}$.\footnote{The mean values of the efficiency of the SNe feedback, ${\rm f_{th}}$, is almost the same in the Ref (1.07) and Recal models (1.06).}

\end{itemize}

The parameters of the subgrid models for feedback from star formation and for gas accretion onto BHs were recalibrated in order to reproduce the observed  $z \sim 0$ GSMF. In Figure \ref{fig:SFRFRes2} we present the weak conversion test for the SFRF and compare the L25N752-Recal and L100N1504-Ref configurations for $z \sim4$, $z \sim2$, $z \sim 0.85$ and $z \sim 0$. Both runs are in agreement with observations and the weak convergence is fulfilled. However, the recalibrated model at redshift $z \sim0$ overproduces the number density of objects with $SFR \sim 1 - 5 \, {\rm M}_{\rm \odot} \, {\rm yr^{-1}}$. The recalibration was performed to match the stellar mass function at $z \sim 0$ and this is reflected in the good convergence of L25N752-Recal and L100N1504-Ref at higher redshifts.

\label{lastpage}
\end{document}